\documentclass[a4paper,12pt]{article}
 \usepackage{jheppub}
 \usepackage[dvipsnames]{xcolor}
 \usepackage[T1]{fontenc}
 \usepackage{amsthm,amsmath,latexsym,amssymb,amsfonts,amscd}
 \usepackage{mathtools,commath}
 \usepackage{graphics,lscape,fancyhdr,array,euscript}
 \usepackage{simplewick}
 \usepackage{empheq}
 
 \usepackage[compat=1.1.0]{tikz-feynman}
 \usepackage{tikz}
 \usetikzlibrary{calc,decorations.markings}
 \usepackage{tikz-cd}
 
 \usepackage{empheq}
 \numberwithin{equation}{section} 
 \usepackage{hyperref}
 \hypersetup{colorlinks=true,linkcolor=magenta,anchorcolor=green,citecolor=cyan,filecolor=black,menucolor=black,urlcolor=brown}
 \makeatletter
 
 \pgfdeclarelayer{bg}    
 \pgfsetlayers{bg,main}  

 \usepackage{float}
\usepackage{dutchcal}
 \allowdisplaybreaks
 \definecolor{linkblue}{rgb}{0.1,0.3,.7}
 \definecolor{forestgreen(web)}{rgb}{0.13, 0.55, 0.13}
 \definecolor{lava}{rgb}{0.81, 0.06, 0.13}
 \hypersetup{
 	breaklinks,
 	colorlinks,
 	citecolor=forestgreen(web),
 	filecolor=linkblue,
 	linkcolor=lava,
 	urlcolor=linkblue
 }

 \newcommand{\besubeqs}{\begin{subequations}}%
 	\newcommand{\esubeqs}{\end{subequations}}

 \def\Li#1(#2){\textrm{Li}_{#1}\left(#2\right)}
 \def\cLi_#1(#2){\mathcal{L}_{#1}\left(#2\right)}
 \def\bLi_#1(#2){\mathbf{L}_{#1}\left(#2\right)}

 \newcommand{\dd}{\mathrm{d}}
 \newcommand{\eul}{\mathrm{e}}
 
 \newcommand{\Op}{\mathcal{O}}
 \newcommand{\zetab}{\bar{\zeta}}
 \newcommand{\cH}{\mathcal{H}}
 \newcommand{\cW}{\mathcal{W}}
 \newcommand{\ket}[1]{\left\vert#1\right\rangle}

 \newcommand{\melement}[3]{\left\langle #1\left\vert #2\right\vert #3\right\rangle}
 \newcommand{\expect}[1]{\left\langle #1\right\rangle}

 \def\u1{u_1}
 
 \def\|#1|{\norm{#1}}
 \def\Mpl#1#2{\textrm{Li}_{#1}\left(#2\right)}

 \preprint{\vbox{\hbox{\hphantom{XX }IPhT-t22/02}\hbox{LMU-ASC 13/22}}}
 
 \title{Analytical evaluation of cosmological correlation functions
 }
 
 \author[a]{Till Heckelbacher,}
 \author[a]{Ivo Sachs,}
 \author[b,c]{Evgeny Skvortsov}
 \author[d]{and Pierre Vanhove}
 
 \affiliation[a]{Arnold-Sommerfeld-Center for Theoretical Physics, Ludwig-Maximilians-Universit\"at \\M\"unchen,
 	Theresienstr. 37, D-80333 Munich, Germany}
 \affiliation[b]{Service de Physique de l'{}Univers, Champs et Gravitation, Universit\'e de Mons, 20 place du Parc, 7000 Mons, Belgium}
 \affiliation[c]{Lebedev Institute of Physics, Leninsky ave. 53, 119991 Moscow, Russia}
 \affiliation[d]{Institut de Physique Theorique, Universit\'e Paris-Saclay, CEA, CNRS, F-91191 Gif-sur-Yvette Cedex, France}
 
\abstract{ 
Using the Schwinger-Keldysh-formalism, reformulated in~\cite{DiPietro:2021sjt} as an
effective field theory in Euclidean anti-de Sitter, we evaluate the one-loop
cosmological four-point function of a conformally coupled interacting
scalar field in de Sitter. Recasting the Witten cosmological
correlator as flat space Feynman integrals, we evaluate the  one-loop cosmological four-point
functions  in de Sitter space  in terms of single-valued multiple polylogarithms. 
From it we derive anomalous dimensions and OPE coefficients of the dual conformal field theory at space-like, future infinity. In particular, we find an interesting degeneracy in the anomalous dimensions relating operators of neighboring spins. }
 
\begin{document}
\maketitle
\flushbottom

\newpage

\section{Introduction}
\label{sec:intro}
De Sitter space-time (dS) is arguably the most relevant and, at the same time, simple model for the early, and late time evolution of the Universe in a cosmological setting. It is a maximally symmetric solution of the Einstein equations with a positive cosmological constant, hence experiencing accelerated expansion. Overwhelming observational evidence points to the fact that in the distant past our universe went through a phase of accelerated expansion called inflation, while the asymptotic future seems to be described by an accelerated expansion as well. Both of these scenarios may be approximately described by a de Sitter space-time.

Furthermore, to explain the spectrum of density fluctuations in the
cosmic microwave background (CMB)~\cite{Mukhanov:1981xt} and structure
formation in the universe, which originate from the early stage of the
universe, it is important to understand quantum field theory in this
background. Nevertheless, despite its relevance, this topic is much
less developed than quantum field theory in Anti-de Sitter space-time
(AdS) let alone Minkowski space-time. This is mainly due to conceptual
and technical difficulties since dS, in contrast to AdS, does not
posses a globally defined time-like Killing vector, which makes the
choice of a vacuum more ambiguous and the definition of an asymptotic
region, relevant for scattering experiments, much more challenging. 

In this paper we would like to advance the study of QFT in dS by
calculating the one-loop corrections to the cosmological correlation
function of a conformally coupled real scalar field with a quartic
self-interaction. One of our main motivations is to make sense of the
notion of holography in the cosmological context. Similar to AdS, one
can define a conformal boundary for dS which, however, is given by a
space-like surface at future infinity, in contrast to AdS. It is therefore
not possible to fix boundary conditions in the same way as in AdS
since this is incompatible with unitary time evolution. Nevertheless,
one expects a CFT description of the bulk theory in dS on the boundary
since the symmetry group of dS acts on the future boundary as the
euclidean conformal group. 

There have been many attempts to implement the concept of holography
in dS, starting with~\cite{Strominger:2001pn}. Most of them focus on
the calculation of the wave function of the
universe~\cite{Hartle:1983ai} for the Bunch-Davies
vacuum~\cite{Maldacena:2002vr}. In this case, there is a
straightforward relation to the situation in AdS. Calculations of the expansion coefficients to the wave
function have been pushed forward recently, using direct integration, unitarity methods, Mellin
space, differential representations and
polytopes~\cite{Hertog:2011ky,Harlow:2011ke,Anninos:2014lwa,Gorbenko:2019rza,Heckelbacher:2020nue,Meltzer:2021zin,Pajer:2020wnj,Goodhew:2021oqg,Jazayeri:2021fvk,Goodhew:2020hob,Gomez:2021qfd,Gomez:2021ujt,Arkani-Hamed:2017fdk}. As
the wave function itself is, however, not an observable these results
are more of a conceptual rather than phenomenological value. In
principle one could obtain a cosmological correlation function
by taking expectation values from this wave functional,
but this approach is impractical in reality since it requires non
perturbative knowledge of the wave function which, for interacting
theories, is technically out of reach at the moment, at least to our knowledge. 

Another approach, which we will follow in this work, is to
evaluate the correlation function directly by performing a path
integral along a closed time contour, the so called Schwinger-Keldysh
or in-in formalism~\cite{Schwinger:1960qe,Keldysh:1964ud}. This
approach has lead to several interesting results and the development of
new
techniques~\cite{Weinberg:2005vy,Akhmedov:2013vka,Arkani-Hamed:2015bza,Arkani-Hamed:2018kmz,Hogervorst:2021uvp,Sleight:2019hfp,Sleight:2020obc,Sleight:2021plv,DiPietro:2021sjt,Fichet:2021xfn}. We
are going to take advantage of progress made
in~\cite{Sleight:2019hfp,Sleight:2020obc,Sleight:2021plv} to express the cosmological correlation functions in the Schwinger-Keldysh formalism as a sum over euclidean AdS (EAdS) Witten diagrams which was expressed in \cite{DiPietro:2021sjt} as an auxiliary EAdS
action. Here we calculate the four-point functions up to one-loop order by direct integration in position space,
applying the formalism developed in~\cite{Heckelbacher:2022fbx} to
evaluate EAdS Witten diagrams. Interestingly, the Witten
diagrams up to this order do not contain any elliptic integrals, in
contrast to EAdS, and therefore can all be expressed in terms of
single-valued multiple polylogarithms. We then compare the late time cosmological correlator to the conformal block expansion which allows us to extract the data of the dual CFT. 

\begin{figure}[h]
    \centering
    \begin{tikzcd}[column sep=tiny]
 \boxed{\text{Cosmological correlator CFT}} & & \textrm{CFT}(\Op^S_{n,l},\Op^A_{n,l},[\Op_1\Op_2]_{n,l}) &  \\
  & & \overbrace{\textrm{GFF}(\Op_1)\times\textrm{GFF}(\Op_2)} \ar[u] \ar[dl] \ar[dr] &  \\
  \boxed{\text{Wave function CFT}} & \textrm{CFT}_{\Psi[\pi]} & \times & \textrm{CFT}_{\Psi[\phi]} 
\end{tikzcd}
    \caption{Deformations of the generalized free field (GFF) CFTs in the wave function CFTs (down) and cosmological correlator CFT (up). $\Op^S_{n,l}$ and $\Op^A_{n,l}$ are orthogonal linear combinations of $[\Op_1\Op_1]_{n,l}$ and $[\Op_2\Op_2]_{n,l}$ (see section~\ref{sec:correl}).}
    \label{fig:CFT_deformations}
\end{figure}
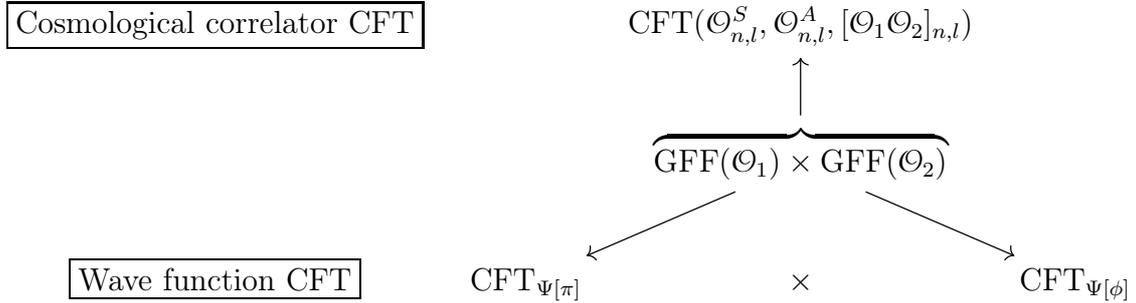

We find that the CFT is given by a deformation of a direct product of generalized free fields. However, in contrast to the CFT corresponding to the expansion of the wave function, the cosmological CFT contains three different trajectories of double trace operators due to the mixing of fields with different boundary conditions pictured in  figure~\ref{fig:CFT_deformations}.

We find that the cosmological correlators obey several CFT consistency
conditions at different loop orders reflecting the fact the boundary
theory is in fact a CFT. The
second order (one-loop) anomalous dimensions
for the double trace operators $:\Op_1\Box^n\partial^l\Op_1:$,
$:\Op_2\Box^n\partial^l\Op_2:$ and $:\Op_1\Box^n\partial^l\Op_2:$,
derived in  section~\ref{sec:CBE}  for all $n$ and $l$,
\begin{eqnarray}
    \gamma^{(2)S}_{n>0,l>0}&=-\frac{\gamma^2}{l(l+1)}; \qquad \gamma^{(2)A}_{n>0,l>0}&=-\frac{\gamma^2}{(2n+l)(2n+l+1)}\nonumber\\*
    \gamma^{(2)}_{n,2l>0}&=\gamma^{(2)S}_{n,2l>0};\qquad
    \gamma^{(2)}_{n,2l+1>0}&=\gamma^{(2)A}_{n,2l+2}\,,
\end{eqnarray}
highlight an interesting symmetry between the anomalous dimensions at different spins.
From the bulk perspective, this is could be a consequence of the symmetry in the EAdS action in eq.~\eqref{eq:EAdS_action_d3}, enforced by the Schwinger-Keldysh formalism and the fact that we take a conformally coupled scalar field. We do not expect this symmetry to hold for general masses. The equations for $\gamma^{(2)S}_{n>0,l>0}$ and $\gamma^{(2)}_{n,l>0\text{ even}}$ show a degeneracy for the conformal dimensions of these operators for all twists $\Delta_{n,l}-l$, which seems quite remarkable.

This paper is organised as follows: In section~\ref{sec:pert_QFT} we briefly review the Schwinger-Keldysh formalism in the context of QFT in dS, define the propagators and give the auxiliary EAdS action first derived in~\cite{DiPietro:2021sjt}. Section~\ref{sec:Correlation_function_calculation} is where we present the calculation of the cosmological correlation function in terms of EAdS Witten diagrams and in section~\ref{sec:CBE} we compare the results to a conformal block expansion on the boundary and extract anomalous dimensions. We conclude in section~\ref{sec:outlook} with a short summary of the results and some suggestions for further investigation. The expression for the Witten diagrams are collected in the appendix~\ref{sec:cross_app} for the cross diagram and appendix~\ref{sec:oneloop_appendix} for the one-loop diagram. The single-valued multiple polylogarithms entering these evaluations are collected in the appendix~\ref{sec:recurring_exp}. The OPE coefficients are conformal blocks for generalized field  are recalled in appendix~\ref{sec:OPE_CB}.

\section{Perturbative QFT in de Sitter space}
\label{sec:pert_QFT}

The main reason why Quantum field theory in dS is less straightforward than in AdS is the fact that it does not have a globally defined time-like Killing vector in all patches relevant for cosmology, leading to a time-dependent classical background. In this section we will describe how to deal with this issue.

\subsection{Schwinger-Keldysh formalism in de Sitter space}
\label{sec:schwinger_keldysh}

\begin{figure}[h]
	\centering
	\begin{tikzpicture}
		\draw (-2,0) to 
		node[midway, below] {$\mathcal{J}^-$}(2,0) 
		to (2,4) to node[midway, above] {$\mathcal{J}^+$} node[pos=.55, below]{$\eta=0$} (-2,4) to  (-2,0);
		\draw (-2,0) to node[midway,below,sloped] {$\eta=-\infty$}(2,4);
		
		\draw[dashed] (-2,3) to [out=0, in=190,looseness=0.6](2,4);
		\draw[dashed] (-2,2) to [out=0, in=200,looseness=0.6] node[at start, left]{$\eta=$const}(2,4);
		\draw[dashed] (-2,1) to [out=0, in=210,looseness=0.6](2,4);
	\end{tikzpicture}
	\caption{Conformal diagram of dS in the Poincar\'e patch.}
	\label{fig:conformal_diagram_dS}
\end{figure}
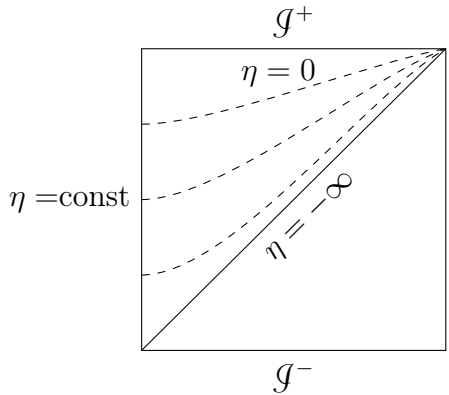

To calculate expectations values of a time dependent background the Schwinger-Keldysh formalism is well suited. For this one specifies the initial vacuum and calculates the expectation value of local field insertions $\phi(X_1)\cdots\phi(X_n)$. Here we will work on the Poincar\'e patch parametrized by coordinates $X=(\vec x,\eta)$ as in figure~\ref{fig:conformal_diagram_dS}, which is given by the lower half space
\begin{equation}
    \mathcal{H}^-_{d+1}:=\{X=(\vec x,\eta):\vec x\in\mathbb{R}^d,\eta<0\}
\end{equation}
equipped with the metric
\begin{equation}
    \dd s^2=\frac{1}{a^2\eta^2}(-\dd\eta^2+\dd\vec x^2)\,.
\end{equation}

In the interaction picture we then have 
\begin{equation}
	\label{eq:correlator_time_evolution}
	\expect{\phi(\vec x_1,\eta)\cdots\phi(\vec x_n,\eta)}_{BD}=\frac{\melement{0_{BD}}{U^\dagger_I(-\infty,\eta)\phi(\vec x_1,\eta)\cdots\phi(\vec x_n,\eta)U_I(-\infty,\eta)}{0_{BD}}}{\melement{0_{BD}}{U^\dagger_I(-\infty,\eta)U_I(-\infty,\eta)}{0_{BD}}}\,.
\end{equation}
Here $\ket{0_{BD}}$ is the Bunch-Davies vacuum to be defined below, while $U_I$ and $U_I^\dagger$ are the time-ordered and anti-time ordered evolution operator in the interaction picture given by
\begin{equation}
	U_I(\eta_0,\eta):=T\left\{\eul^{-i \int_{\eta_0}^{\eta}\dd\tilde{\eta}H_I(\tilde\eta)}\right\},\qquad
	U^\dagger_I(\eta_0,\eta):=\bar T\left\{\eul^{i \int_{\eta_0}^{\eta}\dd\tilde{\eta}H_I(\tilde\eta)}\right\}\,,
\end{equation}
where $H_I$ is the interaction Hamiltonian and $T$ and $\bar T$ denote time- and anti-time ordering respectively. The Bunch-Davies vacuum condition is imposed at $\eta\to-\infty$. The denominator in equation~\eqref{eq:correlator_time_evolution} cancels vacuum bubble contributions, just as in flat space.

There are two ways to perform this calculation. One is to expand the exponentials in $U_I$ and $U_I^\dagger$ and use Wick contraction on the left and right of the insertions to calculate the correlator.  Denoting the fields on the time ordered side of the integral by $\phi_T(X)$, the anti-time order fields by $\phi_A(X)$ and the field insertions on the time slice at future infinity by $\bar{\phi}(\vec{x})$, we have the Wick contractions
\begin{equation}
	\contraction[2ex]{}{\phi_{T/A}}{(X_1)}{\phi_{T/A}} \phi_{T/A}(X_1)\phi_{T/A}(X_2)\to \Lambda_{T/A,T/A}( X_1, X_2)\,,
\end{equation}
where \begin{equation}
    \Lambda_{TT}(X_1,X_2)=\melement{0}{ T\{\phi(X_1)\phi(X_2)\}}{0},\;
    \Lambda_{AA}(X_1,X_2)=\melement{0}{\bar T\{\phi(X_1)\phi(X_2)\}}{0}
\end{equation} 
are time- and anti time-ordered correlators, while $\Lambda_{TA}(X_1,X_2)$ and $\Lambda_{AT}(X_1,X_2)$ are the retarded and advanced Green functions respectively. Similarly,
\begin{equation}
    \contraction[2ex]{}{\phi_{T/A}}{(X_1)}{\bar\phi} \phi_{T/A}(X_1)\bar\phi(\vec x_2)\to \bar\Lambda_{T/A}(X_1,\vec x_2)\,,
\end{equation}
where $\bar\Lambda_{T/A}(X_1,\vec x_2)$ is obtained from $\Lambda_{TT}$ or $\Lambda_{AA}$ by taking $X_2\to \vec x_2$ to the  future space-like conformal boundary of dS.

A free massive scalar field in dS  evolves according to the Klein-Gordon equation
\begin{equation}
	\label{eq:KleinGordon_dS}
	(-\square_{dS}+m^2)\phi=0\,,
\end{equation}
where the d'Alembertian is related to the quadratic Casimir of the $SO(d+1,1)$ isometry group of dS as  $C_2=-\frac{1}{a^2}\square_{dS}$. Comparing this with the weight $\Delta$ and spin $\ell$ representations of the conformal group in $d$ Euclidean dimensions with 
\begin{align}
	\label{eq:casimir_dS}
	C_2=\Delta(\Delta-d)+\ell(\ell+d-2),
\end{align}
we recover the familiar relation between
the mass of a scalar field and the scaling dimension on the boundary
\begin{equation}
	\label{eq:scaling_dim_dS}
	\Delta(\Delta-d)=-\frac{m^2}{a^2}\quad \Longleftrightarrow \quad \Delta_\pm =\frac{d}{2}\pm\sqrt{\frac{d^2}{4}-\frac{m^2}{a^2}}\,.
\end{equation}
These equations are invariant under the shadow transformation ${\Delta\to d-\Delta}$, which relates two unitarily equivalent representations.

We can label the irreducible representations by the spin of the
$SO(d)$ part. We have a Lorentzian field
theory and the states appearing should therefore correspond to unitary
representations of the symmetry group $SO(d+1,1)$. 
The scaling dimension $\Delta$, which can take complex values
restricted by unitarity, is
restricted to fall into different classes. The most relevant ones to our analysis are the principal and complementary series.

The equations of motion guarantee, that any free field transforms in a unitary irreducible representation of the de Sitter group.
Heavy fields with mass $4m^2>a^2d^2$ in dS  correspond to the principal
series which exists for any spin $\ell$. They have a complex valued scaling dimension 
\begin{equation}
	\label{eq:principal_series}
	\Delta = \frac{d}{2}+i\nu \text{ with }\nu\in\mathbb{R}.
\end{equation}
Light fields in dS with mass $0\leq 4m^2<a^2d^2$ correspond to the
complementary series given by the real valued dimension
\begin{equation}
	\label{eq:complementary_series}
	\Delta=\frac{d}{2}+\nu \text{ with }\nu\in\mathbb{R},
\end{equation}
where $-\frac{d}{2}<\nu <\frac{d}{2}$ for $\ell=0$ and $1-\frac{d}{2}<\nu<\frac{d}{2}-1$ for $\ell>0$. As we will discuss later this class of representations will be most relevant to us, since we will consider a conformally coupled field. For more details on this topic we refer the interested reader to~\cite{Dobrev:1976vr}.

EAdS can be constructed from the same ambient Minkowski space as dS with
the same signature of the metric. Therefore, we could conclude that
the Hilbert space of EAdS should be constructed from unitary
irreducible representations of $SO(d+1,1)$. But  this is well-known
not to be the case.
The scaling dimension for EAdS can be obtained by setting $a\to i a$
in equation~\eqref{eq:scaling_dim_dS}. The value for $\Delta$ is
therefore always real and the fields transform
under unitary irreducible representations of 
$SO(d,2)$ the symmetry group of the Lorentzian version of EAdS.
This is not a problem since QFT in EAdS is a euclidean field theory. Only after Wick rotation to Lorentzian AdS the Hilbert space should be given by unitary representations which it clearly does.

The situation for dS is different. In  the four-point function that we analyse in our perturbative calculation, we will see that there are operators appearing in the spectrum with arbitrary dimensions not obeying any $SO(d+1,1)$ unitarity constraints. However, since there is no operator state correspondence in dS, this does not really pose a problem, it just hints at the fact that the relation between the bulk and boundary degrees of freedom is more obscure in dS than in AdS. These points have been raised recently in the context of a proposed cosmological bootstrap in~\cite{Hogervorst:2021uvp,DiPietro:2021sjt}.

Here we will focus on the four-point function of a scalar field theory with interaction term
\begin{equation}
	\int_{-\infty}^{\eta}H_{int}d\eta=\frac{\lambda}{4!}\int_{\mathcal{H}^-_{d+1}}\frac{d^{d+1}X}{(a \eta)^{d+1}}\phi^4(X).
\end{equation}
The four-point function evaluated at future infinity is given by
\begin{equation}
	\lim\limits_{\eta\to\eta_0}\expect{\phi(\eta,\vec x_1)\phi(\eta,\vec x_2)\phi(\eta,\vec x_3)\phi(\eta,\vec x_4)}
\equiv\expect{\phi_0(\vec x_1)\phi_0(\vec x_2)\phi_0(\vec x_3)\phi_0(\vec x_4)}\,.
\end{equation}
Let us begin with the disconnected part 
\begin{multline}
	\expect{\phi_0(\vec x_1)\phi_0(\vec x_2)\phi_0(\vec x_3)\phi_0(\vec x_4)}=\expect{\phi_0(\vec x_1)\phi_0(\vec x_2)}\expect{\phi_0(\vec x_3)\phi_0(\vec x_4)}\cr+\expect{\phi_0(\vec x_1)\phi_0(\vec x_3)}\expect{\phi_0(\vec x_2)\phi_0(\vec x_4)}
	+\expect{\phi_0(\vec x_1)\phi_0(\vec x_4)}\expect{\phi_0(\vec x_2)\phi_0(\vec x_3)}\,,
\end{multline}
where each two-point function is just given by the propagator $\Lambda$ with both legs taken to future infinity.

The first order term in the coupling constant $\lambda$ has two contributions 
\begin{align}
	W_0=&-i\lambda\int_{\cH_{d+1}^-}\frac{\dd^{d+1}X}{(a\eta_T)^{d+1}}\bar\Lambda_T(\vec{x_1},X)\bar\Lambda_T(\vec{x_2},X)\bar\Lambda_T(\vec{x_3},X)\bar\Lambda_T(\vec{x_4},X)\cr
	&+i\lambda\int_{\cH_{d+1}^-}\frac{\dd^{d+1}X}{(a\eta_A)^{d+1}}\bar\Lambda_A(\vec{x_1},X)\bar\Lambda_A(\vec{x_2},X)\bar\Lambda_A(\vec{x_3},X)\bar\Lambda_A(\vec{x_4},X)\,,
\end{align}
from contractions with the time-ordered and the anti-time-ordered Hamiltonian. 
We perform this integral after a Wick rotation for $\eta_T$ and $\eta_A$ individually, such that we do not cross the branch cut,
\begin{equation}
    \label{eq:Wick_rotation}
	\eta_T\to\eul^{-i\frac{\pi}{2}}z;\qquad \eta_A\to\eul^{i\frac{\pi}{2}}z\,.
\end{equation}
With this transformation we can write the cross diagram as
\begin{align}
    \label{eq:four_point_rotated}
	W_0&=-\lambda\int_{\cH_{d+1}^+}\frac{\dd^{d+1}X}{(a z)^{d+1}}\bar\Lambda_T(\vec{x_1},X)\bar\Lambda_T(\vec{x_2},X)\bar\Lambda_T(\vec{x_3},X)\bar\Lambda_T(\vec{x_4},X)\cr
	&-\lambda\int_{\cH_{d+1}^+}\frac{\dd^{d+1}X}{(a z)^{d+1}}\bar\Lambda_A(\vec{x_1},X)\bar\Lambda_A(\vec{x_2},X)\bar\Lambda_A(\vec{x_3},X)\bar\Lambda_A(\vec{x_4},X)\,,
\end{align}
where now, $X:=(\vec x,z)$.

To define the propagator we consider the euclidean version of dS, which is a sphere. Upon Wick rotating back to dS and restricting to the Poincar\'e patch this fixes the vacuum as the Bunch-Davies or euclidean vacuum (see~\cite{Chernikov:1968zm,Schomblond:1976xc,Bunch:1978yq}). The propagator for a
scalar field of mass $m$ between two bulk points $X$ and
$ Y$ on the sphere reads\footnote{The Gau\ss{} hypergeometric function
  is defined as
  \begin{equation}
    {}_2F_1\left(a,b;c;z\right):={\Gamma(c)\over \Gamma(b)\Gamma(c-b)}\int_0^1 t^{b-1} (1-t)^{c-b-1} (1-zt)^{-a} dt   
  \end{equation}
  for $\Re(b)>0$ and $\Re(c)>0$.
}  
\begin{equation}
	\label{eq:sphere_green_function}
	\Lambda_S( X, Y)=\mathcal{N_\mathrm{dS}}\ 
	{}_2F_1\left(\Delta_+,\Delta_-;\frac{d+1}{2};\frac{K( X,Y)-1}{2 K( X, Y)}\right),
\end{equation}
with 
\begin{equation}
	K( X, Y)=\frac{1}{a^2 \sum_{i=0}^{d} X^i Y^i }\,
      \end{equation}
is the inverse of the geodesic distance and 
\begin{equation}
	\mathcal{N}_\mathrm{dS}=\frac{\Gamma(\Delta_+)\Gamma(\Delta_-)}{(4\pi)^{\frac{d+1}{2}}\Gamma\left(\frac{d+1}{2}\right)},
\end{equation}
is a normalization constant. The Green function in dS is obtained from equation~\eqref{eq:sphere_green_function} by Wick rotating back and restricting to the Poincar\'e patch, which we will denote by $\Lambda(K({X},{Y}))$. To obtain the correct time ordering for the Feynman propagator when taking the flat limit, we have to obtain the correct behaviour across the branch cut at $0<K( X, Y)<1$ which coincides with the region of time-like separation. We therefore demand that the commutator between two fields at space-like separation should vanish, while at time-like separation it should be non-vanishing. Expressed in terms of two-point functions of the vacuum state defined by the analytic continuation of~\eqref{eq:sphere_green_function}, this means
\begin{equation}
	\label{eq:commutator_dS}
	\melement{0}{[\phi(X),\phi(Y)]}{0}=\Lambda(K( X, Y)) - \Lambda(K( Y, X)).
\end{equation}
For this expression to be non-vanishing for time-like separation we have to demand that we approach the branch cut from above and below depending on the time-ordering. In the Poincar\'e patch this means doing the replacement $K( X,Y)\to K( X, Y)-i\varepsilon\mathrm{sgn}(\eta_x-\eta_y)$, where $\varepsilon$ is an infinitesimal, positive, real parameter.  The two-point function with the correct behaviour across the branch cut is therefore given by
\begin{equation}
	\label{eq:dS_two_point}
	\Lambda_{TA}( X, Y):=\Lambda\left(K( X, Y)-i\varepsilon\mathrm{sgn}(\eta_x-\eta_y)\right),
      \end{equation}
      with $0\leq\varepsilon\ll1$.
The time ordered Feynman two-point function is given by
\begin{align}
	\Lambda_{TT}( X,Y)&:=\melement{0}{T\{\phi(X_1)\phi(X_2)\}}{0}\cr
	&=\theta(\eta_x-\eta_y)\Lambda_{TA}( X, Y)+\theta(\eta_y-\eta_x)\Lambda_{TA}( Y, X).
\end{align}
This can be written in a more compact form replacing $K( X, Y)\to K( X, Y)+i\varepsilon$ in~\eqref{eq:sphere_green_function}, where we used that $K(X,Y)$ expressed in local Poincar\'e coordinates is given by 
\begin{equation}
    K(X,Y)=\frac{2\eta_x\eta_y}{\eta_x^2+\eta_y^2-(\vec x-\vec y)^2}\,.
\end{equation}
The time ordered Feynman Green function in dS is therefore given by
\begin{equation}
	\label{eq:dS_green_function}
	\Lambda_{TT}( X, Y)=\mathcal{N}_\mathrm{dS}\ 
	{}_2F_1\left(\Delta_+,\Delta_-;\frac{d+1}{2};\frac{K( X, Y)-1}{2 K( X, Y)}-i\varepsilon\right),
\end{equation} 
while the anti-time ordered two-point function is given by 
\begin{align}
	\Lambda_{AA}(X, Y)&:=\melement{0}{\bar T\{\phi(X_1)\phi(X_2)\}}{0},\nonumber\\*
	&=\theta(\eta_x-\eta_y)\Lambda_{TA}( Y, X)+\theta(\eta_y-\eta_x)\Lambda_{TA}( X, Y),\nonumber\\*
	&=\Lambda(K( X, Y)-i\varepsilon).\label{eq:dS_green_function_anti}
\end{align}
These Green functions define the Bunch-Davies or euclidean vacuum. Let
us mention that this is not the unique de Sitter invariant
vacuum. There is an infinite space of de Sitter invariant vacua
parametrised by two continuous parameters~\cite{Allen:1985ux}. All these vacua have singularities at points related by the
antipodal map and therefore do not provide the correct flat limit. The
 Bunch-Davies vacuum is therefore special from a physical perspective. Also, from a cosmological point of view, the Bunch-Davies vacuum seems to be the only reasonable choice, since it gives mode functions for the field that behave like in flat space when going to the infinite past or to wavelengths much smaller than the horizon. From now on we will only work in the Bunch-Davies vacuum.

Note, that, contrary to EAdS, we cannot fix the fall-off behaviour of
the Green function at future infinity to be either $\sim K^{\Delta_+}$
or $\sim K^{\Delta_-}$. This is due to the fact that we can always
rewrite the Bunch-Davies propagator as a sum of the propagators with a
definite fall-off behaviour. By applying the following identities for the hypergeometric function
\begin{equation}
  {}_2F_1\left(a,b;\frac{a+b+1}{2};z\right)=(1-2z)^{-a}{}_2F_1\left(\frac{a}{2},%
\frac{a+1}{2};\frac{a+b+1}{2};\frac{4z(z-1)}{(1-2z)^{2}}%
\right)
\end{equation}
and
\begin{multline}
{}_2F_1(a,b;c,z) = \frac{\Gamma(c)\Gamma(c-a-b)}{\Gamma(c-a)\Gamma(c-b)}{}_2F_1(a,b;a+b+1-c;1-z) \cr
{} + \frac{\Gamma(c)\Gamma(a+b-c)}{\Gamma(a)\Gamma(b)}(1-z)^{c-a-b} {}_2F_1(c-a,c-b;1+c-a-b;1-z)\,,
\end{multline}
we can rewrite the hypergeometric function in~\eqref{eq:sphere_green_function} as
\begin{multline}
    {}_2F_1\Big(\Delta_+,\Delta_-;\frac{d+1}{2};\frac{K-1}{2K}\Big)=
    \frac{\Gamma\left(\frac{d+1}{2}\right)\Gamma\left(\Delta_- -\frac{d}{2}\right)}{\Gamma\left(\frac{\Delta_-}{2}\right)\Gamma\left(\frac{\Delta_- +1}{2}\right)}K^{\Delta_+}\cr
    \times{}_2F_1\left(\frac{\Delta_+}{2},\frac{\Delta_+ +1}{2};\Delta_+-\frac{d-2}{2};K^2\right)+(\Delta_+\leftrightarrow\Delta_-)\,.
\end{multline}
With this formula we can express the time ordered Bunch-Davies propagator~\eqref{eq:dS_green_function} in terms of propagators with fall-off behaviour
\begin{multline}
	\Lambda_{TT}(X, Y)={1\over2\pi}\Gamma\left(\Delta_+-\frac{d}{2}\right)\Gamma\left(\Delta_- -\frac{d}{2}\right)\left(\Delta_+-\frac{d}{2}\right)\cr
\times	\left(\Lambda_{TT}(X, Y,\Delta_-)+\Lambda_{TT}( X, Y,\Delta_+)\right).
\end{multline}
Here we introduced the propagator with a definite fall off behaviour as the Wick rotation of the propagator in EAdS
\begin{multline}
    \label{eq:bulktoboundary_dS_sum}
	\Lambda_{TT}( X, Y,\Delta)=\frac{a^{d-1}}{4\pi^{\frac{d+1}{2}}}
	\frac{\Gamma\left(\frac{\Delta+1}{2}\right)\Gamma\left(\frac{\Delta}{2}\right)}{\Gamma\left(\Delta-\frac{d}{2}+1\right)}\cr
\times	K(X, Y)^\Delta{}_2F_1\left(\frac{\Delta}{2},\frac{\Delta +1}{2};\Delta -\frac{d-2}{2}; K( X, Y)^2-i\varepsilon\right)
\end{multline}
with equivalent expressions for $\Lambda_{AA}$ and $\Lambda_{TA}$.

Following the same conventions as for the  AdS/CFT case we introduce
the bulk-to-boundary propagator as the limit of the bulk-to-bulk
propagator  taking on leg to future infinity
\begin{equation}
  \bar\Lambda_{T/A}(X_1,\vec
  x_2):=\lim\limits_{\eta_0:=\eta_2\to0}\Lambda_{T/A,T/A}(
  X_1,X_2).
\end{equation}
It   does not matter if the boundary limit is taken with a time- or
anti- time ordered point since there is no notion of time ordering at
future infinity.
The bulk-to-boundary propagator reads
\begin{multline}
	\bar\Lambda_{T/A}(X_1,\vec x_2)
	=\frac{\Gamma\left(\Delta_+-\frac{d}{2}\right)\Gamma\left(\Delta_- -\frac{d}{2}\right)\left(\Delta_+-\frac{d}{2}\right)}{2\pi}\times\cr
	\times\left(\eta_0^{\Delta_-}\bar\Lambda(K,\Delta_-)+\eta_0^{\Delta_+}\bar\Lambda(K,\Delta_+)+\cdots\right)\,,\label{eq:bulktoboundary_dS_SK}
\end{multline}
where we introduced the bulk to boundary propagators with definite late-time fall-off behaviour as
\begin{equation}
    \label{eq:Lambda_bar_Delta}
    \bar\Lambda(K,\Delta_\pm)=\mathcal{N}_{\Delta_\pm}\frac{(2\eta_2)^{\Delta_\pm}}{((\vec x_2-\vec x_1)^2-\eta_2^2)^{\Delta_\pm}\mp i\varepsilon}.
\end{equation}

Focusing on the conformally coupled  case with
$\Delta_{\pm}=\frac{d\pm1}{2}$ and using
equation~\eqref{eq:Lambda_bar_Delta} and the Wick rotation~\eqref{eq:Wick_rotation} we can recast the four-point function~\eqref{eq:four_point_rotated} up to the second subleading order in $\eta_0\to0$ as
\begin{align}
	W_0=&-\lambda\int_{\cH_4^+}\frac{d^{d+1}X}{(a z)^{d+1}}\Big(\eta_0^{4\Delta_-}\bar\Lambda(K,\Delta_-)\bar\Lambda(K,\Delta_-)\bar\Lambda(K,\Delta_-)\bar\Lambda(K,\Delta_-)\cr
	&-\eta_0^{2(\Delta_-+\Delta_+)}\bar\Lambda(K,\Delta_+)\bar\Lambda(K,\Delta_+)\bar\Lambda(K,\Delta_-)\bar\Lambda(K,\Delta_-)\cr
	&+\eta_0^{4\Delta_+}\bar\Lambda(K,\Delta_+)\bar\Lambda(K,\Delta_+)\bar\Lambda(K,\Delta_+)\bar\Lambda(K,\Delta_+)+\cdots\Big).
\end{align}
The evaluation of the tree-level four-point function is therefore reduced to a calculation in EAdS, with two different boundary conditions contributing, corresponding to conformal dimensions $\Delta_+$ and $\Delta_-$.

We could proceed with this calculation diagram by diagram, which is the way this relation between cosmological correlator and EAdS Witten diagrams was first written down in \cite{Sleight:2019hfp,Sleight:2020obc,Sleight:2021plv}. However, as shown in~\cite{DiPietro:2021sjt}, there is an elegant way to rewrite the dS action with the Schwinger-Keldysh contour directly in terms of an auxiliary EAdS action, from which the cosmological correlation functions can be extracted by straightforward functional derivation. We will review this formulation in the next subsection. 

\subsection{Auxiliary action for EAdS}
\label{sec:auxiliary_action}
In this section we review the derivation of section~3 of~\cite{DiPietro:2021sjt}, for the auxiliary action for
computing de Sitter correlators.

The closed time evolution between two in-states from the infinite past can be expressed by a path integral with closed time curves. Then a correlation function is given by taking functional derivatives of the time and anti-time ordered sources $j_T$ and $j_A$ of the partition function
\begin{equation}
	Z[j_T,j_A]=\int\mathcal{D}\phi_T\mathcal{D}\phi_A\eul^{iS_c+i\int(\phi_T j_T+\phi_A j_A)},
\end{equation}
with the closed time action given by
\begin{equation}
	iS_c=i\int\limits_{-\infty}^0\frac{\dd\eta\dd^d\vec{x}}{\eta^{d+1}}\left\{-\frac12(\partial\phi_T)^2-\frac12m^2\phi_T^2-V(\phi_T)+\frac12(\partial\phi_A)^2+\frac12m^2\phi_A^2+V(\phi_A)\right\}.
\end{equation}
Performing the Wick rotation $\eta=z\eul^{\pm i\frac{\pi}{2}}$ as described above, the action becomes
\begin{multline}
	iS_c=-\int\limits_0^{\infty}\frac{\dd
		z\dd^d\vec{x}}{z^{d+1}}\Bigg[\eul^{i\pi\frac{d-1}{2}}\left(\frac12(\partial\phi_T)^2-\frac12m^2\phi_T^2-V(\phi_T)\right)\cr
	+
	\eul^{-i\pi\frac{d-1}{2}}\left(\frac12(\partial\phi_A)^2-\frac12m^2\phi_A^2-V(\phi_A)\right)\Bigg]
\end{multline}
As discussed above the classical solution of a free scalar field in de Sitter is given by
$	\phi(\eta,\vec{x})=\phi^+(\eta,\vec{x})+\phi^-(\eta,\vec{x})$ with
\begin{align}
	\phi^+(\eta,\vec{x})&:=\int\dd^3\vec{y}\, \bar{\Lambda}_{\Delta_+}(\eta,\vec{x}-\vec{y})\phi^+_0(\vec{y}),\cr
	\phi^-(\eta,\vec{x})&:=      \int\dd^3\vec{y}\,\bar{\Lambda}_{\Delta_-}(\eta,\vec{x}-\vec{y})\phi^-_0(\vec{y}),
\end{align}
where $\phi^{\pm}(\eta,\vec{x})\to\eta^{\Delta_{\pm}}$, for
$\eta\to0$. Under the Wick rotation we get 
\begin{equation}\label{eq:phi_0pm}
	\phi(z\eul^{\pm i\frac{\pi}{2}},\vec{x})=\eul^{\pm i\frac{\pi}{2}\Delta_+}\phi^+(z,\vec{x})+\eul^{\pm i\frac{\pi}{2}\Delta_-}\phi^-(z,\vec{x}),
\end{equation}
which plugged in the action leads to
\begin{multline}
	iS_c=-\int\limits_0^{\infty}\frac{\dd z\dd^d\vec{x}}{z^{d+1}}\Bigg[{\eul^{-i\pi\left(\Delta_+ -\frac{d-1}{2}\right)}\over2}
	\left((\partial\phi^+)^2-m^2{\phi^+}^2\right)
	+{\eul^{-i\pi\left(\Delta_- -\frac{d-1}{2}\right)}\over2}
	\left((\partial\phi^-)^2-m^2{\phi^-}^2\right)\cr
	+\eul^{-i\frac{\pi}{2}}\left(\partial\phi^-\partial\phi^+-m^2\phi^-\phi^+\right)
	+\frac12\eul^{+i\pi\left(\Delta_+ -\frac{d-1}{2}\right)}\left((\partial\phi^+)^2-m^2{\phi^+}^2\right)\cr
	+\frac12\eul^{i\pi\left(\Delta_- -\frac{d-1}{2}\right)}
	\left((\partial\phi^-)^2-m^2{\phi^-}^2\right)+\eul^{i\frac{\pi}{2}}\left(\partial\phi^-\partial\phi^+-m^2\phi^-\phi^+\right)\cr
	-\eul^{i\pi\frac{d-1}{2}}V\left(\eul^{-i\frac{\pi}{2}\Delta_+}\phi^+ +\eul^{-i\frac{\pi}{2}\Delta_-}\phi^-\right)
	-\eul^{-i\pi\frac{d-1}{2}}V\left(\eul^{i\frac{\pi}{2}\Delta_+}\phi^+
	+\eul^{i\frac{\pi}{2}\Delta_-}\phi^-\right)\Bigg],
\end{multline}
leading to the result, derived in~\cite{DiPietro:2021sjt},
\begin{multline}
    \label{eq:EAdS_action_gen_pot}
	iS_c
	=-\int\limits_0^{\infty}\frac{\dd z\dd^d\vec{x}}{z^{d+1}}\left[-\sin\left(\pi\left(\Delta_+ -\frac{d}{2}\right)\right)\left((\partial\phi^+)^2-m^2{\phi^+}^2\right)\right.\cr
	\left.-\sin\left(\pi\left(\Delta_- -\frac{d}{2}\right)\right)\left((\partial\phi^-)^2-m^2{\phi^-}^2\right)\right.\cr
	\left.-\eul^{i\pi\frac{d-1}{2}}V\left(\eul^{-i\frac{\pi}{2}\Delta_+}\phi^+ +\eul^{-i\frac{\pi}{2}\Delta_-}\phi^-\right)
	-\eul^{-i\pi\frac{d-1}{2}}V\left(\eul^{i\frac{\pi}{2}\Delta_+}\phi^+ +\eul^{i\frac{\pi}{2}\Delta_-}\phi^-\right)\right].
\end{multline}
We want to study a theory in dS with the potential $V(\phi)=\frac{\lambda}{4!}\phi^4$. In that case the action~\eqref{eq:EAdS_action_gen_pot} becomes
\begin{multline}
    \label{eq:EAdS_action_phi4}
    iS_c
	=-\int\limits_0^{\infty}\frac{\dd z\dd^d\vec{x}}{z^{d+1}}\left[-\sin\left(\pi\left(\Delta_+ -\frac{d}{2}\right)\right)\left((\partial\phi^+)^2-m^2{\phi^+}^2\right)+(\phi^+,\Delta_+\leftrightarrow\phi^-,\Delta_-)\right.\cr
	\left.+\frac{2\lambda}{4!}\left({\phi^+}^4\sin\left(\frac{\pi}{2}(3\Delta_+-\Delta_-)\right)
	+6{\phi^+}^2{\phi^-}^2\sin\left(\frac{\pi d}{2}\right)+{\phi^-}^4\sin\left(\frac{\pi}{2}(3\Delta_--\Delta_+)\right)\right.\right.\cr
    \left.\left.+4{\phi^+}^3\phi^-\sin(\pi\Delta_+)+4{\phi^-}^3\phi^+\sin(\pi\Delta_-)\right)\right].
\end{multline}
    
In this work we consider the case of the conformally coupled scalar with $\Delta_+={d+1\over2}$ and $\Delta_-={d-1\over2}$ with odd boundary dimensions $d$. The action~\eqref{eq:EAdS_action_phi4} then becomes
\begin{multline}	\label{eq:EAdS_action}
	iS_c
	=-\int\limits_0^{\infty}\frac{\dd z\dd^{d}\vec{x}}{z^{d+1}}\left[-\left((\partial\phi^+)^2-m^2{\phi^+}^2\right)+\left((\partial\phi^-)^2-m^2{\phi^-}^2\right)\right.\cr
	\left.-(-1)^{d-1\over2} {2\lambda\over 4!} \left({\phi^+}^4-6{\phi^+}^2{\phi^-}^2+{\phi^-}^4\right)\right].
\end{multline}

This action can now be used to calculate correlation functions in dS, showing to all orders in perturbation theory, that cosmological correlators can be expressed in terms of EAdS Witten diagrams. The leading contributions in the late time expansions are given by calculating the EAdS correlators of the field $\phi^-$. Note however, that this field alone will not give a consistent CFT at the boundary, since there will be mixing interaction vertices between $\phi^-$ and $\phi^+$. To be able to describe the CFT on the boundary we have to take into account the subleading terms in the late time expansion of the cosmological correlator as well. We also notice that the kinetic term in the action is not necessarily positive, leading to ghost-like behaviour of one of the fields. This would be a problem if we wanted to interpret this action as describing a bulk theory in EAdS, however, since we only us this action as a tool to describe a theory in dS, we should treat these signs only as a way to keep track of the correct relative prefactors in the expansion.

\section{de Sitter correlation functions from EAdS Witten diagrams}
\label{sec:Correlation_function_calculation}

In this section we focus entirely on the conformally coupled scalar
field.  As we noticed, perturbatively, this can be treated like a
theory of two interacting scalar fields with boundary scaling dimensions
$\Delta_+={d+1\over2}$ and $\Delta_-={d-1\over2}$ in EAdS, governed by the
action~\eqref{eq:EAdS_action}  for odd boundary dimension.
The propagators~\eqref{eq:bulktoboundary_dS_sum} then read
\begin{align}
	\Lambda_{TT}\left( X, Y,{d-1\over2}\right)&=\frac{a^{d-1}\Gamma\left(\frac{d-1}{2}\right)}{2(2\pi)^{\frac{d+1}{2}}}\left(K(X,Y)\over 1-K(X,Y)\right)^{d-1\over2}
\left(1+\left(1-K(X,Y)\over 1+K(X,Y)\right)^{d-1\over2}\right),\\
	\Lambda_{TT}\left( X, Y,{d+1\over2}\right)&=\frac{a^{d-1}\Gamma\left(\frac{d-1}{2}\right)}{2(2\pi)^{\frac{d+1}{2}}}\left(K(X,Y)\over 1-K(X,Y)\right)^{d-1\over2}
\left(1-\left(1-K(X,Y)\over 1+K(X,Y)\right)^{d-1\over2}\right).
\end{align}
We will then be able to use the
formalism of~\cite{Heckelbacher:2022fbx} for evaluating the Witten diagrams. From now, we specialize to the case of $d=3$.

To avoid unnecessary prefactors in the calculation we are changing the normalisation of
the fields as $\phi^\pm \to \phi^\pm/\sqrt{2}$ and the coupling constant as $\lambda\to 2\lambda$
\begin{multline}
	\label{eq:EAdS_action_d3}
	S_{EAdS}=\int\limits_0^{\infty}\frac{\dd z\dd^3\vec{x}}{z^4}\Big[-{1\over2}\left((\partial\phi^+)^2-m^2{\phi^+}^2\right)+{1\over2}\left((\partial\phi^-)^2-m^2{\phi^-}^2\right)\\*	+\frac{\lambda}{4!}\left({\phi^+}^4-6{\phi^+}^2{\phi^-}^2+{\phi^-}^4\right)\Big].
\end{multline}

The $L$-loop  Witten diagrams between sets of fields of dimensions $\Delta_1$
and $\Delta_2$ are denoted by
\begin{equation}
  W^{\Delta_1\Delta_2\Delta_3\Delta_4,D}_{L,dS}(\vec x_1,\vec x_2,\vec
  x_2,\vec x_4).  
\end{equation}
The case $(\Delta_1,\Delta_2,\Delta_3,\Delta_4)=(1,1,1,1)$ is
evaluated in section~\ref{sec:mmmm},
$(\Delta_1,\Delta_2,\Delta_3,\Delta_4)=(2,2,2,2)$ is
evaluated in section~\ref{sec:pppp}, and the mixed correlators with
$(\Delta_1,\Delta_2,\Delta_3,\Delta_4)=(2,2,1,1)$ and permutations are
evaluated in section~\ref{sec:mixed}.

Using the normalization of the
fields and the coupling constant introduced in~\eqref{eq:EAdS_action} and the conformal mappings as described
in~\cite{Heckelbacher:2022fbx}, we can write a generic EAdS four-point Witten diagram with equal external dimensions $\Delta$ as
\begin{equation}
	\label{eq:normalization_dS_EAdS}
	W_{L,\mathrm{dS}}^{\Delta\Delta\Delta\Delta,D}(\vec
  x_1,\dots,\vec x_4)
                       =
	{a^4\over \left(4\pi^2\right)^{2L+4}}\cW_L^{\Delta\Delta\Delta\Delta,D}(\vec
  x_1,\dots,\vec x_4),
\end{equation}
where  $\cW_L^{\Delta\Delta\Delta\Delta,D}$ is the corresponding
Witten diagram in EAdS with standard normalization of the propagator
as defined in~\cite{Heckelbacher:2022fbx}.
The four-point function with mixed boundary conditions will be given by
acting with the differential operator, defined in section~3.2
of~\cite{Heckelbacher:2022fbx}, onto the corresponding legs of the $\Delta=1$ Witten diagrams. All calculations will be done in the loop dependent dimensional regularisation scheme introduced and described in section~3.1.2 of~\cite{Heckelbacher:2022fbx}.

\subsection{Two-point functions}

If we represent the propagators as
\begin{equation}
	\Lambda(X,Y;1)=\begin{tikzpicture}[baseline=(z)]
		\begin{feynman}[inline=(z)]
			\vertex (z);
			\tikzfeynmanset{every vertex=dot}	
			\vertex [left=0.61cm of z, label=90:$X$] (x1);
			\vertex [right=0.61cm of z, label=90:$Y$] (x2);
			\tikzfeynmanset{every vertex={empty dot,minimum size=0mm}}
			\diagram* {
				(x1)--[scalar](x2),
			};
		\end{feynman}
		\begin{pgfonlayer}{bg}
			\draw[blue] (z) circle (1cm);
		\end{pgfonlayer}
	\end{tikzpicture}, \qquad 
	\Lambda(X,Y;2)=\begin{tikzpicture}[baseline=(z)]
		\begin{feynman}[inline=(z)]
			\vertex (z);
			\tikzfeynmanset{every vertex=dot}	
			\vertex [left=0.61cm of z, label=90:$X$] (x1);
			\vertex [right=0.61cm of z, label=90:$Y$] (x2);
			\tikzfeynmanset{every vertex={empty dot,minimum size=0mm}}
			\diagram* {
				(x1)--[ghost](x2),
			};
		\end{feynman}
		\begin{pgfonlayer}{bg}
			\draw[blue] (z) circle (1cm);
		\end{pgfonlayer}
	\end{tikzpicture}
\end{equation}
then the loop corrections to the boundary two-point function up to order $\lambda^2$ for $\Delta=1$ correspond to the diagrams
\begin{align}
    &\begin{tikzpicture}[baseline=(x)]
			\begin{feynman}
				\tikzfeynmanset{every vertex=dot}
				\vertex [label=180:$x_1$] (x1);
				\vertex [right=0.75cm of x1] (x) ;
				\vertex [right=0.75cm of x, label=0:$x_2$] (x2);
				\tikzfeynmanset{every vertex={empty dot,minimum size=0mm}}
				\vertex [above=0.4cm of x] (y);
				\diagram* {
					(x1) --[scalar] (x) --[scalar] (x2),
					(x) --[scalar,out=135, in=180, min distance=0.1cm] 
					(y) --[scalar,out=0, in=45, min distance=0.1cm](x),
					(x1) --[out=85, in=95, min distance=0.9cm,color=blue] (x2) --[out=265, in=275, min distance=0.9cm,color=blue] (x1),
				};
			\end{feynman}
		\end{tikzpicture}
		\begin{tikzpicture}[baseline=(x)]
			\begin{feynman}
				\tikzfeynmanset{every vertex=dot}
				\vertex [label=180:$x_1$] (x1);
				\vertex [right=0.75cm of x1] (x) ;
				\vertex [right=0.75cm of x, label=0:$x_2$] (x2);
				\tikzfeynmanset{every vertex={empty dot,minimum size=0mm}}
				\vertex [above=0.4cm of x] (y);
				\diagram* {
					(x1) --[scalar] (x) --[scalar] (x2),
					(x) --[ghost,out=135, in=180, min distance=0.1cm] 
					(y) --[ghost,out=0, in=45, min distance=0.1cm](x),
					(x1) --[out=85, in=95, min distance=0.9cm,color=blue] (x2) --[out=265, in=275, min distance=0.9cm,color=blue] (x1),
				};
			\end{feynman}
		\end{tikzpicture}
		\begin{tikzpicture}[baseline=(x)]
			\begin{feynman}
				\tikzfeynmanset{every vertex=dot}
				\vertex [label=180:$x_1$] (x1);
				\vertex [right=1cm of x1] (x) ;
				\vertex [right=1cm of x, label=0:$x_2$] (x2);
				\vertex [above=0.4cm of x] (y);
				\tikzfeynmanset{every vertex={empty dot,minimum size=0mm}}
				\vertex [above=0.4cm of y] (z);
				\diagram* {
					(x1) --[scalar] (x) --[scalar] (x2),
					(x) --[scalar,out=135, in=225, min distance=0.1cm] (y)  --[scalar,out=315, in=45, min distance=0.1cm] (x),
					(y) --[scalar,out=135, in=180, min distance=0.1cm] (z) --[scalar,out=0,in=45, min distance=0.1cm] (y),
					(x1) --[out=85, in=95, min distance=1.2cm,color=blue] (x2) --[out=265, in=275, min distance=1.2cm,color=blue] (x1),
				};
			\end{feynman}
		\end{tikzpicture}
		\begin{tikzpicture}[baseline=(x)]
			\begin{feynman}
				\tikzfeynmanset{every vertex=dot}
				\vertex [label=180:$x_1$] (x1);
				\vertex [right=1cm of x1] (x) ;
				\vertex [right=1cm of x, label=0:$x_2$] (x2);
				\vertex [above=0.4cm of x] (y);
				\tikzfeynmanset{every vertex={empty dot,minimum size=0mm}}
				\vertex [above=0.4cm of y] (z);
				\diagram* {
					(x1) --[scalar] (x) --[scalar] (x2),
					(x) --[ghost,out=135, in=225, min distance=0.1cm] (y)  --[ghost,out=315, in=45, min distance=0.1cm] (x),
					(y) --[scalar,out=135, in=180, min distance=0.1cm] (z) --[scalar,out=0,in=45, min distance=0.1cm] (y),
					(x1) --[out=85, in=95, min distance=1.2cm,color=blue] (x2) --[out=265, in=275, min distance=1.2cm,color=blue] (x1),
				};
			\end{feynman}
		\end{tikzpicture}\cr
		&\begin{tikzpicture}[baseline=(x)]
			\begin{feynman}
				\tikzfeynmanset{every vertex=dot}
				\vertex [label=180:$x_1$] (x1);
				\vertex [right=1cm of x1] (x) ;
				\vertex [right=1cm of x, label=0:$x_2$] (x2);
				\vertex [above=0.4cm of x] (y);
				\tikzfeynmanset{every vertex={empty dot,minimum size=0mm}}
				\vertex [above=0.4cm of y] (z);
				\diagram* {
					(x1) --[scalar] (x) --[scalar] (x2),
					(x) --[scalar,out=135, in=225, min distance=0.1cm] (y)  --[scalar,out=315, in=45, min distance=0.1cm] (x),
					(y) --[ghost,out=135, in=180, min distance=0.1cm] (z) --[ghost,out=0,in=45, min distance=0.1cm] (y),
					(x1) --[out=85, in=95, min distance=1.2cm,color=blue] (x2) --[out=265, in=275, min distance=1.2cm,color=blue] (x1),
				};
			\end{feynman}
		\end{tikzpicture}
		\begin{tikzpicture}[baseline=(x)]
			\begin{feynman}
				\tikzfeynmanset{every vertex=dot}
				\vertex [label=180:$x_1$] (x1);
				\vertex [right=1cm of x1] (x) ;
				\vertex [right=1cm of x, label=0:$x_2$] (x2);
				\vertex [above=0.4cm of x] (y);
				\tikzfeynmanset{every vertex={empty dot,minimum size=0mm}}
				\vertex [above=0.4cm of y] (z);
				\diagram* {
					(x1) --[scalar] (x) --[scalar] (x2),
					(x) --[ghost,out=135, in=225, min distance=0.1cm] (y)  --[ghost,out=315, in=45, min distance=0.1cm] (x),
					(y) --[ghost,out=135, in=180, min distance=0.1cm] (z) --[ghost,out=0,in=45, min distance=0.1cm] (y),
					(x1) --[out=85, in=95, min distance=1.2cm,color=blue] (x2) --[out=265, in=275, min distance=1.2cm,color=blue] (x1),
				};
			\end{feynman}
		\end{tikzpicture}
		\begin{tikzpicture}[baseline=(x)]
			\begin{feynman}
				\tikzfeynmanset{every vertex=dot}
				\vertex [label=180:$x_1$] (x1);
				\vertex [right=0.75cm of x1] (x) ;
				\vertex [right=0.5cm of x] (y);
				\vertex [right=0.75cm of y, label=0:$x_2$] (x2);
				\tikzfeynmanset{every vertex={empty dot,minimum size=0mm}}
				\vertex [above=0.4cm of x] (z);
				\vertex [above=0.4cm of y] (w);
				\diagram* {
					(x1) --[scalar] (x) -- [scalar] (y) -- [scalar] (x2),
					(x) --[scalar, out=135, in=180, min distance=0.1cm]
					(z) --[scalar, out=0, in=45, min distance=0.1cm](x),
					(y) --[scalar,out=135, in=180, min distance=0.1cm] (w) --[scalar,out=0, in=45, min distance=0.1cm](y),
					(x1) --[out=85, in=95, min distance=1.2cm,color=blue] (x2) --[out=265, in=275, min distance=1.2cm,color=blue] (x1),
				};
			\end{feynman}
		\end{tikzpicture}
		\begin{tikzpicture}[baseline=(x)]
			\begin{feynman}
				\tikzfeynmanset{every vertex=dot}
				\vertex [label=180:$x_1$] (x1);
				\vertex [right=0.75cm of x1] (x) ;
				\vertex [right=0.5cm of x] (y);
				\vertex [right=0.75cm of y, label=0:$x_2$] (x2);
				\tikzfeynmanset{every vertex={empty dot,minimum size=0mm}}
				\vertex [above=0.4cm of x] (z);
				\vertex [above=0.4cm of y] (w);
				\diagram* {
					(x1) --[scalar] (x) -- [scalar] (y) -- [scalar] (x2),
					(x) --[ghost, out=135, in=180, min distance=0.1cm]
					(z) --[ghost, out=0, in=45, min distance=0.1cm](x),
					(y) --[ghost,out=135, in=180, min distance=0.1cm] (w) --[ghost,out=0, in=45, min distance=0.1cm](y),
					(x1) --[out=85, in=95, min distance=1.2cm,color=blue] (x2) --[out=265, in=275, min distance=1.2cm,color=blue] (x1),
				};
			\end{feynman}
		\end{tikzpicture}\cr
		&\begin{tikzpicture}[baseline=(x)]
			\begin{feynman}
				\tikzfeynmanset{every vertex=dot}
				\vertex [label=180:$x_1$] (x1);
				\vertex [right=0.75cm of x1] (x) ;
				\vertex [right=0.5cm of x] (y);
				\vertex [right=0.75cm of y, label=0:$x_2$] (x2);
				\tikzfeynmanset{every vertex={empty dot,minimum size=0mm}}
				\vertex [above=0.4cm of x] (z);
				\vertex [above=0.4cm of y] (w);
				\diagram* {
					(x1) --[scalar] (x) -- [scalar] (y) -- [scalar] (x2),
					(x) --[scalar, out=135, in=180, min distance=0.1cm]
					(z) --[scalar, out=0, in=45, min distance=0.1cm](x),
					(y) --[ghost,out=135, in=180, min distance=0.1cm] (w) --[ghost,out=0, in=45, min distance=0.1cm](y),
					(x1) --[out=85, in=95, min distance=1.2cm,color=blue] (x2) --[out=265, in=275, min distance=1.2cm,color=blue] (x1),
				};
			\end{feynman}
		\end{tikzpicture}
		\begin{tikzpicture}[baseline=(x)]
			\begin{feynman}
				\tikzfeynmanset{every vertex=dot}
				\vertex [label=180:$x_1$] (x1);
				\vertex [right=0.75cm of x1] (x);
				\vertex [right=0.5cm of x] (y);
				\vertex [right=0.75cm of y, label=0:$x_2$] (x2);
				\diagram* {
					(x1) --[scalar]  (x) -- [scalar]  (y) -- [scalar] (x2),
					(x) -- [scalar,out=85, in=95, min distance=0.3cm] (y) -- [scalar,out=265, in=275, min distance=0.3cm] (x),
					(x1) --[out=85, in=95, min distance=1.2cm,color=blue] (x2) --[out=265, in=275, min distance=1.2cm,color=blue] (x1),
				};
			\end{feynman}
		\end{tikzpicture}
		\begin{tikzpicture}[baseline=(x)]
			\begin{feynman}
				\tikzfeynmanset{every vertex=dot}
				\vertex [label=180:$x_1$] (x1);
				\vertex [right=0.75cm of x1] (x);
				\vertex [right=0.5cm of x] (y);
				\vertex [right=0.75cm of y, label=0:$x_2$] (x2);
				\diagram* {
					(x1) --[scalar]  (x) -- [scalar]  (y) -- [scalar] (x2),
					(x) -- [ghost,out=85, in=95, min distance=0.3cm] (y) -- [ghost,out=265, in=275, min distance=0.3cm] (x),
					(x1) --[out=85, in=95, min distance=1.2cm,color=blue] (x2) --[out=265, in=275, min distance=1.2cm,color=blue] (x1),
				};
			\end{feynman}
		\end{tikzpicture}\label{eq:two_point}\,.
\end{align}
For $\Delta=2$ the diagrams are the same up to replacing the external lines by the $\Delta=2$ bulk to boundary propagator.

Using the results from~\cite{Bertan:2018khc}, it can be checked that the integrals appearing in~\eqref{eq:two_point} all reduce to a divergent piece times a mass-shift term. We can therefore use the same argument that the renormalized mass should be fixed at the value ``measured'' at the boundary, which in our case fixes the leading order fall off behaviour at future infinity to $\Delta=1$. As a result, we can ignore the loop corrections to the two point function in the following calculation of the four-point function, and we will draw the renormalised propagators as
\begin{equation}
	\Lambda(X,Y;1)=\begin{tikzpicture}[baseline=(z)]
		\begin{feynman}[inline=(z)]
			\vertex (z);
			\tikzfeynmanset{every vertex=dot}	
			\vertex [left=0.61cm of z, label=90:$X$] (x1);
			\vertex [right=0.61cm of z, label=90:$Y$] (x2);
			\tikzfeynmanset{every vertex={empty dot,minimum size=0mm}}
			\diagram* {
				(x1)--(x2),
			};
		\end{feynman}
		\begin{pgfonlayer}{bg}
			\draw[blue] (z) circle (1cm);
		\end{pgfonlayer}
	\end{tikzpicture}, \qquad 
	\Lambda(X,Y;2)=\begin{tikzpicture}[baseline=(z)]
		\begin{feynman}[inline=(z)]
			\vertex (z);
			\tikzfeynmanset{every vertex=dot}	
			\vertex [left=0.61cm of z, label=90:$X$] (x1);
			\vertex [right=0.61cm of z, label=90:$Y$] (x2);
			\tikzfeynmanset{every vertex={empty dot,minimum size=0mm}}
			\diagram* {
				(x1)--[double](x2),
			};
		\end{feynman}
		\begin{pgfonlayer}{bg}
			\draw[blue] (z) circle (1cm);
		\end{pgfonlayer}
	\end{tikzpicture}\,.
\end{equation}

\subsection{Four-point functions}
\label{subsec:fourpoint}
Recalling~\eqref{eq:phi_0pm}, the dominant term contribution to the bulk scalar field $\phi$ is contained in $\phi^-$. From this one may conclude that the four-point correlation function at future infinity is given by calculating the correlation functions of the auxiliary field $\phi^-$ at the boundary of EAdS, with  action~\eqref{eq:EAdS_action}. However considering only $\phi^-$ as a boundary field one will not be able retrieve all the information of the dual CFT. This can also be seen form the bulk action~\eqref{eq:EAdS_action} in which $\phi^-$ and $\phi^+$ are coupled. To access the full CFT information we rather have to expand the four-point function to second subleading order in $\eta_0$, that is
\begin{align}
	\label{eq:correlator_expansion}
	&\expect{\phi_0(\vec{x_1})\phi_0(\vec{x_2})\phi_0(\vec{x_3})\phi_0(\vec{x_4})}=
	\eta_0^{4\Delta_-}\expect{\phi^-(\vec{x_1})\phi^-(\vec{x_2})\phi^-(\vec{x_3})\phi^-(\vec{x_4})}\cr
	&+	\eta_0^{2(\Delta_-+\Delta_+)}\left(\expect{\phi^+(\vec{x_1})\phi^+(\vec{x_2})\phi^-(\vec{x_3})\phi^-(\vec{x_4})}+
	\expect{\phi^+(\vec{x_1})\phi^-(\vec{x_2})\phi^+(\vec{x_3})\phi^-(\vec{x_4})}\right.\cr
	&\left.+\expect{\phi^+(\vec{x_1})\phi^-(\vec{x_2})\phi^-(\vec{x_3})\phi^+(\vec{x_4})}\right)
	+\eta_0^{4\Delta_+}\expect{\phi^+(\vec{x_1})\phi^+(\vec{x_2})\phi^+(\vec{x_3})\phi^+(\vec{x_4})}\,.
\end{align}

\subsubsection{\texorpdfstring{$\expect{\phi^-\phi^-\phi^-\phi^-}$}{----}}\label{sec:mmmm}
The contributions to the leading term of the late time expansion of the four-point correlation function in equation~\eqref{eq:correlator_expansion} is given by
\begin{align}
	&\expect{\phi^-(x_1)\phi^-(x_2)\phi^-(x_3)\phi^-(x_4)}=\left(\begin{tikzpicture}[baseline=(z)]
		\begin{feynman}[inline=(z)]
			\vertex (z);
			\tikzfeynmanset{every vertex=dot}	
			\vertex [above left=0.63cm and 0.71cm of z, label=180:$x_1$] (x1);
			\vertex [below left=0.63cm and 0.71cm of z, label=180:$x_2$] (x2);
			\vertex [above right=0.63cm and 0.71cm of z, label=0:$x_3$] (x3);
			\vertex [below right=0.63cm and 0.71cm of z, label=0:$x_4$] (x4);
			\tikzfeynmanset{every vertex={empty dot,minimum size=0mm}}
			\diagram* {
				(x1)--(x3),
				(x2)--(x4),
			};
		\end{feynman}
		\begin{pgfonlayer}{bg}
			\draw[blue] (z) circle (1cm);
		\end{pgfonlayer}
	\end{tikzpicture}+2 \text{ perm.}\right)
	-\lambda\begin{tikzpicture}[baseline=(z)]
		\begin{feynman}[inline=(z)]
			\tikzfeynmanset{every vertex=dot}
			\vertex (z);	
			\vertex [above left=0.71cm and 0.71cm of z, label=180:$x_1$] (x1);
			\vertex [below left=0.71cm and 0.71cm of z, label=180:$x_2$] (x2);
			\vertex [above right=0.71cm and 0.71cm of z, label=0:$x_3$] (x3);
			\vertex [below right=0.71cm and 0.71cm of z, label=0:$x_4$] (x4);
			\tikzfeynmanset{every vertex={empty dot,minimum size=0mm}}
			\diagram* {
				(x1)--(z),
				(x2)--(z),
				(x3)--(z),
				(x4)--(z),
			};
		\end{feynman}
		\begin{pgfonlayer}{bg}
			\draw[blue] (z) circle (1cm);
		\end{pgfonlayer}
	\end{tikzpicture}\cr
	&+\frac{\lambda^2}{2}\left(\begin{tikzpicture}[baseline=(z)]
		\begin{feynman}[inline=(z)]
			\vertex (z);	
			\tikzfeynmanset{every vertex=dot}
			\vertex [above left=0.71cm and 0.71cm of z, label=180:$x_1$] (x1);
			\vertex [below left=0.71cm and 0.71cm of z, label=180:$x_2$] (x2);
			\vertex [above right=0.71cm and 0.71cm of z, label=0:$x_3$] (x3);
			\vertex [below right=0.71cm and 0.71cm of z, label=0:$x_4$] (x4);
			\vertex [left=0.3cm of z] (y1);
			\vertex [right=0.3cm of z] (y2);
			\tikzfeynmanset{every vertex={empty dot,minimum size=0mm}}
			\vertex [above=0.35cm of z] (y3);
			\vertex [below=0.35cm of z] (y4);
			\diagram* {
				(x1)--(y1),
				(x2)--(y1),
				(x3)--(y2),
				(x4)--(y2),
			};
		\end{feynman}
		\begin{pgfonlayer}{bg}
			\draw[blue] (z) circle (1.05cm);
			\draw (z) circle (0.34cm);
		\end{pgfonlayer}
	\end{tikzpicture}+2 \text{ perm.}\right)
	+\frac{\lambda^2}{2}\left(\begin{tikzpicture}[baseline=(z)]
		\begin{feynman}[inline=(z)]
			\vertex (z);	
			\tikzfeynmanset{every vertex=dot}
			\vertex [above left=0.71cm and 0.71cm of z, label=180:$x_1$] (x1);
			\vertex [below left=0.71cm and 0.71cm of z, label=180:$x_2$] (x2);
			\vertex [above right=0.71cm and 0.71cm of z, label=0:$x_3$] (x3);
			\vertex [below right=0.71cm and 0.71cm of z, label=0:$x_4$] (x4);
			\vertex [left=0.3cm of z] (y1);
			\vertex [right=0.3cm of z] (y2);
			\tikzfeynmanset{every vertex={empty dot,minimum size=0mm}}
			\vertex [above=0.35cm of z] (y3);
			\vertex [below=0.35cm of z] (y4);
			\diagram* {
				(x1)--(y1),
				(x2)--(y1),
				(x3)--(y2),
				(x4)--(y2),
			};
		\end{feynman}
		\begin{pgfonlayer}{bg}
			\draw[blue] (z) circle (1.05cm);
			\draw [double] (z) circle (0.34cm);
		\end{pgfonlayer}
	\end{tikzpicture}+2 \text{ perm.}\right)+\mathcal O(\lambda^3).\label{eq:mmmm}
\end{align}
\noindent $\bullet$ The disconnected part is given by the product of
two-point functions
\begin{multline}
    \expect{\phi^-(x_1)\phi^-(x_2)}\expect{\phi^-(x_3)\phi^-(x_4)}
    +\expect{\phi^-(x_1)\phi^-(x_3)}\expect{\phi^-(x_2)\phi^-(x_4)}\cr+\expect{\phi^-(x_1)\phi^-(x_4)}\expect{\phi^-(x_2)\phi^-(x_3)}
    =\frac{2^2a^4}{(4\pi^2)^2}\frac{1}{x_{12}^2x_{34}^2}\left(1+v+\frac{v}{1-Y}\right)\,.
\end{multline}
Here we follow the notation and conventions of~\cite{Heckelbacher:2022fbx}
for the cross-ratio
\begin{equation}
	\label{eq:crossratio}
	v={x_{12}^2x_{34}^2\over
		x_{14}^2x_{23}^2}= \zeta\bar
	\zeta; \qquad
	1-Y={x_{13}^2x_{24}^2\over                   x_{14}^2 x_{23}^2}=(1-\zeta)(1-\bar \zeta)\,.
\end{equation}
where $x_{ij}^2=|\vec{x_i}-\vec{x_j}|^2$.

\noindent $\bullet$ The cross terms is given by the $\Delta=1$ term in EAdS
\begin{equation}\label{eq:Cross_allDelta_int}
	\cW^{1111,D}_0(\zeta,\zetab)=\frac12\frac{v^\Delta}{x^2_{12}x^2_{34}}
	\int\limits_{\mathbb{R}^D}
	\frac{\dd^D X}{\|X-u_1|^{2(D-4)}}\frac{1}{\|X|^2\|X-u_\zeta|^2\|X-u_1|^2}\,,
\end{equation}
where the norm is defined with a euclidean signature
\begin{equation}
    \|X|^2=z^2+\vec x^2
\end{equation}
and the radial coordinate $z$ is expressed with the help of the normal vector to the boundary $u=(0,0,0,1)$ such that
\begin{equation}
    u\cdot X=z.
\end{equation}

\noindent $\bullet$  For the one-loop contributions we use the following expression for the square of the propagator:
\begin{multline}
	\tilde\Lambda(X_1,X_2;\Delta)^2={(u\cdot X_1)^2 \,
		(u\cdot X_2)^2\over \|X_1-X_2|^4}+{(u\cdot X_1)^2 \,
		(u\cdot \sigma(X_2))^2\over \|X_1-\sigma(X_2)|^4}
	\cr-{(-1)^\Delta\over2}\left( {u\cdot X_1 \,
		u\cdot X_2\over \|X_1-X_2|^2}+ {u\cdot X_1 \,
		u\cdot \sigma(X_2)\over \|X_1-\sigma(X_2)|^2} \right),
          \end{multline}
          where $\sigma(X)$ is the antipodal map after Wick rotation
          \begin{equation}
            \sigma(\vec x,z)=(\vec x,-z).            
          \end{equation}
          We have used that the bulk-to-bulk propagator without the
          normalisation factor is given by for $\Delta=1$ and $\Delta=2$

\begin{equation}
\tilde \Lambda(X_1,X_2,\Delta)= {(u\cdot X_1)(u\cdot X_2)\over \|X_1-X_2|^2} - (-1)^\Delta {(u\cdot X_1)(u\cdot \sigma(X_2))\over \|X_1-\sigma(X_2)|^2},
\end{equation}
together with the identity
\begin{equation}
{(u\cdot X_1)(u\cdot X_2)\over \|X_1-X_2|^2}  {(u\cdot X_1)(u\cdot \sigma(X_2))\over \|X_1-\sigma(X_2)|^2}={1\over 4}\left({(u\cdot X_1)(u\cdot X_2)\over \|X_1-X_2|^2}+ {(u\cdot X_1)(u\cdot \sigma(X_2))\over \|X_1-\sigma(X_2)|^2}\right).
\end{equation}
Then, by regrouping contributions from the $\Delta=1$ and $\Delta=2$ fields   propagating in the loops in~\eqref{eq:mmmm}, one can see that for the sum, over $\Delta$, of the propagators squared the cross-terms cancel so that
\begin{equation}
    \label{eq:propagators_squared}
	\tilde\Lambda(X_1,X_2;1)^2+\tilde\Lambda(X_1,X_2;2)^2=2{(u\cdot X_1)^2 \,
		(u\cdot X_2)^2\over \|X_1-X_2|^4}+2{(u\cdot X_1)^2 \,
		(u\cdot \sigma(X_2))^2\over \|X_1-\sigma(X_2)|^4}.
\end{equation}
After unfolding the integral to the whole space $\mathbb R^4$ the
one-loop contribution, in the s-channel, for four external scalars of the same dimension $\Delta$ adds up to
\begin{align}
    \label{eq:equalDelta_oneloop}
  &\begin{tikzpicture}[baseline=(z)]
		\begin{feynman}[inline=(z)]
			\vertex (z);	
			\tikzfeynmanset{every vertex=dot}
			\vertex [above left=0.71cm and 0.71cm of z, label=180:$x_1$] (x1);
			\vertex [below left=0.71cm and 0.71cm of z, label=180:$x_2$] (x2);
			\vertex [above right=0.71cm and 0.71cm of z, label=0:$x_3$] (x3);
			\vertex [below right=0.71cm and 0.71cm of z, label=0:$x_4$] (x4);
			\vertex [left=0.3cm of z] (y1);
			\vertex [right=0.3cm of z] (y2);
			\tikzfeynmanset{every vertex={empty dot,minimum size=0mm}}
			\vertex [above=0.35cm of z] (y3);
			\vertex [below=0.35cm of z] (y4);
			\diagram* {
				(x1)--(y1),
				(x2)--(y1),
				(x3)--(y2),
				(x4)--(y2),
			};
		\end{feynman}
		\begin{pgfonlayer}{bg}
			\draw[blue] (z) circle (1.05cm);
			\draw (z) circle (0.34cm);
		\end{pgfonlayer}
	\end{tikzpicture}+\begin{tikzpicture}[baseline=(z)]
		\begin{feynman}[inline=(z)]
			\vertex (z);	
			\tikzfeynmanset{every vertex=dot}
			\vertex [above left=0.71cm and 0.71cm of z, label=180:$x_1$] (x1);
			\vertex [below left=0.71cm and 0.71cm of z, label=180:$x_2$] (x2);
			\vertex [above right=0.71cm and 0.71cm of z, label=0:$x_3$] (x3);
			\vertex [below right=0.71cm and 0.71cm of z, label=0:$x_4$] (x4);
			\vertex [left=0.3cm of z] (y1);
			\vertex [right=0.3cm of z] (y2);
			\tikzfeynmanset{every vertex={empty dot,minimum size=0mm}}
			\vertex [above=0.35cm of z] (y3);
			\vertex [below=0.35cm of z] (y4);
			\diagram* {
				(x1)--(y1),
				(x2)--(y1),
				(x3)--(y2),
				(x4)--(y2),
			};
		\end{feynman}
		\begin{pgfonlayer}{bg}
			\draw[blue] (z) circle (1.05cm);
			\draw [double] (z) circle (0.34cm);
		\end{pgfonlayer}
	\end{tikzpicture}\\
\nonumber &=\frac{2^{4\Delta}a^4}{(4\pi^2)^6}\int_{(\mathbb R^D)^2} \dd^DX\dd^DY {(u\cdot
            X)^{2\Delta-2}(u\cdot Y)^{2\Delta-2}\over \|X-Y|^4
            \|X-x_1|^2\|X-x_2|^2\|Y-x_3|^2\|Y-x_4|^2} \, ,
\end{align}
with similar expressions for the other channels. 
Finally, performing the conformal mappings as described in~\cite{Heckelbacher:2022fbx} the integrand of equation~\eqref{eq:equalDelta_oneloop} takes the form
\begin{equation}
	\label{eq:W1_fin_div}
	\cW_{1,\mathrm{div}}^{\Delta,4-2\epsilon,s}\!\!\!\!=\frac{(\zeta\zetab)^\Delta}{(x_{12}^2x_{34}^2)^\Delta}\!\!\int\limits_{\mathbb{R}^{2D}}\frac{\dd^{4-2\epsilon} X_1\dd^{4-2\epsilon} X_2(u\cdot X_1)^{2\Delta-2}(u\cdot X_2)^{2\Delta-2}}{\|X_1|^{2\Delta}\|X_1-u_\zeta|^{2\Delta}\|X_2-u_1|^{2\Delta-4\epsilon}\|X_1-u_1|^{-4\epsilon}\|X_1-X_2|^4}\,,
\end{equation}
where the subscript ``div'' indicates that the integral is divergent and $\epsilon=\frac{4-D}{2}$ is a regulator. The contributions to the other channels are given in the appendix in equation~\eqref{eq:W1_fin_div_app}.

These integrals were already calculated in~\cite{Heckelbacher:2022fbx} and the results are given in appendix~\ref{sec:delta1_app}. Note that, because of~\eqref{eq:propagators_squared}, the elliptic
sector, which was present in the one-loop EAdS computation for
$\Delta=1$, cancels out. By consequence, the loop integrals are linearly reducible~\cite{Panzer:2014gra} and thus can be expressed in terms of multiple polylogarithms using the program $\mathtt{HyperInt}$~\cite{Panzer:2014caa}. The entire four-point function then becomes
\begin{multline}
	\expect{\phi^-(x_1)\phi^-(x_2)\phi^-(x_3)\phi^-(x_4)}
	=\frac{2^2a^4}{(4\pi^2)^2}\Bigg[\frac{1}{x_{12}^2x_{34}^2}\left(1+v+\frac{v}{1-Y}\right)   -\frac{2^2\lambda}{(4\pi^2)^2}\cW_0^{1111,4-4\epsilon}(v,Y)\cr
          +\frac{2^2\lambda^2}{(4\pi^2)^4}\left(-\frac{3\pi^2}{\epsilon}\cW_0^{1111,4-4\epsilon}(v,Y)+\frac{\pi^4
   v}{2x_{12}^2x_{34}^2}\sum_{i\in\{s,t,u\}}L_0^{1,i}(v,Y)\right) +\mathcal O(\lambda^3)\Bigg]\,.
\end{multline}
 The integrals $\cW_0^{1111,4-4\epsilon}(v,Y)$ and $L_0^{1,i}$ have been evaluated  in~\cite{Heckelbacher:2022fbx}. We have recalled their expressions in~\eqref{e:L0one} for $L_0^{1,i}$.

\subsubsection{\texorpdfstring{$\expect{\phi^+\phi^+\phi^+\phi^+}$}{++++}}\label{sec:pppp}
The contributions to the $\phi^+\phi^+\phi^+\phi^+$ term of the late time expansion of the four point correlation function in equation~\eqref{eq:correlator_expansion} are given by
\begin{align}
	&\expect{\phi^+(x_1)\phi^+(x_2)\phi^+(x_3)\phi^+(x_4)}=\left(\begin{tikzpicture}[baseline=(z)]
		\begin{feynman}[inline=(z)]
			\vertex (z);
			\tikzfeynmanset{every vertex=dot}	
			\vertex [above left=0.63cm and 0.71cm of z, label=180:$x_1$] (x1);
			\vertex [below left=0.63cm and 0.71cm of z, label=180:$x_2$] (x2);
			\vertex [above right=0.63cm and 0.71cm of z, label=0:$x_3$] (x3);
			\vertex [below right=0.63cm and 0.71cm of z, label=0:$x_4$] (x4);
			\tikzfeynmanset{every vertex={empty dot,minimum size=0mm}}
			\diagram* {
				(x1)--[double](x3),
				(x2)--[double](x4),
			};
		\end{feynman}
		\begin{pgfonlayer}{bg}
			\draw[blue] (z) circle (1cm);
		\end{pgfonlayer}
	\end{tikzpicture}+2 \text{ perm.}\right)
	-\lambda\begin{tikzpicture}[baseline=(z)]
		\begin{feynman}[inline=(z)]
			\tikzfeynmanset{every vertex=dot}
			\vertex (z);	
			\vertex [above left=0.71cm and 0.71cm of z, label=180:$x_1$] (x1);
			\vertex [below left=0.71cm and 0.71cm of z, label=180:$x_2$] (x2);
			\vertex [above right=0.71cm and 0.71cm of z, label=0:$x_3$] (x3);
			\vertex [below right=0.71cm and 0.71cm of z, label=0:$x_4$] (x4);
			\tikzfeynmanset{every vertex={empty dot,minimum size=0mm}}
			\diagram* {
				(x1)--[double](z),
				(x2)--[double](z),
				(x3)--[double](z),
				(x4)--[double](z),
			};
		\end{feynman}
		\begin{pgfonlayer}{bg}
			\draw[blue] (z) circle (1cm);
		\end{pgfonlayer}
	\end{tikzpicture}\cr
	&+\frac{\lambda^2}{2}\left(\begin{tikzpicture}[baseline=(z)]
		\begin{feynman}[inline=(z)]
			\vertex (z);	
			\tikzfeynmanset{every vertex=dot}
			\vertex [above left=0.71cm and 0.71cm of z, label=180:$x_1$] (x1);
			\vertex [below left=0.71cm and 0.71cm of z, label=180:$x_2$] (x2);
			\vertex [above right=0.71cm and 0.71cm of z, label=0:$x_3$] (x3);
			\vertex [below right=0.71cm and 0.71cm of z, label=0:$x_4$] (x4);
			\vertex [left=0.3cm of z] (y1);
			\vertex [right=0.3cm of z] (y2);
			\tikzfeynmanset{every vertex={empty dot,minimum size=0mm}}
			\vertex [above=0.35cm of z] (y3);
			\vertex [below=0.35cm of z] (y4);
			\diagram* {
				(x1)--[double](y1),
				(x2)--[double](y1),
				(x3)--[double](y2),
				(x4)--[double](y2),
			};
		\end{feynman}
		\begin{pgfonlayer}{bg}
			\draw[blue] (z) circle (1.05cm);
			\draw (z) circle (0.34cm);
		\end{pgfonlayer}
	\end{tikzpicture}+2 \text{ perm.}\right)
	+\frac{\lambda^2}{2}\left(\begin{tikzpicture}[baseline=(z)]
		\begin{feynman}[inline=(z)]
			\vertex (z);	
			\tikzfeynmanset{every vertex=dot}
			\vertex [above left=0.71cm and 0.71cm of z, label=180:$x_1$] (x1);
			\vertex [below left=0.71cm and 0.71cm of z, label=180:$x_2$] (x2);
			\vertex [above right=0.71cm and 0.71cm of z, label=0:$x_3$] (x3);
			\vertex [below right=0.71cm and 0.71cm of z, label=0:$x_4$] (x4);
			\vertex [left=0.3cm of z] (y1);
			\vertex [right=0.3cm of z] (y2);
			\tikzfeynmanset{every vertex={empty dot,minimum size=0mm}}
			\vertex [above=0.35cm of z] (y3);
			\vertex [below=0.35cm of z] (y4);
			\diagram* {
				(x1)--[double](y1),
				(x2)--[double](y1),
				(x3)--[double](y2),
				(x4)--[double](y2),
			};
		\end{feynman}
		\begin{pgfonlayer}{bg}
			\draw[blue] (z) circle (1.05cm);
			\draw [double] (z) circle (0.34cm);
		\end{pgfonlayer}
	\end{tikzpicture}+2 \text{ perm.}\right)+\mathcal O(\lambda^3).
\end{align}
\noindent $\bullet$ The cross term is again just given by the same expression as the $\Delta=2$ cross in EAdS, given in appendix~\ref{sec:cross_app}.

\noindent $\bullet$ Since the squares of the bulk-to-bulk propagators are the same, similar arguments as for the $\Delta=1$ case hold, i.e. the result can be written as a sum of the divergent and finite parts of the one-loop Witten diagrams with $\Delta=2$. The details are given in appendix~\ref{sec:delta2_app}

The entire four-point function at this order is therefore given by
\begin{multline}
	\expect{\phi^+(x_1)\phi^+(x_2)\phi^+(x_3)\phi^+(x_4)}=
	\frac{2^4a^4}{(4\pi^2)^2}\Bigg[\frac{1}{x_{12}^4x_{34}^4}\left(1+v^2+\frac{v^2}{(1-Y)^2}\right)\cr-\frac{2^4\lambda}{(4\pi^2)^2}\cW_0^{2,4-4\epsilon}-\frac{2^4\lambda^2}{(4\pi^2)^4}
	\Bigg(-\frac{3\pi^2}{\epsilon}\cW_0^{2222,4-4\epsilon}(v,Y)\cr
+3\pi^2\cW_0^{2222,4}(v,Y)+\frac12\sum\limits_{j\in\{s,t,u\}}\cW_{1,\mathrm{fin}}^{2222,j}(v,Y)+\frac{\pi^4
  v}{2x_{12}^2x_{34}^2}\sum_{i\in\{s,t,u\}}L_0^{2,i}(v,Y)\Bigg)+\mathcal O(\lambda^3) \Bigg]
\end{multline}
where $\cW_{1,\mathrm{fin}}^{2222,j}$ and $L_0^{2,i}$ have been
calculated in~\cite{Heckelbacher:2022fbx} and recalled in~\eqref{e:W2fin} and~\eqref{e:L02} respectively. In fact, as described in~\cite{Heckelbacher:2022fbx} there is a differential relation between the correlators with $\phi^+$ and $\phi^-$ external legs. We will make 
use of this in the next subsection.

\subsubsection{Mixed correlators}\label{sec:mixed}

Additionally, we have the correlation functions of $\phi^+$ with $\phi^-$, which are sub-leading in the late-time expansion. They are equivalent up to permutation of the operators $\phi^-(x_i),\phi^+(x_j)$ so we will only calculate $\expect{\phi^+(x_1)\phi^+(x_2)\phi^-(x_3)\phi^-(x_4)}$ and discuss the other combinations at the end of section~\ref{sec:renormalization}. 

The diagrams we calculate are given by
\begin{multline}
	\expect{\phi^+(x_1)\phi^+(x_2)\phi^-(x_3)\phi^-(x_4)}=\begin{tikzpicture}[baseline=(z)]
		\begin{feynman}[inline=(z)]
			\vertex (z);
			\tikzfeynmanset{every vertex=dot}	
			\vertex [above left=0.63cm and 0.71cm of z, label=180:$x_1$] (x1);
			\vertex [below left=0.63cm and 0.71cm of z, label=180:$x_2$] (x2);
			\vertex [above right=0.63cm and 0.71cm of z, label=0:$x_3$] (x3);
			\vertex [below right=0.63cm and 0.71cm of z, label=0:$x_4$] (x4);
			\tikzfeynmanset{every vertex={empty dot,minimum size=0mm}}
			\diagram* {
				(x1)--[double](x2),
				(x3)--(x4),
			};
		\end{feynman}
		\begin{pgfonlayer}{bg}
			\draw[blue] (z) circle (1cm);
		\end{pgfonlayer}
	\end{tikzpicture}+\lambda\begin{tikzpicture}[baseline=(z)]
	\begin{feynman}[inline=(z)]
		\tikzfeynmanset{every vertex=dot}
		\vertex (z);	
		\vertex [above left=0.71cm and 0.71cm of z, label=180:$x_1$] (x1);
		\vertex [below left=0.71cm and 0.71cm of z, label=180:$x_2$] (x2);
		\vertex [above right=0.71cm and 0.71cm of z, label=0:$x_3$] (x3);
		\vertex [below right=0.71cm and 0.71cm of z, label=0:$x_4$] (x4);
		\tikzfeynmanset{every vertex={empty dot,minimum size=0mm}}
		\diagram* {
			(x1)--[double](z),
			(x2)--[double](z),
			(x3)--(z),
			(x4)--(z),
		};
	\end{feynman}
	\begin{pgfonlayer}{bg}
		\draw[blue] (z) circle (1cm);
	\end{pgfonlayer}
\end{tikzpicture}\cr
-\lambda^2\left(\frac12\begin{tikzpicture}[baseline=(z)]
	\begin{feynman}[inline=(z)]
		\vertex (z);	
		\tikzfeynmanset{every vertex=dot}
		\vertex [above left=0.71cm and 0.71cm of z, label=180:$x_1$] (x1);
		\vertex [below left=0.71cm and 0.71cm of z, label=180:$x_2$] (x2);
		\vertex [above right=0.71cm and 0.71cm of z, label=0:$x_3$] (x3);
		\vertex [below right=0.71cm and 0.71cm of z, label=0:$x_4$] (x4);
		\vertex [left=0.3cm of z] (y1);
		\vertex [right=0.3cm of z] (y2);
		\tikzfeynmanset{every vertex={empty dot,minimum size=0mm}}
		\vertex [above=0.35cm of z] (y3);
		\vertex [below=0.35cm of z] (y4);
		\diagram* {
			(x1)--[double](y1),
			(x2)--[double](y1),
			(x3)--(y2),
			(x4)--(y2),
		};
	\end{feynman}
	\begin{pgfonlayer}{bg}
		\draw[blue] (z) circle (1.05cm);
		\draw (z) circle (0.34cm);
	\end{pgfonlayer}
\end{tikzpicture}+\frac12\begin{tikzpicture}[baseline=(z)]
\begin{feynman}[inline=(z)]
	\vertex (z);	
	\tikzfeynmanset{every vertex=dot}
	\vertex [above left=0.71cm and 0.71cm of z, label=180:$x_1$] (x1);
	\vertex [below left=0.71cm and 0.71cm of z, label=180:$x_2$] (x2);
	\vertex [above right=0.71cm and 0.71cm of z, label=0:$x_3$] (x3);
	\vertex [below right=0.71cm and 0.71cm of z, label=0:$x_4$] (x4);
	\vertex [left=0.3cm of z] (y1);
	\vertex [right=0.3cm of z] (y2);
	\tikzfeynmanset{every vertex={empty dot,minimum size=0mm}}
	\vertex [above=0.35cm of z] (y3);
	\vertex [below=0.35cm of z] (y4);
	\diagram* {
		(x1)--[double](y1),
		(x2)--[double](y1),
		(x3)--(y2),
		(x4)--(y2),
	};
\end{feynman}
\begin{pgfonlayer}{bg}
	\draw[blue] (z) circle (1.05cm);
	\draw [double] (z) circle (0.34cm);
\end{pgfonlayer}
\end{tikzpicture}
+\begin{tikzpicture}[baseline=(z)]
\begin{feynman}[inline=(z)]
	\vertex (z);	
	\tikzfeynmanset{every vertex=dot}
	\vertex [above left=0.71cm and 0.71cm of z, label=180:$x_1$] (x1);
	\vertex [below left=0.71cm and 0.71cm of z, label=180:$x_2$] (x2);
	\vertex [above right=0.71cm and 0.71cm of z, label=0:$x_3$] (x3);
	\vertex [below right=0.71cm and 0.71cm of z, label=0:$x_4$] (x4);
	\vertex [above=0.3cm of z] (y1);
	\vertex [below=0.3cm of z] (y2);
	\tikzfeynmanset{every vertex={empty dot,minimum size=0mm}}
	\vertex [above=0.35cm of z] (y3);
	\vertex [below=0.35cm of z] (y4);
	\diagram* {
		(x1)--[double](y1),
		(x2)--[double](y2),
		(x3)--(y1),
		(x4)--(y2),
		(y1)--[out=350,in=10, min distance=0.4cm](y2),
		(y1)--[double,out=190,in=170, min distance=0.4cm](y2),
	};
\end{feynman}
\begin{pgfonlayer}{bg}
	\draw[blue] (z) circle (1.05cm);
\end{pgfonlayer}
\end{tikzpicture}+\begin{tikzpicture}[baseline=(z)]
\begin{feynman}[inline=(z)]
	\vertex (z);	
	\tikzfeynmanset{every vertex=dot}
	\vertex [above left=0.71cm and 0.71cm of z, label=180:$x_1$] (x1);
	\vertex [below left=0.71cm and 0.71cm of z, label=180:$x_2$] (x2);
	\vertex [above right=0.71cm and 0.71cm of z, label=0:$x_4$] (x4);
	\vertex [below right=0.71cm and 0.71cm of z, label=0:$x_3$] (x3);
	\vertex [above=0.3cm of z] (y1);
	\vertex [below=0.3cm of z] (y2);
	\tikzfeynmanset{every vertex={empty dot,minimum size=0mm}}
	\vertex [above=0.35cm of z] (y3);
	\vertex [below=0.35cm of z] (y4);
	\diagram* {
		(x1)--[double](y1),
		(x2)--[double](y2),
		(x4)--(y1),
		(x3)--(y2),
		(y1)--[out=350,in=10, min distance=0.4cm](y2),
		(y1)--[double,out=190,in=170, min distance=0.4cm](y2),
	};
\end{feynman}
\begin{pgfonlayer}{bg}
	\draw[blue] (z) circle (1.05cm);
\end{pgfonlayer}
\end{tikzpicture}\right )\cr+\mathcal O(\lambda^3).
\end{multline}
\noindent $\bullet$ The disconnected part only contains the product of two propagators and is therefore given by
\begin{align}
	\expect{\phi^+(x_1)\phi^+(x_2)\phi^-(x_3)\phi^-(x_4)}=\frac{2^3a^4}{2^2(4\pi^2)^2}\frac{1}{x_{12}^4x_{34}^2}
\end{align}
\noindent $\bullet$ The tree-level contribution can be inferred from~\eqref{eq:Cross_allDelta_int} by acting on the latter with 
\begin{equation}
    \cH_{12}=\frac{1}{x_{12}^2}\left(2\Delta-2v\frac{\partial}{\partial v}\right)\,.
\end{equation}
Thus,
\begin{equation}\label{eq:dS_12}
	W_{0,\mathrm{dS}}^{2211,4-4\epsilon}(\vec x_1,\dots,\vec x_4)=\frac{2^6 a^4}{(4\pi^2)^4}\frac14\cH_{12}\cW_0^{1111,4-4\epsilon}(\vec x_1,\dots,\vec x_4).
\end{equation}
To compute the right-hand-side we express $\cW_0^{1111,4-4\epsilon}(\vec x_1,\dots,\vec x_4)$ in parametric representation and act with $\cH_{12}$ before expanding the result in $\epsilon$. See section~3.2 of~\cite{Heckelbacher:2022fbx} and appendix~\ref{sec:cross_app} for more details with $\cW_0^{2211,4-4\epsilon}(\vec x_1,\dots,\vec x_4)$ there, related to~\eqref{eq:dS_12} as 
\begin{equation}
	W_{0,\mathrm{dS}}^{2211,4-4\epsilon}(\vec x_1,\dots,\vec x_4)=\frac{2^6a^4}{(4\pi^2)^4}\cW_0^{2211,4-4\epsilon}(\vec x_1,\dots,\vec x_4)\,.
\end{equation}
The one-loop contributions can be obtained in the same way. We observe that the sum of the first two terms contains a term like equation~\eqref{eq:propagators_squared}. The same arguments apply therefore for the cancellation of the mixed terms and, we get
\begin{align}
    &\frac12\begin{tikzpicture}[baseline=(z)]
	\begin{feynman}[inline=(z)]
		\vertex (z);	
		\tikzfeynmanset{every vertex=dot}
		\vertex [above left=0.71cm and 0.71cm of z, label=180:$x_1$] (x1);
		\vertex [below left=0.71cm and 0.71cm of z, label=180:$x_2$] (x2);
		\vertex [above right=0.71cm and 0.71cm of z, label=0:$x_3$] (x3);
		\vertex [below right=0.71cm and 0.71cm of z, label=0:$x_4$] (x4);
		\vertex [left=0.3cm of z] (y1);
		\vertex [right=0.3cm of z] (y2);
		\tikzfeynmanset{every vertex={empty dot,minimum size=0mm}}
		\vertex [above=0.35cm of z] (y3);
		\vertex [below=0.35cm of z] (y4);
		\diagram* {
			(x1)--[double](y1),
			(x2)--[double](y1),
			(x3)--(y2),
			(x4)--(y2),
		};
	\end{feynman}
	\begin{pgfonlayer}{bg}
		\draw[blue] (z) circle (1.05cm);
		\draw (z) circle (0.34cm);
	\end{pgfonlayer}
\end{tikzpicture}+\frac12\begin{tikzpicture}[baseline=(z)]
\begin{feynman}[inline=(z)]
	\vertex (z);	
	\tikzfeynmanset{every vertex=dot}
	\vertex [above left=0.71cm and 0.71cm of z, label=180:$x_1$] (x1);
	\vertex [below left=0.71cm and 0.71cm of z, label=180:$x_2$] (x2);
	\vertex [above right=0.71cm and 0.71cm of z, label=0:$x_3$] (x3);
	\vertex [below right=0.71cm and 0.71cm of z, label=0:$x_4$] (x4);
	\vertex [left=0.3cm of z] (y1);
	\vertex [right=0.3cm of z] (y2);
	\tikzfeynmanset{every vertex={empty dot,minimum size=0mm}}
	\vertex [above=0.35cm of z] (y3);
	\vertex [below=0.35cm of z] (y4);
	\diagram* {
		(x1)--[double](y1),
		(x2)--[double](y1),
		(x3)--(y2),
		(x4)--(y2),
	};
\end{feynman}
\begin{pgfonlayer}{bg}
	\draw[blue] (z) circle (1.05cm);
	\draw [double] (z) circle (0.34cm);
\end{pgfonlayer}
\end{tikzpicture}\cr
&=\frac12\frac{2^{6}a^4}{(4\pi^2)^6}\int_{(\mathbb R^D)^2} \dd^DX\dd^DY {(u\cdot
            X)^2\over \|X-Y|^4
            \|X-x_1|^4\|X-x_2|^4\|Y-x_3|^2\|Y-x_4|^2} \, .\cr
\end{align}
It is not hard to see that this integral is given by acting with $\cH_{12}$ on $\cW_{1,\mathrm{div}}^{1,4-2\epsilon,s}$ giving
\begin{align}
	W_{1,\mathrm{dS}}^{2211,4-2\epsilon,s}(\vec x_1,\dots,\vec x_4)&=\frac12\frac{2^6 a^4}{(4\pi^2)^6}\frac14\cH_{12}\cW_{1,\mathrm{div}}^{1,4-2\epsilon,s}(\vec x_1,\dots,\vec x_4),\cr
&=:\frac{2^6 a^4}{(4\pi^2)^6}\cW_{1}^{2211,4-2\epsilon,s}(\vec x_1,\dots,\vec x_4)
\end{align}
where $\cW_{1}^{2211,4-2\epsilon,s}$ is given in appendix~\ref{sec:mixed_app}.

For the last two terms we use the fact that the propagators can be expressed as (see section~2.3 of~\cite{Heckelbacher:2022fbx} for details):
\begin{align}
	\label{eq:LambdaSpecial}
	\Lambda(X,Y;1)&=-\left(a\over 2\pi\right)^2
	\left(\frac{zw}{\|X-Y|^2}+\frac{zw}{\|X-\sigma(Y)|^2}\right),\cr
	\Lambda(X,Y;2)&=-\left(a\over 2\pi\right)^2
	\left(\frac{zw}{\|X-Y|^2}-\frac{zw}{\|X-\sigma(Y)|^2}\right)\,.
\end{align}
Therefore the product appearing in the Witten diagrams is given by:
\begin{align}
	\label{eq:prop_square}
	\Lambda(X,Y;1)\Lambda(X,Y;2)&=\left(a\over 2\pi\right)^4\left[\frac{(zw)^2}{\|X-Y|^4}-\frac{(zw)^2}{\|X-\sigma(Y)|^4}\right]
\end{align}
We can unfold the region of integration of the last two diagrams from $(\cH_D^+)^2$ to $\mathbb{R}^{2D}$ by using that the measure of integration is odd under the action of the antipodal map, like the product of propagators in~\eqref{eq:prop_square}. We then have 
\begin{multline}\label{e:graphmixed}
    \begin{tikzpicture}[baseline=(z)]
\begin{feynman}[inline=(z)]
	\vertex (z);	
	\tikzfeynmanset{every vertex=dot}
	\vertex [above left=0.71cm and 0.71cm of z, label=180:$x_1$] (x1);
	\vertex [below left=0.71cm and 0.71cm of z, label=180:$x_2$] (x2);
	\vertex [above right=0.71cm and 0.71cm of z, label=0:$x_3$] (x3);
	\vertex [below right=0.71cm and 0.71cm of z, label=0:$x_4$] (x4);
	\vertex [above=0.3cm of z] (y1);
	\vertex [below=0.3cm of z] (y2);
	\tikzfeynmanset{every vertex={empty dot,minimum size=0mm}}
	\vertex [above=0.35cm of z] (y3);
	\vertex [below=0.35cm of z] (y4);
	\diagram* {
		(x1)--[double](y1),
		(x2)--[double](y2),
		(x3)--(y1),
		(x4)--(y2),
		(y1)--[out=350,in=10, min distance=0.4cm](y2),
		(y1)--[double,out=190,in=170, min distance=0.4cm](y2),
	};
\end{feynman}
\begin{pgfonlayer}{bg}
	\draw[blue] (z) circle (1.05cm);
\end{pgfonlayer}
\end{tikzpicture}\propto \int_{(\mathcal H_D^+)^2} {\dd^DX \dd^DY\over z^4w^4} \left({(zw)^2\over \|X-Y|^4}- {(zw)^2\over \|X-\sigma(Y)|^4}\right) \cr\times{(zw)^3\over \|X-x_1|^4\|Y-x_2|^4 \|X-x_3|^2\|Y-x_4|^2}
\end{multline}
since $\|X-\sigma(Y)|^2=(\vec x-\vec y)^2+(z+w)^2$ we unfold the $Y$ integral to the full space $\mathbb R^D$ to get
\begin{equation}
   ~\eqref{e:graphmixed}= \int_{\mathcal H_D^+} {\dd^DX\over z^4} \int_{\mathbb R^D} {\dd^DY\over w^4} {(zw)^2\over (\|X-Y|^2)^2} {(zw)^3\over (\|X-x_1|^2\|Y-x_2|^2)^2 \|X-x_3|^2\|Y-x_4|^2}.
\end{equation}
We then unfold the $z$ integration to the full space $\mathbb R^D$
to get
\begin{equation}\label{e:graphmixed2}
   ~\eqref{e:graphmixed}= \frac12 \int_{\mathbb R^D} {\dd^DX\over z^4} \int_{\mathbb R^D} {\dd^DY\over w^4} {(zw)^2\over (\|X-Y|^2)^2} {(zw)^3\over (\|X-x_1|^2\|Y-x_2|^2)^2 \|X-x_3|^2\|Y-x_4|^2}.
\end{equation}
Including the correct normalization we end up with
\begin{align}
    \begin{tikzpicture}[baseline=(z)]
\begin{feynman}[inline=(z)]
	\vertex (z);	
	\tikzfeynmanset{every vertex=dot}
	\vertex [above left=0.71cm and 0.71cm of z, label=180:$x_1$] (x1);
	\vertex [below left=0.71cm and 0.71cm of z, label=180:$x_2$] (x2);
	\vertex [above right=0.71cm and 0.71cm of z, label=0:$x_3$] (x3);
	\vertex [below right=0.71cm and 0.71cm of z, label=0:$x_4$] (x4);
	\vertex [above=0.3cm of z] (y1);
	\vertex [below=0.3cm of z] (y2);
	\tikzfeynmanset{every vertex={empty dot,minimum size=0mm}}
	\vertex [above=0.35cm of z] (y3);
	\vertex [below=0.35cm of z] (y4);
	\diagram* {
		(x1)--[double](y1),
		(x2)--[double](y2),
		(x3)--(y1),
		(x4)--(y2),
		(y1)--[out=350,in=10, min distance=0.4cm](y2),
		(y1)--[double,out=190,in=170, min distance=0.4cm](y2),
	};
\end{feynman}
\begin{pgfonlayer}{bg}
	\draw[blue] (z) circle (1.05cm);
\end{pgfonlayer}
\end{tikzpicture}=&\frac12\frac{2^{6}a^4}{(4\pi^2)^6}\int_{(\mathbb R^D)^2} \dd^DX\dd^DY \cr
&\times{(u\cdot X)(u\cdot Y)\over \|X-Y|^4
            \|X-x_1|^4\|Y-x_2|^4\|X-x_3|^2\|Y-x_4|^2} \, .\cr
\end{align}
Again, this integral is given by acting with $\cH_{12}$ on $\cW_{1,\mathrm{div}}^{1,4-2\epsilon,t}$ in equation~\eqref{eq:W1_fin_div_app}. The same applies to the last diagram with respect to $\cW_{1,\mathrm{div}}^{1,4-2\epsilon,u}$ and we obtain for these two contributions
\begin{align}
    W_{1,\mathrm{dS}}^{2211,4-2\epsilon,i}(\vec x_1,\dots,\vec x_4)&=\frac12\frac{2^6 a^4}{(4\pi^2)^6}\frac14\cH_{12}\cW_{1,\mathrm{div}}^{1,4-2\epsilon,i}(\vec x_1,\dots,\vec x_4),\nonumber\\*
    &=:\frac{2^6 a^4}{(4\pi^2)^6}\cW_{1}^{2211,4-2\epsilon,i}(\vec x_1,\dots,\vec x_4)
\end{align}
where $\cW_{1}^{2211,4-2\epsilon,i}$ with $i\in\{s,t,u\}$ is given in appendix~\ref{sec:mixed_app}. The complete four-point function is therefore given by
\begin{multline}\label{e:mixeds}
    \expect{\phi^+(x_1)\phi^+(x_2)\phi^-(x_3)\phi^-(x_4)}=
    \frac{2^3a^4}{(4\pi^2)^2}\Bigg[\frac{1}{x_{12}^4x_{34}^2}+\frac{2^3\lambda}{(4\pi^2)^2}
    \cW_0^{2211,4-4\epsilon}\cr
    -\frac{2^3\lambda^2}{(4\pi^2)^4}\left(-\frac{3\pi^2}{\epsilon}\cW_0^{2211,4-4\epsilon}
    +\sum_{i\in\{s,t,u\}}\cW_{1,\mathrm{finite}}^{2211,4,i}\right)\Bigg]\,.
\end{multline}
with $\cW_{1,\mathrm{finite}}^{2211,4,i}$ given in equations~\eqref{eq:W2211finite}. The correlation functions\break  $\expect{\phi^+(x_1)\phi^-(x_2)\phi^+(x_3)\phi^-(x_4)}$ and $\expect{\phi^+(x_1)\phi^-(x_2)\phi^-(x_3)\phi^+(x_4)}$ can be obtained from this result by exchanging external points accordingly. This, however, only works after regularisation as we will discuss in the next section.

\subsection{Renormalization and finite result}\label{sec:renormalization}

To simplify the calculation in EAdS we changed the normalisation of
the fields $\phi^\pm$ and the coupling constant $\lambda$ in the
auxiliary action~\eqref{eq:EAdS_action}. However if we want to
interpret our result in terms of a de Sitter calculation we have to
reverse that procedure, especially if we want to compare the $\beta$
function with the well-known flat-space result. At leading order they
should coincide, since the leading short distance divergence does not depend on the global geometry.

Following the same arguments as in section~4.2.1 in~\cite{Heckelbacher:2022fbx}, we introduce the renormalized coupling constant $\lambda_R$ through the divergent bare coupling as $\lambda=\lambda_R(a\mu)\mu^{2\epsilon}+\delta\lambda$. Then,  up to finite terms, the connected part of the four-point functions is given by
\begin{align}
\label{eq:substraction}
    \frac{2^{\sum_i\Delta_i}a^4}{(8\pi^2)^4}
    &(\mu a)^{4\epsilon}\Big(2\lambda_R\cW_0^{\Delta_1\Delta_2\Delta_3\Delta_4,4-4\epsilon}+\frac{\lambda_R^2}{16\pi^4}\frac{3\pi^2}{\epsilon}\cW_0^{\Delta_1\Delta_2\Delta_3\Delta_4,4-4\epsilon}\Big)\nonumber\\*
    &=\frac{2^{\sum_i\Delta_i}a^4 2}{(8\pi^2)^4}(\mu a)^{4\epsilon}
      \left(\lambda_R+\frac{3\lambda_R^2}{32\pi^2\epsilon}\right)\cW_0^{\Delta_1\Delta_2\Delta_3\Delta_4,4-4\epsilon}\nonumber\\*
      &=:
    \frac{2^{\Delta_1+\cdots+\Delta_4}a^4 2}{(8\pi^2)^4}\mu^{2\epsilon}\lambda\cW_0^{\Delta_1\Delta_2\Delta_3\Delta_4,4-4\epsilon}.
\end{align}
This determines the counter-term 
\begin{equation}
    \delta\lambda=-\frac{3\lambda_R^2\mu^{2\epsilon}}{32\pi^2\epsilon}
\end{equation}
while the finite $\log\mu$ contribution to $\lambda$ gives rise to the Callan-Symanzik equation
\begin{equation}
    0=\frac{\dd}{\dd\log\mu}\lambda,
\end{equation}
which leads to the leading order contribution to the beta function
\begin{equation}
    \beta=\frac{3\lambda_R^2}{16\pi^2}+\Op(\lambda_R^3)
\end{equation}
coinciding with the flat space result.

After renormalisation with a minimal subtraction scheme and restoring
the canonical normalisation of the fields and coupling constant, from
a dS point of view, we obtain the following finite results for the
four-point functions with equal external dimensions $\Delta_-=1$ or $\Delta_+=2$
\begin{multline}
    \expect{\phi^\pm(x_1)\phi^\pm(x_2)\phi^\pm(x_3)\phi^\pm(x_4)}=\frac{2^{2\Delta_\pm}a^4}{(8\pi^2)^2}\Bigg[\frac{1}{x_{12}^{2\Delta_\pm}x_{34}^{2\Delta_\pm}}\Bigg(1+v^{\Delta_\pm}+\frac{v^{\Delta_\pm}}{(1-Y)^{\Delta_\pm}}\Bigg)\cr
    -\frac{2^{2\Delta_\pm}2\lambda_R}{(8\pi^2)^2}\cW_0^{\Delta_\pm\Delta_\pm\Delta_\pm\Delta_\pm,4}+\frac{2^{2\Delta_\pm}4\lambda_R^2}{(8\pi^2)^4}\sum\limits_{i\in\{s,t,u\}}\cW_{1,\mathrm{finite}}^{\Delta_\pm\Delta_\pm\Delta_\pm\Delta_\pm,i}\Bigg],
  \end{multline}
  where $\cW_0^{1111,4}$ is given in~\eqref{eq:crossDelta1},
  $\cW_0^{2222,4}$ is given in~\eqref{eq:crossDelta2},  
  $\cW_{1,\mathrm{finite}}^{1111,i}$ are given in~\eqref{eq:Delta1_result_L0} and
  $\cW_{1,\mathrm{finite}}^{2222,i}$  are given in~\eqref{eq:Delta2_result_L0fin}.
The mixed correlator is given by
\begin{equation}
    \label{eq:mixed_ren}
    \expect{\phi^+(x_1)\phi^+(x_2)\phi^-(x_3)\phi^-(x_4)}=\frac{a^4}{8\pi^4}\Bigg[\frac{1}{x_{12}^4x_{34}^2}+\frac{ \lambda_R}{4\pi^4}\cW_0^{2211,4}+\frac{ \lambda_R^2}{128\pi^8}\sum\limits_{i\in\{s,t,u\}}\cW_{1,\mathrm{finite}}^{2211,i}\Bigg],
\end{equation}
where the term $\cW_0^{2211,4}$  is given in~\eqref{e:W2211-cross} and 
$\cW_{1,\mathrm{finite}}^{2211,i}$ are given in~\eqref{eq:W2211finite}.

Note, that we considered the tree-level four-point function in $D=4-4\epsilon$ dimensions in equation~\eqref{eq:substraction}, meaning that the counter term contains a finite piece, given by the coefficient of the $\mathcal{O}(\epsilon)$ contribution to $\cW_0^{\Delta_1\Delta_2\Delta_3\Delta_4,4-4\epsilon}$. As discussed in section~4.2.1 of~\cite{Heckelbacher:2022fbx} this is done to restore the global AdS symmetry in the bulk, guaranteeing that the renormalized four-point function transforms homogeneously under dilatations on the boundary. As a consequence one should be able to obtain the four-point functions $\expect{\phi^+(x_1)\phi^-(x_2)\phi^+(x_3)\phi^-(x_4)}$ and $\expect{\phi^+(x_1)\phi^-(x_2)\phi^-(x_3)\phi^+(x_4)}$ by simple permutation of the external points in equation~\eqref{eq:mixed_ren}, resulting in transformations on the conformal cross-ratios. 

Concretely, the correlation function
$\expect{\phi^+(x_1)\phi^-(x_2)\phi^+(x_3)\phi^-(x_4)}$ is obtained
from~\eqref{eq:mixed_ren} by  making the replacements $x_2\leftrightarrow
x_3$ which corresponds to $\zeta\to
1-\zeta,\zetab\to1-\zetab$ or $(v,1-Y)\to (1-Y,v)$. Similarly, 
$\expect{\phi^+(x_1)\phi^-(x_2)\phi^-(x_3)\phi^+(x_4)}$ is obtained from~\eqref{eq:mixed_ren} by making
the replacements $x_2\leftrightarrow x_4$ which corresponds to
$\zeta\to\frac{1}{\zeta},\zetab\to\frac{1}{\zetab}$ or $(v,1-Y)\to(1/v
, (1-Y)/v)$. We checked explicitly, that this holds for our result, providing an additional test for the loop dependent regularisation scheme introduced in~\cite{Heckelbacher:2022fbx} to restore the conformal symmetry on the boundary, which is a priori broken by naive dimensional regularisation.

\section{Conformal block expansion}
\label{sec:CBE}

We have seen in the last section that we can interpret the leading- and subleading expansion coefficients of the field at late times as operators, $\Op_1$ and $\Op_2$, of dimension $\Delta=1$ and $\Delta=2$ respectively, living on the euclidean $\mathbb{R}^3$ hypersurface at future infinity. Furthermore, since we have an auxiliary EAdS action for the correlation functions of the latter, we conclude that the theory on the boundary should, at least perturbatively, be described by a dual CFT.

In total there are five different four-point functions to be considered for describing this CFT. We write the possible OPEs between the operators $\Op_1$ and $\Op_2$ schematically as 
\begin{align}\label{eq:OPE+gen}
	\Op_1(x_1)\times\Op_1(x_2)&\sim\sum_{\tilde \Op} a^{11}_{\tilde\Op}\tilde{\Op}(x_2),\nonumber\\*
	\Op_2(x_1)\times\Op_2(x_2)&\sim\sum_{\tilde \Op}a^{22}_{\tilde\Op}\tilde{\Op}(x_2),\\*
\nonumber	
\Op_1(x_1)\times\Op_2(x_2)&\sim\sum_{\tilde \Op}a^{12}_{\tilde\Op}\tilde{\Op}(x_2)\,,
\end{align}
where $a^{ij}_{\tilde\Op}$ are  OPE coefficients and ``$\sim$'' means that the contributions of descendant operators are implicit.

In terms of conformal blocks~\cite{Dolan:2000ut}, the general form of the five four-point functions we have to consider is 
\begin{subequations}
\begin{align}
	\expect{\Op_1(x_1)\Op_1(x_2)\Op_1(x_3)\Op_1(x_4)}&=\frac{1}{x_{12}^2x_{34}^2}\sum_{\tilde \Op,l}(a^{11}_{\tilde\Op})^2\mathcal{G}_{\tilde{\Op},l},\label{eq:CBE1111}\\
	\expect{\Op_2(x_1)\Op_2(x_2)\Op_1(x_3)\Op_1(x_4)}&=\frac{1}{x_{12}^4x_{34}^2}\sum_{\tilde \Op,l}a^{22}_{\tilde\Op}a^{11}_{\tilde\Op}\mathcal{G}_{\tilde{\Op},l},\label{eq:CBE2211}\\
	\expect{\Op_2(x_1)\Op_1(x_2)\Op_2(x_3)\Op_1(x_4)}&=\frac{1}{(x_{12}^2x_{34}^2)^{\frac32}}\left(\frac{x^2_{24}}{x^2_{13}}\right)^{\frac12}\sum_{\tilde \Op,l}(a^{12}_{\tilde\Op})^2\mathcal{G}_{\tilde{\Op},l},\label{eq:CBE2121}\\
	\expect{\Op_2(x_1)\Op_1(x_2)\Op_1(x_3)\Op_2(x_4)}&=\frac{1}{(x_{12}^2x_{34}^2)^{\frac32}}\frac{(x^2_{24}x^2_{13})^{\frac12}}{x^2_{14}}\sum_{\tilde \Op,l}(a^{12}_{\tilde\Op})^2\mathcal{G}_{\tilde{\Op},l},\label{eq:CBE2112}\\
	\expect{\Op_2(x_1)\Op_2(x_2)\Op_2(x_3)\Op_2(x_4)}&=\frac{1}{x_{12}^4x_{34}^4}\sum_{\tilde \Op,l}(a^{22}_{\tilde\Op})^2\mathcal{G}_{\tilde{\Op},l}.\label{eq:CBE2222}
\end{align}
\end{subequations}
where $\mathcal{G}_{\tilde{\Op},l}$ is the conformal block  for the primary field $\tilde\Op$.
In the following we will denote the square of the OPE coefficients by capital letters, that is 
\begin{equation}
    A^{\Delta_1\Delta_2}_{\Op}:= (a^{\Delta_1\Delta_2}_{\Op})^2.
\end{equation}
Since we have no three-point functions due to the quartic vertex none of the ``single trace'' operators $\Op_1$ and $\Op_2$ will appear in the OPE.

The spectrum of ``double trace'' operators for the disconnected part can be read off from the corresponding four-point functions by conglomeration as described in~\cite{Fitzpatrick:2011dm}. The possible primary operators are given  by
\begin{equation}
    \label{eq:double_trace}
    :\Op_1\square^n\partial^l\Op_1:,\ :\Op_2\square^n\partial^l\Op_2:,\ :\Op_2\square^n\partial^l\Op_1:
  \end{equation}
  which we will denote by
\begin{equation}
    [\Op_1\Op_1]_{n,l}\,, [\Op_2\Op_2]_{n,l}\,, [\Op_2\Op_1]_{n,l}
  \end{equation}
  respectively.
  They have the corresponding scaling dimension $2+2n+l$, $4+2n+l$
and $3+2n+l$ with $n,l\in\mathbb{N}$. Recall that in the scalar four-point function we can only distinguish operators by their scaling dimension, which may be the same for different values
of $n$ and $l$. Furthermore, while the dimensions of $\Op_1$ and $\Op_2$ are determined by the (renormalized) mass $m$, which is fixed for a conformally coupled bulk scalar, we may expect that the ``double trace'' operators pick up anomalous dimensions due to the bulk interaction term.

\subsection{Correlation functions with degenerate conformal block expansion}\label{sec:correl}

Let us first consider the four-point functions~\eqref{eq:CBE1111},~\eqref{eq:CBE2211} and~\eqref{eq:CBE2222}. By examining the bulk diagrams we notice, that we will have mixing between the double trace operators in the double OPE. If the two-point function between the operators $[\Op_1\Op_1]_{n+1,l}$ and $[\Op_2\Op_2]_{n,l}$ does not vanish they are not a good basis for the conformal block expansion. Instead, we choose a basis of operators $\Op^S_{n,l}$ and $\Op^A_{n,l}$ both with scaling dimension $\Delta^{S/A}_{n,l}=2+2n+l+\Op(\lambda)$ and spin $l$ such that they are orthogonal, i.e. at $\Op(\lambda^0)$ they have the two point functions
\begin{align}\label{eq:OPE+SA}
    &\expect{\Op^S_{n,l}(x_1)\Op^A_{n,l}(x_2)}=0;\nonumber\\* &\expect{\Op^S_{n,l}(x_1)\Op^S_{n,l}(x_2)}=\expect{\Op^A_{n,l}(x_1)\Op^A_{n,l}(x_2)}=\frac12\expect{[\Op_1\Op_1]_{n,l}(x_1)[\Op_1\Op_1]_{n,l}(x_2)},
\end{align} 
where the additional factor of $1/2$ guarantees canonical normalization of the final result. Combining~\eqref{eq:OPE+gen},~\eqref{eq:double_trace} and~\eqref{eq:OPE+SA} we then write
\begin{align}
\label{eq:GFF_OPE}
    \Op_1\times\Op_1&\sim 1+\sum\limits_{n,\frac{l}{2}\in\mathbb{N}}a^{1,1}_{[\Op_1\Op_1]_{n,l}}[\Op_1\Op_1]_{n,l}
    \equiv 1+\sum\limits_{n,\frac{l}{2}\in\mathbb{N}}(a^{1,1}_{\Op^S_{n,l}}\Op^S_{n,l}+a^{1,1}_{\Op^A_{n,l}}\Op^A_{n,l})\\
    \Op_2\times\Op_2&\sim1+\sum\limits_{n,\frac{l}{2}\in\mathbb{N}}a^{2,2}_{[\Op_2\Op_2]_{n,l}}[\Op_2\Op_2]_{n,l}
    \equiv  1+\sum\limits_{n,\frac{l}{2}\in\mathbb{N}}(a^{2,2}_{\Op^S_{n,l}}\Op^S_{n,l}+a^{2,2}_{\Op^A_{n,l}}\Op^A_{n,l})\,,
\end{align}
where the OPE coefficients $a^{\Delta,\Delta}_{[\Op_\Delta\Op_\Delta]_{n,l}}$ for the generalized free field are given in the appendix~\ref{sec:OPE_CB}. 

 To find the OPE coefficients of the operators in the orthogonal basis we can express the four-point functions of the generalized free field in terms of conformal blocks as
\begin{subequations}
\label{eq:fourpoint_equalD}
\begin{align}
    \left.\expect{\Op_1(x_1)\Op_1(x_2)\Op_1(x_3)\Op_1(x_4)}\right\vert_{\lambda^0}&=\frac{1}{x_{12}^2x_{34}^2}\left(1+\sum_{n,\frac{l}{2}\in\mathbb{N}}\left(A_{\Op_{n,l}^{S}}^{1,1}+A_{\Op_{n,l}^{A}}^{1,1}\right)\frac12G^{0,0}_{\Delta_{n,l}}\right),\\
    \left.\expect{\Op_2(x_1)\Op_2(x_2)\Op_2(x_3)\Op_2(x_4)}\right\vert_{\lambda^0}&=\frac{1}{x_{12}^4x_{34}^4}\left(1+\sum_{n,\frac{l}{2}\in\mathbb{N}}\left(A_{\Op_{n,l}^{S}}^{2,2}+A_{\Op_{n,l}^{A}}^{2,2}\right)\frac12G^{0,0}_{\Delta_{n,l}}\right),\\
    \left.\expect{\Op_2(x_1)\Op_2(x_2)\Op_1(x_3)\Op_1(x_4)}\right\vert_{\lambda^0}&=\frac{1}{x_{12}^4x_{34}^2}\left(1+\sum_{n,\frac{l}{2}\in\mathbb{N}}\left(a_{\Op_{n,l}^{S}}^{2,2}a_{\Op_{n,l}^{S}}^{1,1}+a_{\Op_{n,l}^{A}}^{2,2}a_{\Op_{n,l}^{A}}^{1,1}\right)\frac12G^{0,0}_{\Delta_{n,l}}\right),\label{eq:GFF_2211}
\end{align}
\end{subequations}
with the equation for the conformal blocks $G^{a,b}_{\Delta_{n,l}}$ given in the appendix~\ref{sec:OPE_CB}. We used the fact that the conformal blocks for operators with the same dimension and spin are identical, meaning they coincide for $\Op^S_{n,l}$ and $\Op^A_{n,l}$. Comparing this expansion to the generalized free field, we see immediately that the OPE coefficients in the new basis must obey the following conditions
\begin{align}
    \label{eq:OPEcoefficients_0order}
    A^{1,1}_{\Op^S_{n,l}}+A^{1,1}_{\Op^A_{n,l}}=&2A^{1,1}_{[\Op_1\Op_1]_{n,l}};\quad A^{2,2}_{\Op^S_{n,l}}+A^{2,2}_{\Op^A_{n,l}}=2A^{2,2}_{[\Op_2\Op_2]_{n-1,l}};\nonumber\\
    &a_{\Op_{n,l}^{S}}^{2,2}a_{\Op_{n,l}^{S}}^{1,1}+a_{\Op_{n,l}^{A}}^{2,2}a_{\Op_{n,l}^{A}}^{1,1}=0\,.
\end{align}
Note, that from the second condition it follows that $a^{2,2}_{\Op^S_{0,l}}=a^{2,2}_{\Op^A_{0,l}}=0$, since $A^{2,2}_{[\Op_2\Op_2]_{-1,l}}=0$.

Eqn.~\eqref{eq:OPEcoefficients_0order} does not determine the zeroth order OPE coefficients uniquely. We have to proceed to first order in $\lambda$ to obtain additional conditions to fix them.
We expect the operators $\Op^S_{n,l}$ and $\Op^A_{n,l}$ to receive anomalous dimensions from the interaction term in the bulk
\begin{equation}
    \Delta^{S/A}=2+2n+l+\sum_{i=0}^\infty\gamma^{(i)S/A}_{n,l}
\end{equation}
with $\gamma^{(i)S/A}_{n,l}$ of order $\lambda^i$ in the coupling constant. A convenient parametrization is to expand the squared OPE coefficients and conformal blocks in $\gamma$~\cite{Heemskerk:2009pn,Fitzpatrick:2011dm}:
\begin{align}
	\mathcal{A}^{\Delta,\Delta}_{\Op_{n,l}^{S/A}}=&A^{\Delta,\Delta}_{\Op_{n,l}^{S/A}}
	+(\gamma^{(1)S/A}_{n,l}+\gamma^{(2)S/A}_{n,l})A^{\Delta,\Delta(1)}_{\Op_{n,l}^{S/A}}
	+\frac12(\gamma^{(1)S/A}_{n,l})^2A^{(2)\Delta,\Delta}_{\Op_{n,l}^{S/A}}+ \cdots \\
\nonumber	
    \mathcal{a}^{1,1}_{\Op^S_{n,l}}\mathcal{a}^{2,2}_{\Op^S_{n,l}}&=a^{1,1}_{\Op^{S/A}_{n,l}}a^{2,2}_{\Op^{S/A}_{n,l}}+(\gamma^{(1)S/A}_{n,l}+\gamma^{(2)S/A}_{n,l})a^{1,1(1)}_{\Op^{S/A}_{n,l}}a^{2,2(1)}_{\Op^{S/A}_{n,l}}
	+\frac12(\gamma^{(1)S/A}_{n,l})^2a^{1,1(2)}_{\Op^{S/A}_{n,l}}a^{2,2(2)}_{\Op^{S/A}_{n,l}}+ \cdots\\
\mathcal{G}^{0,0}_{\Delta_{(n,l)},l}=&G^{0,0}_{\Delta(n,l),l}+(\gamma^{(1)S/A}_{n,l}+\gamma^{(2)S/A}_{n,l})\underbrace{\left.\frac{\partial G^{0,0}_{\Delta,l}}{\partial\Delta}\right\vert_{\Delta(n,l)}}_{{G'}^{0,0}_{\Delta(n,l),l}}
	+\frac12(\gamma^{(1)S/A}_{n,l})^2\underbrace{\left.\frac{\partial^2G^{0,0}_{\Delta,l}}{\partial\Delta^2}\right\vert_{\Delta(n,l)}}_{{G''}^{0,0}_{\Delta(n,l),l}}+ \cdots\,, 
\end{align}
where the expansion of $\mathcal{a}^{1,1}_{\Op^S_{n,l}}\mathcal{a}^{2,2}_{\Op^S_{n,l}}$ can be obtained by expanding $\sqrt{\mathcal{A}^{2,2}_{\Op_{n,l}^{S/A}}\mathcal{A}^{1,1}_{\Op_{n,l}^{S/A}}}$, providing us with an additional consistency check for our calculation.

In the following we will go in detail through the process of extracting the anomalous dimensions and OPE coefficients up to second order in $\lambda$. Since this part is quite technical, we highlighted the main result, which are the first and second order anomalous dimensions.

\paragraph{First order calculation} 

The first order contributions in $\lambda$ to the four-point functions~\eqref{eq:fourpoint_equalD} are then given by
\begin{subequations}
\label{eq:fourpoint_equalD_firstorder}
\begin{align}
    &\left.\expect{\Op_\Delta(x_1)\Op_\Delta(x_2)\Op_\Delta(x_3)\Op_\Delta(x_4)}\right\vert_{\lambda^1}=
    \frac{1}{\left(x_{12}^2x_{34}^2\right)^\Delta}\times\nonumber\\*
    &\sum_{n,\frac{l}{2}\in\mathbb{N}}
    \left((\gamma^{(1)S}_{n,l}A^{\Delta,\Delta}_{\Op^S_{n,l}}+\gamma^{(1)A}_{n,l}A^{\Delta,\Delta}_{\Op^A_{n,l}}){G'}^{0,0}_{\Delta_{(n,l)},l}
    +\left(\gamma^{(1)S}_{n,l}A^{\Delta,\Delta(1)}_{\Op^S_{n,l}}+\gamma^{(1)A}_{n,l}A^{\Delta,\Delta(1)}_{\Op^A_{n,l}}\right)G^{0,0}_{\Delta_{(n,l)},l}\right)\label{eq:GFF1111_firstorder}\\
    &\left.\expect{\Op_2(x_1)\Op_2(x_2)\Op_1(x_3)\Op_1(x_4)}\right\vert_{\lambda^1}=
    \frac{1}{x_{12}^4x_{34}^2}
    \sum_{n,\frac{l}{2}\in\mathbb{N}}\left((\gamma^{(1)S}_{n,l}a^{1,1}_{\Op^S_{n,l}}a^{2,2}_{\Op^S_{n,l}}
    +\gamma^{(1)A}_{n,l}a^{1,1}_{\Op^A_{n,l}}a^{2,2}_{\Op^A_{n,l}}){G'}^{0,0}_{\Delta_{(n,l)},l}\right.\nonumber\\*
    &\qquad\qquad\qquad\qquad\qquad\left.+\left(\gamma^{(1)S}_{n,l}a^{2,2(1)}_{\Op^S_{n,l}}a^{1,1(1)}_{\Op^S_{n,l}}
    +\gamma^{(1)A}_{n,l}a^{2,2(1)}_{\Op^A_{n,l}}a^{1,1(1)}_{\Op^A_{n,l}}\right)
    G^{0,0}_{\Delta_{(n,l)},l}\right).\label{eq:GFF_2211_firstorder}
\end{align}
\end{subequations}
We compare this expansion to the bulk calculation. Keeping in mind
that the derivative of a conformal block produces a term $\propto\log
v$ we notice that the logarithmic terms in the four-point functions
give us three additional conditions on the free OPE coefficients
$a^{1,1}_{\Op^{S/A}_{n,l}}$ and $a^{2,2}_{\Op^{S/A}_{n,l}}$, while
also introducing two new unknown quantities in the first order
anomalous dimensions $\gamma^{(1)S}_{n,l}$ and
$\gamma^{(1)A}_{n,l}$. Comparing to the bulk results, the additional
conditions for $l=0$ are
\begin{align}
    \gamma^{(1)S}_{n,l}A^{1,1}_{\Op^S_{n,l}}+\gamma^{(1)A}_{n,l}A^{1,1}_{\Op^A_{n,l}}
    &=\frac{\lambda}{16\pi^2}A^{1,1}_{[\Op_1\Op_1]_{n,l}},\nonumber\\
    \gamma^{(1)S}_{n,l}A^{2,2}_{\Op^S_{n,l}}+\gamma^{(1)A}_{n,l}A^{2,2}_{\Op^A_{n,l}}
    &=\frac{\lambda}{16\pi^2}A^{2,2}_{[\Op_2\Op_2]_{n-1,l}}\label{eq:OPEcoefficients_1order}\\
    \gamma^{(1)S}_{n,l}a^{2,2}_{\Op^S_{n,l}}a^{1,1}_{\Op^S_{n,l}}
    +\gamma^{(1)A}_{n,l}a^{2,2}_{\Op^A_{n,l}}a^{1,1}_{\Op^A_{n,l}}
    &=\frac{\lambda}{16\pi^2}a^{1,1}_{[\Op_1\Op_1]_{n,l}}a^{2,2}_{[\Op_2\Op_2]_{n-1,l}}\,.\nonumber
\end{align}
For $n>0$, equations~\eqref{eq:OPEcoefficients_0order} and~\eqref{eq:OPEcoefficients_1order} require either $\gamma^{(1)S}_{n,l}$ or $\gamma^{(1)A}_{n,l}$ to vanish. This choice is a matter of convention as $\Op^S$ and $\Op^A$ have not been defined separately so far. We choose $\gamma^{(1)A}_{n>0,l}=0$. Then the solution for the zeroth order OPE coefficients and first order anomalous dimensions is
\begin{align}
  &A^{1,1}_{\Op^S_{n,l}}=A^{1,1}_{\Op^A_{n,l}}=A^{1,1}_{[\Op_1\Op_1]_{n,l}};\quad A^{2,2}_{\Op^S_{n,l}}=A^{2,2}_{\Op^A_{n,l}}=A^{2,2}_{[\Op_2\Op_2]_{n-1,l}},\\
  &a^{1,1}_{\Op^S_{n,l}}a^{2,2}_{\Op^S_{n,l}}=-a^{1,1}_{\Op^A_{n,l}}a^{2,2}_{\Op^A_{n,l}}
  =\sqrt{A^{1,1}_{[\Op_1\Op_1]_{n,l}}A^{2,2}_{[\Op_2\Op_2]_{n-1,l}}},
\end{align}
\begin{align}
  \boxed{\gamma^{(1)S}_{n,l}=\gamma\delta_{0,l};\qquad\gamma^{(1)A}_{n,l}=0\qquad
  \text{with }\gamma:=\frac{\lambda}{16\pi^2}\,.}
\end{align}
From the pieces without logarithmic terms we can access information about the first order OPE coefficients. Since we chose $\gamma^{(1)A}_{n,l}=0$ this determines only the OPE coefficients for $\Op^S_{n,l}$:
\begin{equation}
    A^{1,1(1)}_{\Op^S_{n,0}}=\frac12\frac{\partial}{\partial n}A^{1,1}_{\Op^S_{n,0}};\qquad
    A^{2,2(1)}_{\Op^S_{n,0}}=\frac12\frac{\partial}{\partial n}A^{2,2}_{\Op^S_{n,0}}\quad \text{for }n\geq 1\,.
\end{equation}
Note that the first order OPE coefficients of the four-point function with mixed external dimensions are determined by the four-point functions with equal dimensions as
\begin{equation}
    a^{2,2(1)}_{\Op^{S/A}_{n,0}}a^{1,1(1)}_{\Op^{S/A}_{n,0}}=\frac{A^{2,2(1)}_{\Op^{S/A}_{n,0}}A^{1,1}_{\Op^{S/A}_{n,0}}+A^{2,2}_{\Op^{S/A}_{n,0}}A^{1,1(1)}_{\Op^{S/A}_{n,0}}}{2\sqrt{A^{1,1}_{\Op^{S/A}_{n,0}}A^{2,2}_{\Op^{S/A}_{n,0}}}}\,,\quad\text{for }n\geq1\,,
\end{equation}
therefore providing an additional consistency check for the calculation, which our result passes.

For $n=0$ the situation is a bit more complicated. Since $a^{2,2}_{\Op^{S/A}_{0,l}}=0$ we do not have the additional condition on the difference of the anomalous dimensions coming from equation~\eqref{eq:GFF_2211_firstorder}. We therefore find from equation~\eqref{eq:GFF1111_firstorder} that
\begin{equation}
    \gamma^{(1)S}_{0,0}A^{1,1}_{\Op^S_{0,0}}+\gamma^{(1)A}_{0,0}A^{1,1}_{\Op^A_{0,0}}
    =2\gamma A^{1,1}_{[\Op_1\Op_1]_{0,0}}\,,
\end{equation}
and since the expansion of equation~\eqref{eq:GFF1111_firstorder} with $\Delta=2$ starts with $\Op(v^2)$ we need to have
\begin{equation}
    \gamma^{(1)S}_{0,0}A^{2,2(1)}_{\Op^S_{0,0}}+\gamma^{(1)A}_{0,0}A^{2,2(1)}_{\Op^A_{0,0}}=0.\label{eq:OPEfirstorder}
\end{equation}
The expansion of the bulk result for equation~\eqref{eq:GFF_2211_firstorder} starts already at order $\Op(v)$ but since it does not contain any $\log(v)$ terms at that order we get the additional condition
\begin{equation}
    \label{eq:OPEfirstorder_mixed}
    \gamma^{(1)S}_{0,0}a^{2,2(1)}_{\Op^S_{0,0}}a^{1,1(1)}_{\Op^S_{0,0}}
    +\gamma^{(1)A}_{0,0}a^{2,2(1)}_{\Op^A_{0,0}}a^{1,1(1)}_{\Op^A_{0,0}}=2\gamma.
\end{equation}

\paragraph{Second order calculation} 
 
At second order in $\lambda$, the contributions from the conformal block expansion are given by
\begin{subequations}
\label{eq:fourpoint_equalD_secondorder}
\begin{align}
    &\left.\expect{\Op_\Delta(x_1)\Op_\Delta(x_2)\Op_\Delta(x_3)\Op_\Delta(x_4)}\right\vert_{\lambda^2}=\frac{1}{(x_{12}^2x_{34}^2)^\Delta}\sum_{n,\frac{l}{2}\in\mathbb{N}}
    \left(\frac12\left((\gamma^{(1)S}_{n,l})^2+(\gamma^{(1)A}_{n,l})^2\right)A^{\Delta,\Delta}_{\Op^S_{n,l}}{G''}^{0,0}_{\Delta_{(n,l)},l}\right.\nonumber\\*
    &+\left((\gamma^{(1)S}_{n,l})^2A^{\Delta,\Delta(1)}_{\Op^S_{n,l}}+(\gamma^{(1)A}_{n,l})^2A^{\Delta,\Delta(1)}_{\Op^A_{n,l}}\right){G'}^{0,0}_{\Delta_{(n,l)},l}
    +\frac12\left((\gamma^{(1)S}_{n,l})^2A^{\Delta,\Delta(2)}_{\Op^S_{n,l}}+(\gamma^{(1)A}_{n,l})^2A^{\Delta,\Delta(2)}_{\Op^A_{n,l}}\right){G}^{0,0}_{\Delta_{(n,l)},l}\nonumber\\*
    &+\left.(\gamma^{(2)S}_{n,l}+\gamma^{(2)A}_{n,l})A^{\Delta,\Delta}_{\Op^S_{n,l}}{G'}^{0,0}_{\Delta_{(n,l)},l}+\left(\gamma^{(2)S}_{n,l}A^{\Delta,\Delta(1)}_{\Op^S_{n,l}}+\gamma^{(2)A}_{n,l}A^{\Delta,\Delta(1)}_{\Op^A_{n,l}}\right)G^{0,0}_{\Delta_{(n,l)},l}\right)\\
    &\nonumber\\
    &\text{and}\nonumber\\
    &\left.\expect{\Op_2(x_1)\Op_2(x_2)\Op_1(x_3)\Op_1(x_4)}\right\vert_{\lambda^2}=
    \frac{1}{x_{12}^4x_{34}^2}\sum_{n,\frac{l}{2}\in\mathbb{N}}
    \left(\frac12\left((\gamma^{(1)S}_{n,l})^2-(\gamma^{(1)A}_{n,l})^2\right)a^{1,1}_{\Op^S_{n,l}}a^{2,2}_{\Op^S_{n,l}}{G''}^{0,0}_{\Delta_{(n,l)},l}\right.\nonumber\\*
    &+\left((\gamma^{(1)S}_{n,l})^2a^{1,1(1)}_{\Op^S_{n,l}}a^{2,2(1)}_{\Op^S_{n,l}}+(\gamma^{(1)A}_{n,l})^2a^{1,1(1)}_{\Op^A_{n,l}}a^{2,2(1)}_{\Op^A_{n,l}}\right){G'}^{0,0}_{\Delta_{(n,l)},l}
    \nonumber\\*
    &+(\gamma^{(2)S}_{n,l}-\gamma^{(2)A}_{n,l})a^{1,1}_{\Op^S_{n,l}}a^{2,2}_{\Op^S_{n,l}}{G'}^{0,0}_{\Delta_{(n,l)},l}+\left(\gamma^{(2)S}_{n,l}a^{1,1(1)}_{\Op^A_{n,l}}a^{2,2(1)}_{\Op^S_{n,l}}+\gamma^{(2)A}_{n,l}a^{1,1(1)}_{\Op^A_{n,l}}a^{2,2(1)}_{\Op^A_{n,l}}\right)G^{0,0}_{\Delta_{(n,l)},l}\nonumber\\*
    &+\left.\frac12\left((\gamma^{(1)S}_{n,l})^2a^{1,1(2)}_{\Op^S_{n,l}}a^{2,2(2)}_{\Op^S_{n,l}}+(\gamma^{(1)A}_{n,l})^2a^{1,1(2)}_{\Op^A_{n,l}}a^{2,2(2)}_{\Op^A_{n,l}}\right){G}^{0,0}_{\Delta_{(n,l)},l}\right)\,,\label{eq:GFF_2211_secondorder}
\end{align}
\end{subequations}
where all single trace primaries have the same weight in the first equation.  
Again we compare this to the results from the bulk calculation. The terms proportional to  $\log(v)^2$ provide us with a consistency check between the first and second order calculation. We find
\begin{align}
    &\frac{\left(\gamma^{(1)S}_{0,0}\right)^2A^{1,1}_{\Op^S_{0,0}}+\left(\gamma^{(1)A}_{0,0}\right)^2A^{1,1}_{\Op^A_{0,0}}}
    {\left(\gamma^{(1)S}_{0,0}A^{1,1}_{\Op^S_{0,0}}+\gamma^{(1)A}_{0,0}A^{1,1}_{\Op^A_{0,0}}\right)^2}
    =\frac{1}{2A^{1,1}_{[\Op_1\Op_1]_{0,0}}};\\
    &\frac{\left(\gamma^{(1)S}_{n>0,0}\right)^2+\left(\gamma^{(1)A}_{n>0,0}\right)^2}{\left(\gamma^{(1)S}_{n>0,0}+\gamma^{(1)A}_{n>0,0}\right)^2}=\frac{\left(\gamma^{(1)S}_{n>0,0}\right)^2-\left(\gamma^{(1)A}_{n>0,0}\right)^2}{\left(\gamma^{(1)S}_{n>0,0}-\gamma^{(1)A}_{n>0,0}\right)^2}=1
\end{align}
from which it follows that $\gamma^{(1)A}_{n>0,0}=0$ in consistency with the first order calculation, while for $n=0$ we find that
\begin{equation}
    \gamma^{(1)S}_{0,0}=\gamma^{(1)A}_{0,0}=\gamma;\qquad
    A^{1,1}_{\Op^S_{0,0}}+A^{1,1}_{\Op^A_{0,0}}=2A^{1,1}_{[\Op_1\Op_1]_{0,0}}.
\end{equation}
From condition~\eqref{eq:OPEfirstorder} it follows then, that 
\begin{equation}
    A^{2,2(1)}_{\Op^S_{0,0}}+A^{2,2(1)}_{\Op^A_{0,0}}=0,
\end{equation}
and from equation~\eqref{eq:OPEfirstorder_mixed} we get
\begin{equation}
    a^{2,2(1)}_{\Op^S_{0,0}}a^{1,1(1)}_{\Op^S_{0,0}}
    +a^{2,2(1)}_{\Op^A_{0,0}}a^{1,1(1)}_{\Op^A_{0,0}}=2.
\end{equation}

The expansion of equation~\eqref{eq:GFF_2211_secondorder} starts at order $\Op(v)$, where the terms at that order contain $\log(v)$ terms and terms purely polynomial in $v,Y$. The logarithmic terms can be absorbed by imposing equation~\eqref{eq:OPEfirstorder_mixed} providing an additional consistency check between the first and second order calculation. The polynomial parts give  $\gamma^{(2)S}_{0,0}a^{2,2(1)}_{\Op^S_{0,0}}a^{1,1(1)}_{\Op^S_{0,0}}
    +\gamma^{(2)A}_{0,0}a^{2,2(1)}_{\Op^A_{0,0}}a^{1,1(1)}_{\Op^A_{0,0}}$, which can only be solved, if we go to the next order in $\lambda$.
    
The expansion of $\left.\expect{\Op_2(x_1)\Op_2(x_2)\Op_2(x_3)\Op_2(x_4)}\right\vert_{\lambda^2}$ starts at $\Op(v)$, where the terms at this order are purely polynomial in $v$ and $Y$. These terms can be absorbed by choosing
\begin{equation}
    A^{2,2(2)}_{\Op^S_{0,0}}+A^{2,2(2)}_{\Op^A_{0,0}}=1.
\end{equation}
The coefficients of the $\log(v)$ terms give us access to the sum and difference between the second order anomalous dimensions. We obtain the following results
\begin{align}
    \gamma^{(2)S}_{n>0,l>0}+\gamma^{(2)A}_{n>0,l>0}=-\frac{\gamma^2}{l(l+1)}-\frac{\gamma^2}{2n+l}+\frac{\gamma^2}{2n+l+1},\nonumber\\
  \gamma^{(2)S}_{n>0,l>0}-\gamma^{(2)A}_{n>0,l>0}=-\frac{\gamma^2}{l(l+1)}+\frac{\gamma^2}{2n+l}-\frac{\gamma^2}{2n+l+1},
\end{align}
If $l=0$ we find the following
\begin{align}
    \gamma^{(2)S}_{n>0,0}+\gamma^{(2)A}_{n>0,0}=3H^{(1)}_{2n}\gamma^2-\frac{\gamma^2}{2n(2n+1)}-\gamma^2,\nonumber\\
  \gamma^{(2)S}_{n>0,0}-\gamma^{(2)A}_{n>0,0}=3H^{(1)}_{2n}\gamma^2+\frac{\gamma^2}{2n(2n+1)}-7\gamma^2,
\end{align}
with $H^{(1)}_n=\sum_{m=1}^n 1/m$ the harmonic number, 
which implies that
\begin{empheq}[box=\fbox]{align}
  \gamma^{(2)S}_{n>0,l>0}&=-\frac{\gamma^2}{l(l+1)};\qquad \gamma^{(2)A}_{n>0,l>0}=-\frac{\gamma^2}{(2n+l)(2n+l+1)}\,,\nonumber\\*
  \gamma^{(2)S}_{n>0,0}&=3\gamma^2\sum_{m=1}^{2n}{1\over m}-4\gamma^2;\qquad \gamma^{(2)A}_{n>0,0}=-\frac{\gamma^2}{2n(2n+1)}+3\gamma^2.
\end{empheq}
Remarkably the anomalous dimensions for $\Op^S_{n>0,l>0}$ seem to be completely degenerate for all values of $n$ and the dimension for $\Op^A_{n,l>0}$ can be brought into the general form
\begin{align}
    \label{eq:DeltaAboxed}
    \boxed{\Delta^A_{n,l>0}=\bar\Delta^A_{n,l}-\frac{\gamma^2}{(\bar\Delta^A_{n,l}-2)(\bar\Delta^A_{n,l}-1)}+\Op(\gamma^3)}
\end{align}
where $\bar\Delta^A_{n,l}=\Delta^A_{n,l}\vert_{\lambda=0}=2+2n+l$.
For the $n=0$ trajectory we can again only make a statement about the sum
\begin{equation}
    \boxed{\gamma^{(2)S}_{0,0}+\gamma^{(2)A}_{0,0}=-2\gamma^2; \quad\gamma^{(2)S}_{0,l>0}+\gamma^{(2)A}_{0,l>0}=-\frac{\gamma^2}{l(l+1)}.}
\end{equation}

\subsection{Correlation functions with non-degenerate conformal block expansion}

The four-point functions $\expect{\Op_2(x_1)\Op_1(x_2)\Op_2(x_3)\Op_1(x_4)}$ and $\expect{\Op_2(x_1)\Op_1(x_2)\Op_1(x_3)\Op_2(x_4)}$ provide us with the OPE of 
\begin{equation}
\label{eq:GFF_OPE_12}
    \Op_1\times\Op_2\sim \sum\limits_{n,l\in\mathbb{N}}a^{1,2}_{[\Op_1\Op_2]_{n,l}}[\Op_1\Op_2]_{n,l}.
\end{equation}
Since the two-point function between these operators vanishes, the OPE will be regular. The double trace operators appearing in the free four-point function are the double trace operators $[\Op_1\Op_2]_{n,l}$ with scaling dimension $\Delta_{n,l}=3+2n+l$. Since they have odd dimensions for even spin and even dimensions for odd spin, they can be distinguished from the operators $\Op^S_{n,l}$ and $\Op^A_{n,l}$ in the OPE and the conformal block expansion will be non-degenerate. 

The free four-point functions are given by
\begin{subequations}\begin{align}
    	\left.\expect{\Op_2(x_1)\Op_1(x_2)\Op_2(x_3)\Op_1(x_4)}\right\vert_{\lambda^0}&=\frac{1}{(x_{12}^2x_{34}^2)^{\frac32}}\left(\frac{x^2_{24}}{x^2_{13}}\right)^{\frac12}\left(\frac{v}{1-Y}\right)^{\frac32},\label{eq:CBE2121free}\\
    	\left.\expect{\Op_2(x_1)\Op_1(x_2)\Op_1(x_3)\Op_2(x_4)}\right\vert_{\lambda^0}&=\frac{1}{(x_{12}^2x_{34}^2)^{\frac32}}\frac{\left(x^2_{24}x^2_{13}\right)^{\frac12}}{x^2_{14}}\frac{v^{\frac32}}{\sqrt{1-Y}}.\label{eq:CBE2112free}
\end{align}
\end{subequations}
Expanding in terms of conformal blocks gives
\begin{subequations}\begin{align}
    	\left.\expect{\Op_2(x_1)\Op_1(x_2)\Op_2(x_3)\Op_1(x_4)}\right\vert_{\lambda^0}&=\frac{1}{(x_{12}^2x_{34}^2)^{\frac32}}\left(\frac{x^2_{24}}{x^2_{13}}\right)^{\frac12}\sum\limits_{n,l\in\mathbb{N}}A^{2,1}_{[\Op_2\Op_1]_{n,l}}
    	G^{\frac12,\frac12}_{\Delta_{n,l}},\\
    	\left.\expect{\Op_2(x_1)\Op_1(x_2)\Op_1(x_3)\Op_2(x_4)}\right\vert_{\lambda^0}&=\frac{1}{(x_{12}^2x_{34}^2)^{\frac32}}\frac{\left(x^2_{24}x^2_{13}\right)^{\frac12}}{x^2_{14}}\sum\limits_{n,l\in\mathbb{N}}A^{2,1}_{[\Op_2\Op_1]_{n,l}}
    	G^{\frac12,-\frac12}_{\Delta_{n,l}},
\end{align}
\end{subequations}
where the squared OPE coefficients are given in the appendix~\ref{sec:OPE_CB}. A major difference with respect to the OPE of the correlation functions in the previous section is the fact that now also operators with odd spin $l$ contribute.

At first order in the bulk coupling $\lambda$ we can determine the first order anomalous dimensions and OPE coefficients through 
\begin{subequations}\begin{align}
 &  \left.\expect{\Op_2(x_1)\Op_1(x_2)\Op_2(x_3)\Op_1(x_4)}\right\vert_{\lambda^1}=\frac{1}{(x_{12}^2x_{34}^2)^{\frac32}}\left(\frac{x^2_{24}}{x^2_{13}}\right)^{\frac12}\times\nonumber\\*
    &\sum\limits_{n,l\in\mathbb{N}}
    \gamma^{(1)}_{n,l}\left(A^{2,1}_{[\Op_2\Op_1]_{n,l}}
    	{G'}^{\frac12,\frac12}_{\Delta_{n,l}}+A^{2,1 (1)}_{[\Op_2\Op_1]_{n,l}}
    	{G}^{\frac12,\frac12}_{\Delta_{n,l}}\right),\\
 &   \left.\expect{\Op_2(x_1)\Op_1(x_2)\Op_2(x_3)\Op_1(x_4)}\right\vert_{\lambda^1}=\frac{1}{(x_{12}^2x_{34}^2)^{\frac32}}\frac{\left(x^2_{24}x^2_{13}\right)^{\frac12}}{x^2_{14}}\times\nonumber\\*
    &\sum\limits_{n,l\in\mathbb{N}}
    \gamma^{(1)}_{n,l}\left(A^{2,1}_{[\Op_2\Op_1]_{n,l}}
    	{G'}^{\frac12,-\frac12}_{\Delta_{n,l}}+A^{2,1 (1)}_{[\Op_2\Op_1]_{n,l}}
    	{G}^{\frac12,-\frac12}_{\Delta_{n,l}}\right)\,.
\end{align}\end{subequations}
Comparing with the bulk calculation gives the result
\begin{equation}
    \gamma^{(1)}_{n,l}=\gamma\delta_{0,l};\qquad A^{2,1(1)}_{[\Op_2\Op_1]_{n,0}}=
    \frac12\frac{\partial}{\partial n}A^{2,1}_{[\Op_2\Op_1]_{n,0}}\,.
\end{equation}
The result is the same for both of the above four-point functions showing the consistency of the calculation.  

At second order in $\lambda$ we get the following conformal block expansion
\begin{subequations}\begin{align}
    &\left.\expect{\Op_2(x_1)\Op_1(x_2)\Op_2(x_3)\Op_1(x_4)}\right\vert_{\lambda^2}=\frac{1}{(x_{12}^2x_{34}^2)^{\frac32}}\left(\frac{x^2_{24}}{x^2_{13}}\right)^{\frac12}\times\nonumber\\*
    &\sum\limits_{n,l\in\mathbb{N}}\left[
    \gamma^{(2)}_{n,l}\left(A^{2,1}_{[\Op_2\Op_1]_{n,l}}
    	{G'}^{\frac12,\frac12}_{\Delta_{n,l}}+A^{2,1 (1)}_{[\Op_2\Op_1]_{n,l}}
    	{G}^{\frac12,\frac12}_{\Delta_{n,l}}\right)
    	\right.\nonumber\\*
    	&\left.+\frac12\left(\gamma^{(1)}_{n,l}\right)^2\left(A^{2,1}_{[\Op_2\Op_1]_{n,l}}
    	{G''}^{\frac12,\frac12}_{\Delta_{n,l}}+2A^{2,1 (1)}_{[\Op_2\Op_1]_{n,l}}
    	{G'}^{\frac12,\frac12}_{\Delta_{n,l}}+
    	A^{2,1 (2)}_{[\Op_2\Op_1]_{n,l}}
    	{G}^{\frac12,\frac12}_{\Delta_{n,l}}\right)\right],\label{eq:2121}\\
    &\left.\expect{\Op_2(x_1)\Op_1(x_2)\Op_1(x_3)\Op_2(x_4)}\right\vert_{\lambda^2}=\frac{1}{(x_{12}^2x_{34}^2)^{\frac32}}\frac{\left(x^2_{24}x^2_{13}\right)^{\frac12}}{x^2_{14}}\times\nonumber\\*
    &\sum\limits_{n,l\in\mathbb{N}}\left[
    \gamma^{(2)}_{n,l}\left(A^{2,1}_{[\Op_2\Op_1]_{n,l}}
    	{G'}^{\frac12,-\frac12}_{\Delta_{n,l}}+A^{2,1 (1)}_{[\Op_2\Op_1]_{n,l}}
    	{G}^{\frac12,-\frac12}_{\Delta_{n,l}}\right)
    	\right.\nonumber\\*
    	&\left.+\frac12\left(\gamma^{(1)}_{n,l}\right)^2\left(A^{2,1}_{[\Op_2\Op_1]_{n,l}}
    	{G''}^{\frac12,-\frac12}_{\Delta_{n,l}}+2A^{2,1 (1)}_{[\Op_2\Op_1]_{n,l}}
    	{G'}^{\frac12,-\frac12}_{\Delta_{n,l}}+
    	A^{2,1 (2)}_{[\Op_2\Op_1]_{n,l}}
    	{G}^{\frac12,-\frac12}_{\Delta_{n,l}}\right)\right]\,.\label{eq:2112}
\end{align}\end{subequations}
Again the coefficient of the $\log(v)^2$ term provides us with a consistency check between the first and second order calculation which our results pass. From either~\eqref{eq:2121} or~\eqref{eq:2112} we can determine the second order anomalous dimensions. As it should be they lead to identical results given by the following formulas:
\begin{empheq}[box=\fbox]{align}
    \gamma^{(2)}_{n,0}&=3\gamma^2\sum_{m=1}^{2n+1}{1\over m}-7\gamma^2,\nonumber\\*
    \gamma^{(2)}_{n,l>0}&=\begin{cases}
       -\frac{\gamma^2}{l(1+l)}  & \text{for } l\mod{2}=0 \cr
        -\frac{\gamma^2}{(l+2n+2)(l+2n+1)} & \text{for } l\mod{2}=1. \cr
    \end{cases}
\end{empheq}

Comparing to the results from the previous section we notice a striking similarity. The anomalous dimensions of $[\Op_1\Op_2]_{n,2l>0}$ and $\Op^S_{n,2l>0}$ are the same while for $[\Op_1\Op_2]_{n,2l+1}$ we find a form similar to $\Op^A_{n,2l}$
\begin{equation}
    \label{eq:Deltaoddboxed}
    \boxed{\Delta_{n,2l+1}=\bar\Delta_{n,2l+1}-\frac{\gamma^2}{(\bar\Delta_{n,2l+1}-2)(\bar\Delta_{n,2l+1}-1)}+\Op(\gamma^3),\qquad l\geq0}
\end{equation}
with $\bar\Delta_{n,l}=\Delta_{n,l}\vert_{\lambda=0}=3+2n+l$. Note that $\Delta^A_{n,l>0}$ in~\eqref{eq:DeltaAboxed} only has contributions for even spin, while~\eqref{eq:Deltaoddboxed} applies to odd spins.

\subsection{The whole picture}

Let us summarize the results of this rather technical section: We confirmed the proposal in~\cite{DiPietro:2021sjt,Sleight:2019hfp,Sleight:2020obc,Sleight:2021plv}, that cosmological four-point functions can be described by a CFT dual to an effective field theory in Euclidean AdS, by describing  explicitly the CFT dual to conformally coupled scalar $\phi^4$ theory at loop level. The CFT consists of two scalar single-trace operators $\Op_1$ and $\Op_2$ with scaling dimension $\Delta\in\{1,2\}$ and an infinite tower of three types of double-trace operators $\Op^S_{n,l}$, $\Op^A_{n,l}$ with dimension $\bar\Delta^{S/A}_{n,l}=2+2n+l$ and $[\Op_1\Op_2]_{n,l}$ with dimension $\bar\Delta_{n,l}=3+2n+l$. For $\Op^S_{n,l}$ and $\Op^A_{n,l}$ the spin $l$ can only take even integer values, while for $[\Op_1\Op_2]_{n,l}$ it can take all integer values.

The operator $\Op^S_{n,l}$ receives anomalous dimensions encoded in the four-point functions 
$\expect{\Op_\Delta(x_1)\Op_\Delta(x_2)\Op_\Delta(x_3)\Op_\Delta(x_4)}$ and 
$\expect{\Op_2(x_1)\Op_2(x_2)\Op_1(x_3)\Op_1(x_4)}$ and so does $\Op^A_{n,l}$. Similarly, 
the operator $[\Op_1\Op_2]_{n,l}$ receives anomalous dimensions from the four-point function  
$\expect{\Op_2(x_1)\Op_1(x_2)\Op_2(x_3)\Op_1(x_4)}$ or, equivalently,
$\expect{\Op_2(x_1)\Op_1(x_2)\Op_1(x_3)\Op_2(x_4)}$. However, the
spectrum contains operators with all integer spins instead of only
even spins, which was the case for $\Op^S_{n,l}$ and
$\Op^A_{n,l}$. Interestingly, there is a simple relation between the
anomalous dimensions of $\Op^S_{n,l}$, $\Op^A_{n,l}$ and
$[\Op_1\Op_2]_{n,l}$ given by
\begin{empheq}[box=\fbox]{equation}
  \gamma^{(2)}_{n,2l>0}=\gamma^{(2)S}_{n,2l>0}, \qquad
   \gamma^{(2)}_{n,2l+1>0}=\gamma^{(2)A}_{n,2l+2} \qquad l>0.
\end{empheq}
This relation seems to suggest a symmetry between the operators
$\Op^A_{n,l}, \Op^S_{n,l}$ and $[\Op_1\Op_2]_{n,l}$, which could have
several origins. One possible explanation is the special choice for
the scaling dimension of the single-trace operators,
$\Delta_\pm\in\{1,2\}$. It is easily checked that for different values
of $\Delta_\pm$ the relative coefficients between the vertices
in~\eqref{eq:EAdS_action_phi4} change and even new vertices of the
form ${\phi^+}^3\phi^-$ are generated. The cancellation of the
elliptic sector, discussed in section~\ref{subsec:fourpoint}, does not
occur anymore, and we expect the integrals to have a very different structure. As we do not have a simple form for the propagator for general values of $\Delta$ the technical implementation of the explicit loop calculation, necessary to check this claim, is much more involved, and we leave it for future studies. For conformal coupling in odd $d$ the propagator simplifies to a rational function of $K$ and the auxiliary EAdS action~\eqref{eq:EAdS_action_d3} is always the same. We therefore expect the general structure of the results, including the apparent symmetry to hold for those cases as well.

On the other hand, for generic scaling dimension of the single trace operators, the action~\eqref{eq:EAdS_action} still displays a symmetry due to the fact that all vertices have the same coupling constant $\lambda$, which look fine-tuned in the general class of $\phi^4$ theories in EAdS. Possibly, the apparent symmetry in the anomalous dimensions of the double trace operators is related to this.

Comparing with previous work~\cite{Heckelbacher:2020nue,Heckelbacher:2022fbx} we can draw the following picture.
Starting from the theory in the bulk we can calculate either the Bunch-Davies wave function ~\cite{Heckelbacher:2020nue} or the cosmological correlation functions as we did here. The Bunch-Davis wavefunction is defined as 
\begin{align}
	\Psi[\phi_0(x)]&=\lim\limits_{\eta'\rightarrow-\infty(1+i\varepsilon)}\int\limits_{\substack{\phi(0,x)=\phi_0(x)\\\phi(\eta',x)=0}}\mathcal{D}\phi\eul^{iS[\phi]}\quad\text{or}\cr
	\tilde\Psi[\pi_0(x)]&=\int\mathcal{D}\phi_0\eul^{i\int\dd^3 x\phi_0(x)\pi_0(x)}\Psi[\phi_0]\,,\label{eq:BD_wavefunction}
\end{align}
where $\phi_0$ and $\pi_0(x)$ denote the value of the bulk field and its canonically conjugate momentum at the boundary respectively. From a dS point of view $\Psi[\pi_0]$ corresponds to choosing Neumann instead of Dirichlet boundary conditions at future infinity.

Performing a semiclassical expansion of~\eqref{eq:BD_wavefunction} one finds  that the Bunch-Davis wave function has an interpretation as a generating functional for a CFT at future infinity. 
A conformally coupled scalar field in dS, without self-interactions, will give rise to a direct product of CFTs of two generalized free fields, where $\Psi[\phi_0(x)]$ corresponds to the external dimension $\Delta=2$ while $\Psi[\pi_0(x)]$ to $\Delta=1$. Introducing interactions in the bulk theory deforms the theory on the boundary. However, no non-trivial OPEs between $\Op_1$ and $\Op_2$ are introduced. Thus, the deformations will only affect the $\Delta=1$ and $\Delta=2$ sector separately and the theory keeps its product structure. In~\cite{Heckelbacher:2020nue} it was shown that the deformed CFT obtained in this way is identical to that obtained from a bulk theory in EAdS considered in~\cite{Bertan:2018afl,Bertan:2018khc,Heckelbacher:2022fbx}.

The cosmological correlator CFT introduces non-trivial OPE's between $\Op_1$ and $\Op_2$. Thus, the deformed CFT looses its product structure. Additionally, a new tower of double trace operators $[\Op_1\Op_2]_{n,l}$ receives anomalous dimensions due to the new mixing vertex introduced by the Schwinger-Keldysh formalism. Curiously we noticed, that the anomalous dimensions generated for these new operators are the same as the ones already found for $\Op^S_{n,l}$ and $\Op^A_{n,l}$.

There is, however, a relation between the CFT of the Bunch-Davies wave function and that of cosmological correlators. This can be seen by expressing a cosmological correlation function as 
\begin{equation}
    \label{eq:cosmo_corr_wavefunction}
    \expect{\phi_0(x_1)\phi_0(x_2)\phi_0(x_3)\phi_0(x_4)}=\int\mathcal{D}\phi_0\Psi^*[\phi_0]\Psi[\phi_0]\phi_0(x_1)\phi_0(x_2)
    \phi_0(x_3)\phi_0(x_4)
\end{equation}
or, equivalently,
\begin{multline}
    \label{eq:cosmo_corr_wavefunction_pi}
    \expect{\phi_0(x_1)\phi_0(x_2)\phi_0(x_3)\phi_0(x_4)}=\cr\int\mathcal{D}\phi_0\mathcal{D}\pi_0\eul^{i\int\dd^3 x\phi_0(x)\pi_0(x)}\tilde\Psi[\pi_0]\Psi[\phi_0]\phi_0(x_1)\phi_0(x_2)
    \phi_0(x_3)\phi_0(x_4)\,,
\end{multline}
where in the second step we used the inverse Fourier transformation of~\eqref{eq:BD_wavefunction} as is explained in~\cite{DiPietro:2021sjt}. Analogous expressions exist for $\pi_0(x)$. The CFT of cosmological correlators can therefore be understood as a functional integral over the wavefunction CFTs with all possible boundary conditions, where the mixing between the two kinds boundary conditions contained in the Fourier exponential. This is analogous to the mixing vertex that was introduced in section~\ref{sec:schwinger_keldysh} resulting from the Schwinger-Keldysh contour. 

Finally, let us note, that the expression~\eqref{eq:cosmo_corr_wavefunction} is merely of conceptual value since it requires the exact knowledge of the wavefunctionals to perform the integral. From~\eqref{eq:cosmo_corr_wavefunction} it is not even clear that the result of the functional integration should preserve conformal symmetry. Computationally, the way to go is through the Schwinger-Keldysh formalism and the auxiliary EAdS action, introduced in~\cite{DiPietro:2021sjt} and reviewed in section~\ref{sec:auxiliary_action}.  The two different ways to deform the generalized free field is schematically depicted in figure~\ref{fig:CFT_deformations}.

\section{Outlook}
\label{sec:outlook}
The goal of this work was to extend the technique of mapping EAdS Witten diagrams to flat space Feynman integrals, developed in~\cite{Heckelbacher:2022fbx}, to calculate cosmological correlation functions in a de Sitter background. We achieved this goal for a conformally coupled,  field with quartic self-interaction, by applying the Schwinger-Keldysh formalism in the form of~\cite{DiPietro:2021sjt}, where it was shown that the calculation can be mapped to an equivalent problem for an auxiliary EAdS action.

We succeeded to extract anomalous dimensions of ``double trace''
operators appearing in the conformal block expansion of the four point
functions up to one-loop order. As suspected, we find that the
cosmological correlator CFT  differs from the Bunch-Davies wave
function CFT. Furthermore, there is no straightforward way to obtain the conformal data of the latter from the former.

Interestingly, we find an apparent symmetry between different operators in the OPE's. We expect this to be explained by either the special choice of the field masses, the constraints coming from the Schwinger-Keldysh contour, or a combination of both. To further investigate this phenomenon, one would have to consider different masses of the fields which, however, is very nontrivial due to the complicated structure of the propagator in those cases.

Another way to proceed would be to test if this symmetry still holds
for higher-loop contributions. The cancellations in the loop integrals
discussed in section~\ref{subsec:fourpoint}, points to some
simplifications regarding the corresponding calculation in
EAdS. Especially,  the diagrams given by multiple bubbles attached after
one another, which are expressible in terms of single-valued multiple polylogarithms at any loop order.

One can also try to make contact with the cosmological bootstrap
program by expressing our results for the two- and four-point function
in momentum space (with respect to the three-dimensional space-like hypersurface). This can be of use since, to our knowledge, loop corrections have not been available in that formalism so far. It would be interesting to analyze the connection with the position space results of this work. 

Another interesting avenue is to make contact with inflationary cosmology which deviates form the de Sitter geometry but, for the two-point function in momentum space the violation of scale-invariance manifests itself only in the spectral index. Perhaps, there is a similarly tractable pattern for three- and four-point functions along the lines of~\cite{Arkani-Hamed:2018kmz}.

	\section*{Acknowledgments}
	
	We thank Shota Komatsu, Alexander Zhiboedov and Arthur Lipstein for discussions. The research of P.V. has received funding from the ANR
	grant ``Amplitudes'' ANR-17- CE31-0001-01, and the ANR grant ``SMAGP''
	ANR-20-CE40-0026-01. The work of E.S. was partially supported by the European Research Council (ERC) under the European Union’s Horizon 2020 research and innovation programme (grant agreement No 101002551) and by the Fonds de la Recherche Scientifique --- FNRS under Grant No. F.4544.21 The work of T.H. and I.S. was funded by the Excellence Cluster Origins of the DFG under Germany’s Excellence Strategy EXC-2094 390783311.
	
\appendix

\section{Cross diagrams}
\label{sec:cross_app}
The leading term is given by the $\Delta=1$ result from EAdS, which in dimensional regularisation in $D=4-4\epsilon$ have been evaluated in appendix~B of~\cite{Heckelbacher:2022fbx} and  is given by
\begin{equation}
		\label{eq:crossDelta1}
		\cW_0^{1111,4-4\epsilon}(\zeta,\zetab)=\frac{\pi^2}{x_{12}^2x_{34}^2}\left(\zeta\zetab\frac{ 2 i D(\zeta,\zetab)}{\zeta-\zetab}+\epsilon \cW_{0,\epsilon}^{1111,4}(\zeta,\zetab) +\mathcal O(\epsilon^2)\right).
	\end{equation}
	with
		\begin{equation}
		\cW_{0,\epsilon}^{1111,4}(\zeta,\zetab)=\frac{\zeta\zetab \pi^{2}}{x_{12}^2x_{34}^2}\left(\frac{f_1(\zeta,\zetab)}{\zeta-\zetab}-\frac{2iD(\zeta,\zetab)}{\zeta-\zetab}\log(\zeta\zetab)+
			\frac{2iD(\zeta,\zetab)}{\zeta-\zetab}\log((1-\zeta)(1-\zetab))\right)\,,
	\end{equation}
	where  $D(\zeta,\bar\zeta)$ is the Bloch-Wigner diloagarithm in~\eqref{eq:BlochWigner} and $f_1(\zeta,\bar\zeta)$ is given in~\eqref{e:f1}.
The sub-leading terms are given by either acting with $\cH_{12},\cH_{13}$ or $\cH_{14}$ on the cross term for $\Delta=1$, before expanding in $D=4-4\epsilon$. We obtain the following parametric representations
\begin{multline}
	\cW_0^{2211,4-4\epsilon}=\frac{\pi^{2-2\epsilon}(\zeta\zetab)^2\Gamma(2-2\epsilon)}{4\Gamma(1-4\epsilon)x_{12}^4x_{34}^2}\cr\times\int\limits_{(\mathbb{RP}^+)^2}
	\frac{d\alpha_1 d\alpha_2 d\alpha_3\alpha_1\alpha_2^{-4\epsilon}\alpha_3}
	{(\alpha_1+\alpha_2+\alpha_3)\left(\alpha_2\alpha_3(1-\zeta)(1-\zetab)+\alpha_1(\alpha_2+\alpha_3\zeta\zetab)\right)^{2-2\epsilon}}.
\end{multline}
We obtain for the $\Op(1)$ terms
\begin{equation}\label{e:W2211-cross}
	\frac{2x_{12}^4x_{34}^2}{\pi^2(\zeta\zetab)^2}\cW_0^{2211,4}=\frac{(\zeta+\zetab-2)2iD(\zeta,\zetab)}{(\zeta-\zetab)^3}
	-\frac{\zeta+\zetab-2\zeta\zetab}{2\zeta\zetab(\zeta-\zetab)^2}\log((1-\zeta)(1-\zetab))-\frac{\log(\zeta\zetab)}{(\zeta-\zetab)^2}.
\end{equation}
The sub-sub-leading term is given by the $\Delta=2$ result from EAdS, which in dimensional regularisation in $D=4-4\epsilon$ is given by
{\small\begin{multline}
		\label{eq:crossDelta2}
		\cW_0^{2222,4}(\zeta,\zetab)=\frac{3\pi^2}{4 x_{12}^4x_{34}^4}\cr 
		\times\Bigg((\zeta\zetab)^2\Big(\frac{4 \zeta ^2 \zetab^2-(\zeta +\zetab)^3+2 \zeta  \zetab (\zeta +\zetab)^2+2 (\zeta +\zetab)^2-8 \zeta  \zetab (\zeta +\zetab)+4\zeta  \zetab}{(\zeta-\zetab)^4}\frac{2i D(\zeta,\zetab)}{\zeta-\zetab}\cr
		+\frac{(\zeta +\zetab)^2-3 \zeta  \zetab (\zeta +\zetab)+2 \zeta  \zetab}{(\zeta-\zetab)^4}\log(\zeta\zetab)\cr
		+\frac{3 \zeta  \zetab (\zeta +\zetab)-2 (\zeta +\zetab)^2+3 (\zeta +\zetab)-4 \zeta  \zetab}{(\zeta-\zetab)^4}\log((1-\zeta)(1-\zetab))+\frac{1}{(\zeta-\zetab)^2}\Big)+\epsilon \cW_{0,\epsilon}^{2222,4}+\mathcal O(\epsilon^2)\Bigg)\,.
	\end{multline}}
	with the coefficient of the order $\epsilon$ term
		{\scriptsize\begin{align}
			\label{eq:crossD2_epsilon}
			&\cW_{0,\epsilon}^{2222,4}=\frac{3(\zeta\zetab)^2\left(-\left(\zeta+\zetab\right)^{3}+2 \left(\zeta+\zetab\right)^{2} \zeta \zetab+2 \left(\zeta+\zetab\right)^{2}-8 \zeta \zetab \left(\zeta+\zetab\right)+4 \zeta^{2} \zetab^{2}+4 \zeta \zetab\right)}{2(\zeta-\zetab)^5}f_2\cr
			&-\frac{4i(\zeta\zetab)^2\left(-3 \left(\zeta+\zetab\right)^{3}+5 \left(\zeta+\zetab\right)^{2} \zeta \zetab+5 \left(\zeta+\zetab\right)^{2}-12 \zeta \zetab \left(\zeta+\zetab\right)+4 \zeta^{2} \zetab^{2}+4 \zeta \zetab\right)D(\zeta,\zetab)}{(\zeta-\zetab)^5}\cr
			&+\frac{3(\zeta\zetab)^2\left(-2 \left(\zeta+\zetab\right)^{2}+3 \zeta \zetab \left(\zeta+\zetab\right)+3 \zeta+3 \zetab-4 \zeta \zetab\right)}{2(\zeta-\zetab)^4}\left(\Mpl{1}{\zeta}\Mpl{1}{\zetab}+\Mpl{1,1}{1,\zeta}+\Mpl{1,1}{1,\zetab}\right)\cr
			&-\frac{3\zeta\zetab\left(-\left(\zeta+\zetab\right)^{2} \zeta \zetab+3 \left(\zeta+\zetab\right) \zeta^{2} \zetab^{2}-2 \zeta^{2} \zetab^{2}\right)}{2(\zeta-\zetab)^4}\log(\zeta\zetab)\log((1-\zeta)(1-\zetab))\cr
			&-\frac{\zeta\zetab\left(\left(\zeta+\zetab\right)^{3} \zeta \zetab+\left(\zeta+\zetab\right)^{3}-18 \left(\zeta+\zetab\right)^{2} \zeta \zetab+8 \left(\zeta+\zetab\right) \zeta^{2} \zetab^{2}+8 \zeta \zetab \left(\zeta+\zetab\right)+24 \zeta^{2} \zetab^{2}\right)}{4(\zeta-\zetab)^4}\log((1-\zeta)(1-\zetab))\cr
			&+\frac{3(\zeta\zetab)^2\left(-\left(\zeta+\zetab\right)^{2}+3 \zeta \zetab \left(\zeta+\zetab\right)-2 \zeta \zetab\right)}{4(\zeta-\zetab)^4}\log^2(\zeta\zetab)+\cr
			&+\frac{(\zeta\zetab)^2\left(-\left(\zeta+\zetab\right)^{4}+\left(\zeta+\zetab\right)^{3} \zeta \zetab+10 \left(\zeta+\zetab\right)^{3}-18 \left(\zeta+\zetab\right)^{2} \zeta \zetab+8 \left(\zeta+\zetab\right) \zeta^{2} \zetab^{2}-8 \left(\zeta+\zetab\right)^{2}-4 \zeta \zetab \left(\zeta+\zetab\right)+16 \zeta^{2} \zetab^{2}+8 \zeta \zetab\right)\log(\zeta\zetab)}{4(\zeta-\zetab)^4(1-\zeta)(1-\zetab)}\cr
	\end{align}}

\section{One-loop diagrams}
\label{sec:oneloop_appendix}

\subsection{Leading term $\Delta_1=\Delta_2=\Delta_3=\Delta_4=1$}
\label{sec:delta1_app}
The leading term is given by the correlation function of the $\Delta=1$ scalar field in EAdS, with only the divergent part contributing. 
\begin{align}
	\label{eq:W1_fin_div_app}
	\cW_{1,\mathrm{div}}^{\Delta,4-2\epsilon,s}\!\!\!\!=&\frac{(\zeta\zetab)^\Delta}{(x_{12}^2x_{34}^2)^\Delta}\!\!\int\limits_{\mathbb{R}^{2D}}\frac{\dd^{4-2\epsilon} X_1\dd^{4-2\epsilon} X_2(u\cdot X_1)^{2\Delta-2}(u\cdot X_2)^{2\Delta-2}}{\|X_1|^{2\Delta}\|X_1-u_\zeta|^{2\Delta}\|X_2-u_1|^{2\Delta-4\epsilon}\|X_1-u_1|^{-4\epsilon}\|X_1-X_2|^4}\,,\cr
	\cW_{1,\mathrm{div}}^{\Delta,4-2\epsilon,t}\!\!\!\!=&\frac{(\zeta\zetab)^\Delta}{(x_{12}^2x_{34}^2)^\Delta}\!\!\int\limits_{\mathbb{R}^{2D}}\frac{\dd^{4-2\epsilon} X_1\dd^{4-2\epsilon} X_2(u\cdot X_1)^{2\Delta-2}(u\cdot X_2)^{2\Delta-2}}{\|X_1|^{2\Delta}\|X_2-u_\zeta|^{2\Delta}\|X_2-u_1|^{2\Delta-4\epsilon}\|X_1-u_1|^{-4\epsilon}\|X_1-X_2|^4}\cr
	\cW_{1,\mathrm{div}}^{\Delta,4-2\epsilon,u}\!\!\!\!=&\frac{(\zeta\zetab)^\Delta}{(x_{12}^2x_{34}^2)^\Delta}\!\!\int\limits_{\mathbb{R}^{2D}}\frac{\dd^{4-2\epsilon} X_1\dd^{4-2\epsilon} X_2(u\cdot X_1)^{2\Delta-2}(u\cdot X_2)^{2\Delta-2}}{\|X_1|^{2\Delta}\|X_2-u_\zeta|^{2\Delta}\|X_1-u_1|^{2\Delta-4\epsilon}\|X_2-u_1|^{-4\epsilon}\|X_1-X_2|^4}
\end{align}

The parametric representation is given by

\noindent $\bullet$  For the $s$-channel
\begin{equation}
		\label{eq:W1div_param}
		\cW_{1,\mathrm{div}}^{1,4-2\epsilon,s} =\frac{2\pi^{4-2\epsilon}\zeta\zetab}{\Gamma(-2\epsilon)x_{12}^2x_{34}^2}
		\int\limits_{(\mathbb{RP}^+)^4}\prod\limits_{i=1}^5\dd\alpha_i
             \frac{\alpha_3^{-1-2\epsilon}\alpha_1^{-2\epsilon}\alpha_5(U^s)^{-1-\epsilon}}{\left(F^s\right)^{1-2\epsilon}}
              \end{equation}
              with
              \begin{align}
                U^s&:=
                     (\alpha_2+\alpha_3+\alpha_4)\alpha_5+(\alpha_2+\alpha_3+\alpha_4+\alpha_5)\alpha_1
                \cr
                F^s&:=\alpha_4(\alpha_3\alpha_5+\alpha_1(\alpha_3+\alpha_5))(1-\zeta)(1-\zetab)+\alpha_2\alpha_4(\alpha_1+\alpha_5)\zeta\zetab\cr
                     &+\alpha_2(\alpha_3\alpha_5+\alpha_1(\alpha_3+\alpha_5))
  \label{e:UFs}            \end{align}

\noindent $\bullet$  For the $t$-channel              
          \begin{equation}
		\cW_{1,\mathrm{div}}^{1,4-2\epsilon,t} =\frac{2\pi^{4-2\epsilon}\zeta\zetab}{\Gamma(-2\epsilon)x_{12}^2x_{34}^2}
		\int\limits_{(\mathbb{RP}^+)^4}\prod\limits_{i=1}^5\dd\alpha_i
               \frac{\alpha_2^{-1-2\epsilon}\alpha_3^{-2\epsilon}\alpha_5(U^t)^{-1-\epsilon}}{\left(F^t\right)^{1-2\epsilon}}
              \end{equation}
              with
              \begin{align}
                U^t&:=(\alpha_1+\alpha_2)(\alpha_3+\alpha_4)+(\alpha_1+\alpha_2+\alpha_3+\alpha_4)\alpha_5\cr
                     F^t&:=\alpha_4((\alpha_1+\alpha_2)\alpha_3+(\alpha_2+\alpha_3)\alpha_5)(1-\zeta)(1-\zetab)+\alpha_1\alpha_2(\alpha_3+\alpha_4+\alpha_5)\cr
                          &+\alpha_1\alpha_5(\alpha_3+\alpha_4\zeta\zetab)
 \label{e:UFt}             \end{align}

\noindent $\bullet$  For the $u$-channel              
          \begin{equation}
		\cW_{1,\mathrm{div}}^{1,4-2\epsilon,u} =\frac{2\pi^{4-2\epsilon}\zeta\zetab}{\Gamma(-2\epsilon)x_{12}^2x_{34}^2}
		\int\limits_{(\mathbb{RP}^+)^4}\prod\limits_{i=1}^5\dd\alpha_i\frac{\alpha_1^{-1-2\epsilon}\alpha_4^{-2\epsilon}\alpha_5(U^u)^{-1-\epsilon}}{\left(Fû\right)^{1-2\epsilon}}
              \end{equation}
              with
                \begin{align}
                U^u&:=(\alpha_1+\alpha_2)(\alpha_3+\alpha_4)+(\alpha_1+\alpha_2+\alpha_3+\alpha_4)\alpha_5\cr
                     F^u&:=\alpha_3\alpha_4\alpha_5+\alpha_1\alpha_3(\alpha_4+\alpha_5)+\alpha_2(\alpha_4\alpha_5+\alpha_1(\alpha_3+\alpha_4+\alpha_5))(1-\zeta)(1-\zetab)\cr
                          &+\alpha_2\alpha_3(\alpha_4+\alpha_5\zeta\zetab).
  \label{e:UFu}            \end{align}
The result is given by
\begin{equation}
	\label{eq:Delta1_result}
	\cW_{1,\mathrm{div}}^{1,4-2\epsilon,i}=-\frac{\pi^2}{\epsilon}\cW_0^{1,4-4\epsilon}+\cW_{1,\mathrm{finite}}^{1111,i}\quad
        \text{with} i=s,t,u\,,
\end{equation}
where $\cW_{1,\mathrm{finite}}^{1111,i}$ for each channel is given by
\begin{equation}
	\label{eq:Delta1_result_L0}
	\cW_{1,\mathrm{finite}}^{1111,i}=\frac{\pi^4 v}{2x^2_{12}x^2_{34}}L_0^{1,i}
\end{equation}
where the integrals $L_0^{1,i}$ for $i\in\{s,t,u\}$ are known from EAdS calculations and
obtained in appendix~C.1.3 of~\cite{Heckelbacher:2022fbx}, with the
result
\begin{align}\label{e:L0one}
	L_0^{1,s}(\zeta,\bar\zeta)&={f_1(\zeta,\bar\zeta)-2i
		\log(\zeta\bar\zeta) D(\zeta,\bar\zeta)\over
		\zeta-\bar\zeta}  \\
	L_0^{1,t}(\zeta,\bar\zeta)&={f_1(\zeta,\bar\zeta)-2i
		\log((1-\zeta)(1-\bar\zeta)) D(\zeta,\bar\zeta)\over
		\zeta-\bar\zeta}\\
	L_0^{1,u}(\zeta,\bar\zeta)&={f_1(\zeta,\bar\zeta)\over
		\zeta-\bar\zeta} \,.
\end{align}

\subsection{Sub-leading term  $\Delta_1=\Delta_2=2$ $\Delta_3=\Delta_4=1$}
\label{sec:mixed_app}
The subleading terms are given by either acting with $\cH_{12},\cH_{13}$ or $\cH_{14}$ on the divergent part of the $\Delta=1$ result. In the parametric representation we obtain

\noindent $\bullet$ For the $s$-channel
\begin{equation}
	\cW_1^{2211,4-2\epsilon,s} =\frac{\pi^{4-2\epsilon}(\zeta\zetab)^2(1-2\epsilon)}{4\Gamma(-2\epsilon)x_{12}^4x_{34}^2}\int\limits_{(\mathbb{RP}^+)^4}\prod_{i=1}^{4}d\alpha_i
	 \frac{\alpha_2 \alpha_4 \alpha_5 (\alpha_1+\alpha_5) \alpha_1^{-2 \epsilon } \alpha_3^{-2 \epsilon -1} (U^s)^{-\epsilon -1}}
	{ (F^s)^{-2 (\epsilon -1)}}
            \end{equation}
            with $U^s$ and $F^s$ given in~\eqref{e:UFs}.

 \noindent $\bullet$ For the $t$-channel
         \begin{equation}
 	\cW_1^{2211,4-2\epsilon,t} =\frac{\pi^{4-2\epsilon}(\zeta\zetab)^2}{4\Gamma(-2\epsilon))x_{12}^4x_{34}^2}\int\limits_{(\mathbb{RP}^+)^4}\prod_{i=1}^{4}d\alpha_i
 	 \frac{\alpha_1\alpha_2^{-1-2\epsilon}\alpha_4\alpha_5^2(U^t)^{-1-\epsilon}(1-2\epsilon)}
 	{(F^t)^{-2 (\epsilon -1)} }
      \end{equation}
      with  $U^t$ and $F^t$ given in~\eqref{e:UFt}.

\noindent $\bullet$   For the $u$-channel   
          \begin{equation}
	\cW_1^{2211,4-2\epsilon,u} =\frac{\pi^{4-2\epsilon}(\zeta\zetab)^2}{4\Gamma(-2\epsilon))x_{12}^4x_{34}^2}\int\limits_{(\mathbb{RP}^+)^4}\prod_{i=1}^{4}d\alpha_i
	 \frac{\alpha_1^{-1-2\epsilon}\alpha_2\alpha_3\alpha_4^{-2\epsilon}\alpha_5^2(Uû)^{-1-\epsilon}(1-2\epsilon)}{(F^u)^{-2 (\epsilon -1)}}
       \end{equation}
 with  $U^u$ and $F^u$ given in~\eqref{e:UFu}.

	Integrating over the Feynman  parameters and expanding in $\epsilon$ we find the following structure for each diagram
	\begin{equation}
		\cW_1^{2211,4-2\epsilon,i}=-\frac{\pi^2}{\epsilon}\cW_0^{2211,4-4\epsilon}+\cW_{1,\mathrm{finite}}^{2211,i}+\Op(\epsilon^2)
                ~\text{with}~ i=s,t,u
	\end{equation}
	where the finite part $\cW_{1,\mathrm{finite}}$ for each diagram is given by
	\begin{subequations}
	\label{eq:W2211finite}
	\begin{multline}
		\frac{4x_{12}^4x_{34}^2}{\pi^4(\zeta\zetab)^2}\cW_{1,\mathrm{finite}}^{2211,s} =
		\frac{\zeta+\zetab-2}{(\zeta-\zetab)^3}f_1-\frac{(\zeta+\zetab-2)\log(\zeta\zetab)}{(\zeta-\zetab)^3}2iD(\zeta,\zetab)+
		\frac{4i D(\zeta,\zetab)}{\zeta\zetab(\zeta-\zetab)}\cr
		 -\frac{\log(\zeta\zetab)(\log((1-\zeta)(1-\zetab))-\log(\zeta\zetab)+4)}{(\zeta-\zetab)^2}\cr
		 -\frac{2(\zeta+\zetab-2\zeta\zetab)\log((1-\zeta)(1-\zetab))}{\zeta\zetab(\zeta-\zetab)^2}
                    \end{multline}
          \begin{equation}
		\frac{4x_{12}^4x_{34}^2}{\pi^4(\zeta\zetab)^2}\cW_{1,\mathrm{finite}}^{2211,t} =
		\frac{\zeta+\zetab-2}{(\zeta-\zetab)^3}(f_1+2f_2)+\frac{2iD(\zeta,\zetab)}{\zeta\zetab(\zeta-\zetab)}-\frac{4\log(\zeta\zetab)}{(\zeta-\zetab)^2}+\frac{\zeta+\zetab-2\zeta\zetab}{\zeta\zetab(\zeta-\zetab)^2}f_3
                   \end{equation}
          \begin{multline}
		\frac{4x_{12}^4x_{34}^2}{\pi^4(\zeta\zetab)^2}\cW_{1,\mathrm{finite}}^{2211,u} =
		\frac{\zeta+\zetab-2}{(\zeta-\zetab)^3}f_1+\frac{2iD(\zeta,\zetab)}{\zeta\zetab(\zeta-\zetab)}
		+\frac{2(2\zeta\zetab-\zeta-\zetab)}{\zeta\zetab(\zeta-\zetab)^2}\log((1-\zeta)(1-\zetab))\cr
		 -\frac{\zeta+\zetab}{\zeta\zetab(\zeta-\zetab)^2}\log(\zeta\zetab)\log((1-\zeta)(1-\zetab))-\frac{4\log(\zeta\zetab)}{(\zeta-\zetab)^2}
	\end{multline}
	\end{subequations}
	where $D(\zeta,\zetab),f_1,f_2$ and $f_3$ are given in appendix~\ref{sec:recurring_exp}.

\subsection{Sub-sub-leading term $\Delta_1=\Delta_2=\Delta_3=\Delta_4=2$ }\label{sec:delta2_app}
The sub-sub-leading term is given by the correlation function of the
$\Delta=2$ scalar field in EAdS, with only the divergent part
contributing. The parametric representation is given

\noindent $\bullet$ For the $s$-channel \begin{multline}
		\label{eq:W2div_param}
		\cW_{1,\mathrm{div}}^{2222,4-2\epsilon,s} =\frac{4\pi^{4-2\epsilon}(\zeta\zetab)^2}{16\Gamma(-2\epsilon)x_{12}^4x_{34}^4}
		\int\limits_{(\mathbb{RP}^+)^4}\prod\limits_{i=1}^5\dd\alpha_i
                \frac{\alpha_3^{-1-2\epsilon}\alpha_1^{-2\epsilon}\alpha_5(U^s)^{-1-\epsilon}}{\left(F^s\right)^{3-2\epsilon}}\cr\times
                F_s(\zeta,\zetab,\epsilon;\alpha_1,\alpha_2,\alpha_3,\alpha_4,\alpha_5),
              \end{multline}
              with  $U^s$ and $F^s$ given in~\eqref{e:UFt}.
 
\noindent $\bullet$ For the $t$-channel         \begin{multline}
		\cW_{1,\mathrm{div}}^{2222,4-2\epsilon,t} =\frac{4\pi^{4-2\epsilon}(\zeta\zetab)^2}{16\Gamma(-2\epsilon)x_{12}^4x_{34}^4}
		\int\limits_{(\mathbb{RP}^+)^4}\prod\limits_{i=1}^5\dd\alpha_i
                \frac{\alpha_2^{-1-2\epsilon}\alpha_3^{-2\epsilon}\alpha_5(U^t)^{-1-\epsilon}}{\left(F^t\right)^{3-2\epsilon}}\nonumber\\*
		 \times F_t(\zeta,\zetab,\epsilon;\alpha_1,\alpha_2,\alpha_3,\alpha_4,\alpha_5),
                   \end{multline}
 with $U^t$ and $F^t$ given in~\eqref{e:UFt}.\\
\noindent $\bullet$  For the $u$-channel          \begin{multline}
		\cW_{1,\mathrm{div}}^{2222,4-2\epsilon,u} =\frac{4\pi^{4-2\epsilon}(\zeta\zetab)^2}{16\Gamma(-2\epsilon)x_{12}^4x_{34}^4}
		\int\limits_{(\mathbb{RP}^+)^4}\prod\limits_{i=1}^5\dd\alpha_i \frac{\alpha_1^{-1-2\epsilon}\alpha_4^{-2\epsilon}\alpha_5(U^u)^{-1-\epsilon}}{\left(F^u\right)^{3-2\epsilon}}\cr
		 \times F_u(\zeta,\zetab,\epsilon;\alpha_1,\alpha_2,\alpha_3,\alpha_4,\alpha_5),
               \end{multline}
with                $U^u$ and $F^u$ given in~\eqref{e:UFu}.

The expansion of the prefactors starts at $\Op(\epsilon)$ so only integrals that diverge at least with $\epsilon^{-1}$ contribute to the final result. When only keeping those terms, the functions $F_s,F_t$ and $F_u$ are given by:
\begin{multline}
	F_s =C_1(\alpha_1^2\alpha_2\alpha_4\alpha_5^2+2\alpha_1\alpha_2\alpha_3\alpha_4\alpha_5^2+\alpha_2\alpha_3^2\alpha_4\alpha_5^2)\cr
	 +C_2(\alpha_1\alpha_2^2\alpha_4\alpha_5^2+\alpha_2^2\alpha_3\alpha_4\alpha_5^2+\alpha_1^2\alpha_2^2\alpha_4\alpha_5)+C_3(\alpha_1^2\alpha_4^2\alpha_5^2+2\alpha_1\alpha_3\alpha_4^2\alpha_5^2+\alpha_3^2\alpha_4^2\alpha_5^2)\cr
	 +C_4(\alpha_1^2\alpha_2\alpha_4^2\alpha_5+\alpha_1\alpha_2\alpha_4^2\alpha_5^2+\alpha_2\alpha_3\alpha_4^2\alpha_5^2)+C_5(2\alpha_1\alpha_2^2\alpha_4^2\alpha_5+\alpha_2^2\alpha_4^2\alpha_5^2+\alpha_1^2\alpha_2^2\alpha_4^2)\cr
	 +\alpha_1^2\alpha_2^2\alpha_5^2+2\alpha_1\alpha_2^2\alpha_3\alpha_5^2+\alpha_2^2\alpha_3^2\alpha_5^2
        \end{multline}
          \begin{multline}
	F_t =C_1(\alpha_1^2\alpha_3^2\alpha_4\alpha_5+\alpha_1\alpha_2^2\alpha_4\alpha_5^2+2\alpha_1\alpha_2\alpha_3\alpha_4\alpha_5^2+\alpha_1\alpha_3^2\alpha_4\alpha_5^2)+C_2(\alpha_1^2\alpha_2\alpha_4\alpha_5^2+\alpha_1^2\alpha_3\alpha_4\alpha_5^2)\cr
	 +C_3(\alpha_2^2\alpha_4^2\alpha_5^2+2\alpha_2\alpha_3\alpha_4^2\alpha_5^2+\alpha_3^2\alpha_4^2\alpha_5^2+\alpha_1^2\alpha_3^2\alpha_4^2+2\alpha_1\alpha_3^2\alpha_4^2\alpha_5)\cr
	 +C_4(\alpha_1^2\alpha_3\alpha_4^2\alpha_5+\alpha_1\alpha_2\alpha_4^2\alpha_5^2+\alpha_1\alpha_3\alpha_4^2\alpha_5^2)+C_5\alpha_1^2\alpha_4^2\alpha_5^2\cr
	 +\alpha_1^2\alpha_2^2\alpha_5^2+2\alpha_1^2\alpha_2\alpha_3\alpha_5^2+\alpha_1^2\alpha_3^2\alpha_5^2
           \end{multline}
          \begin{multline}
	F_u =C_1(\alpha_1^2\alpha_2\alpha_3\alpha_5+2\alpha_1\alpha_2\alpha_3\alpha_4\alpha_5^2+\alpha_2^2\alpha_3\alpha_4^2\alpha_5+\alpha_2\alpha_3\alpha_4^2\alpha_5^2)\cr
	 +C_2(\alpha_1\alpha_2\alpha_3^2\alpha_5^2+\alpha_2^2\alpha_3^2\alpha_4\alpha_5+\alpha_2\alpha_3^2\alpha_4\alpha_5^2)+C_3(\alpha_1^2\alpha_2^2\alpha_5^2+2\alpha_1\alpha_2^2\alpha_4\alpha_5^2+\alpha_2^2\alpha_4^2\alpha_5^2)\cr
	 +C_4(\alpha_1\alpha_2^2\alpha_3\alpha_5^2+\alpha_2^2\alpha_3\alpha_4\alpha_5^2)+C_5\alpha_2^2\alpha_3^2\alpha_5^2+\alpha_1^2\alpha_3^2\alpha_5^2+2\alpha_1\alpha_3^2\alpha_4\alpha_5^2+\alpha_2^2\alpha_3^2\alpha_4^2\cr
	 +2\alpha_2\alpha_3^2\alpha_4^2\alpha_5+\alpha_3^2\alpha_4^2\alpha_5^2
\end{multline}
with the coefficients $C_i$ given by
\begin{align}
	\label{eq:Ccoefficients}
	C_1&= (1-6 \epsilon ) (\zeta +\zetab-\zeta  \zetab)+8 \epsilon -2,\\
	C_2&=-(1-6\epsilon)\zeta\zetab-1+2\epsilon,\cr
	C_3&=(4 \zeta  \zetab \epsilon ^2-4 \epsilon ^2 (\zeta +\zetab)+8 \epsilon ^2-4 \epsilon +1)(1-\zeta)(1-\zetab),\cr
	C_4&=8 \zeta ^2 \zetab^2 \epsilon ^2-8 \zeta  \zetab \epsilon ^2 (\zeta +\zetab)+\zeta  \zetab \left(8 \epsilon ^2+4 \epsilon -2\right)+(1-2 \epsilon ) (\zeta +\zetab)+2 \epsilon -1,\cr
\nonumber	C_5&=4 \zeta ^2 \zetab^2 \epsilon ^2+\zeta  \zetab \left(4 \epsilon ^2-4 \epsilon +1\right).
\end{align}
Integrating over the Feynman parameters we obtain the result for the
channels $i=s,t,u$
\begin{equation}
	\label{eq:Delta2_result_L0}
	\cW_1^{2222,4-2\epsilon,i}=-\frac{\pi^2}{\epsilon}\cW_0^{2222,4-4\epsilon}+3\pi^2\cW_0^{2222,4}+\frac{3\pi^4}{8x_{12}^4x_{34}^4} L_0^{2,i}+\frac12\cW_{\mathrm{fin}}^{2222,i}+\Op(\epsilon^2)
\end{equation}
where $L_0^{2,i}$ and $\cW_{\mathrm{fin}}^{2,4,i}$ are known from EAdS calculations and given in~\cite{Heckelbacher:2022fbx}. After a minimal substraction scheme, i.e. subtracting the term 
$-\frac{\pi^2}{\epsilon}\cW_0^{2,4-4\epsilon}$ the remaining finite piece is given by
\begin{equation}	\label{eq:Delta2_result_L0fin}
    \cW_{1,\mathrm{finite}}^{2222,i}=3\pi^2\cW_0^{2222,4}+\frac{3\pi^4}{8x_{12}^4x_{34}^4} L_0^{2,i}+\frac12\cW_{\mathrm{fin}}^{2222,i}\,.
  \end{equation}
  where $\cW_0^{2222,4}$ is given in~\eqref{eq:crossDelta2},  the
  contributions $\cW_{\mathrm{fin}}^{2222,i}$ were denoted
  $\cW_{\mathrm{fin}}^{2,4,i}$ and  evaluated in appendix~C.1.2
  of~\cite{Heckelbacher:2022fbx} and $  L_0^{2,i}$  were evaluated in appendix~C.1.3
  of~\cite{Heckelbacher:2022fbx}.
  We recall the results here for completeness

  	{\footnotesize\begin{align}\label{e:W2fin}
			\cW_{1,\mathrm{fin}}^{2222,s}&=\frac{\pi^4}{8}\frac{(\zeta\zetab)^2}{(x_{12}x_{34})^4}\left(\frac{(\zeta+\zetab-2)8iD(\zeta,\zetab)}{(\zeta-\zetab)^3}+\frac{(4\zeta-2)\zetab-2\zeta}{\zeta\zetab(\zeta-\zetab)^2}\log((1-\zeta)(1-\zetab))
			-\frac{4\log(\zeta\zetab)}{(\zeta-\zetab)^2}\right)\cr
			\cW_{1,\mathrm{fin}}^{2222,t}&=\frac{\pi^4}{8}\frac{(\zeta\zetab)^2}{(x_{12}x_{34})^4}\left(-\frac{(\zeta+\zetab)8iD(\zeta,\zetab)}{(\zeta-\zetab)^3}+\frac{(4\zeta-2)\zetab-2\zeta}{(1-\zeta)(1-\zetab)(\zeta-\zetab)^2}\log(\zeta\zetab)
			-\frac{4\log((1-\zeta)(1-\zetab))}{(\zeta-\zetab)^2}\right)\cr
                                                       \cW_{1,\mathrm{fin}}^{2222,u}&=\frac{\pi^4}{8}\frac{(\zeta\zetab)^2}{(x_{12}x_{34})^4}\left(-\frac{((4\zeta-2)\zetab-2\zeta)4iD(\zeta,\zetab)}{(\zeta-\zetab)^3}+\frac{2(\zeta+\zetab)}{(\zeta-\zetab)^2}\log(\zeta\zetab)\right.\cr
                                                                                      &\left.-\frac{2(\zeta+\zetab-2)\log((1-\zeta)(1-\zetab))}{(\zeta-\zetab)^2}\right)
	\end{align}}

{\footnotesize\begin{multline}\label{e:L02}
	L_0^{2,s}(\zeta,\bar\zeta)\cdot(\zeta-\bar\zeta)^5= \left(\left( \zeta+\bar\zeta \right) ^{2}-3\, \left( \zeta+\bar\zeta
	\right) \zeta\,\bar\zeta+2\,\zeta\,\bar\zeta\right) f_{3}(\zeta,\bar\zeta)\cr
	+ \left(- \left( \zeta+\bar\zeta \right) ^{3}+2\, \left( \zeta+\bar\zeta
	\right) ^{2}\zeta\,\bar\zeta+2\, \left( \zeta+\bar\zeta \right) ^
	{2}-8\, \left( \zeta+\bar\zeta \right) \zeta\,\bar\zeta+4\,{\zeta}
	^{2}{\bar\zeta}^{2}+4\,\zeta\,\bar\zeta
	\right) f_{1}(\zeta,\bar\zeta)\cr
	-2\,i \left( 2\,{\zeta}
	^{3}\bar\zeta+8\,{\zeta}^{2}{\bar\zeta}^{2}+2\,\zeta\,{\bar\zeta
	}^{3}-{\zeta}^{3}-11\,{\zeta}^{2}\bar\zeta-11\,\zeta\,{\bar\zeta}^
	{2}-{\bar\zeta}^{3}+2\,{\zeta}^{2}+8\,\zeta\,\bar\zeta+2\,{\bar\zeta}^{2}
	\right) \ln  \left( \zeta\,\bar\zeta \right) D(\zeta,\bar\zeta) \cr
	-4\,i
	\left( {\zeta}^{3}\bar\zeta+6\,{\zeta}^{2}{\bar\zeta}^{2}+\zeta\,
	{\bar\zeta}^{3}-{\zeta}^{3}-7\,{\zeta}^{2}\bar\zeta-7\,\zeta\,{\bar\zeta}^{2}-{\bar\zeta}^{3}+2\,{\zeta}^{2}+4\,\zeta\,\bar\zeta
	+2\,{\bar\zeta}^{2} \right) D(\zeta,\bar\zeta)\cr
	-2\, \left( \zeta-\bar\zeta \right) \zeta\,\bar\zeta\, \left( \zeta+\bar\zeta-2
	\right) \ln  \left( \zeta\,\bar\zeta \right) \cr
	+ \left( 2\,\zeta\,\bar\zeta-\zeta-\bar\zeta \right)  \left( \zeta-\bar\zeta
	\right)  \left( \zeta+\bar\zeta-2 \right) \ln  \left(  \left( -1+
	\zeta \right)  \left( -1+\bar\zeta \right)  \right) +2\, \left( 
	\zeta-\bar\zeta \right) ^{3}
    \end{multline}}
  {\footnotesize
\begin{multline}
	L_0^{2,t}(\zeta,\bar\zeta)\cdot(\zeta-\bar\zeta)^5=\left(  \left( 3\,\zeta-2 \right) {\bar\zeta}^{2}+ \left( 3\,{\zeta
	}^{2}-8\,\zeta+3 \right) \bar\zeta-2\,{\zeta}^{2}+3\,\zeta \right)
	f_4 (\zeta,\bar\zeta)\cr
	+ \left(  \left( 2\,\zeta-1 \right) {\bar\zeta}^{3}+ \left( 8
	\,{\zeta}^{2}-11\,\zeta+2 \right) {\bar\zeta}^{2}+ \left( 2\,{\zeta}
	^{3}-11\,{\zeta}^{2}+8\,\zeta \right) \bar\zeta-{\zeta}^{3}+2\,{
		\zeta}^{2} \right) f_{1}(\zeta,\bar\zeta)\cr
	+  2\,i\big(  \left( -2\zeta+1
	\right) {\bar\zeta}^{3}-\left( 8\,{\zeta}^{2}-11\,\zeta+2
	\right) {\bar\zeta}^{2}-\left( 2\,{\zeta}^{3}-11\,{\zeta}^{2}+
	8\,\zeta \right) \bar\zeta\cr
	+\left( -2+\zeta \right) {\zeta}^{2}
	\big) \ln  \left(  \left( 1-\zeta \right)  \left( 1-\bar\zeta
	\right)  \right)  D(\zeta,\bar\zeta)\cr
	-4\,i\left(\zeta\,{\bar\zeta}^{3}+\left( 6\,{\zeta}
	^{2}-8\,\zeta+1 \right) {\bar\zeta}^{2}+\zeta\, \left( {\zeta}^{
		2}-8\,\zeta+6 \right) \bar\zeta+{\zeta}^{2} \right) D(\zeta,\bar\zeta)\cr
	- \left( 2\,\zeta\,\bar\zeta-\zeta-\bar\zeta \right)  \left( {
		\zeta}^{2}-{\bar\zeta}^{2} \right) \ln  \left( \zeta\,\bar\zeta
	\right) +2\, \left( 1-\bar\zeta \right)  \left( 1-\zeta \right) 
	\left( \zeta-\bar\zeta \right)  \left( \zeta+\bar\zeta \right) 
	\ln  \left(  \left( 1-\zeta \right)  \left( 1-\bar\zeta \right) 
	\right) \cr
	+2\, \left( \zeta-\bar\zeta \right) ^{3}
      \end{multline}}

    {\footnotesize
\begin{multline}
	L_0^{2,u}(\zeta,\bar\zeta)\cdot(\zeta-\bar\zeta)^5= \left( {\zeta}^{2}+4\,\zeta\,\bar\zeta+{\bar\zeta}^{2}-3\,\zeta-3
	\,\bar\zeta \right) f_{5}(\zeta,\bar\zeta)\cr
	+ \left( 2\,{\zeta}^{3}\bar\zeta+8\,{
		\zeta}^{2}{\bar\zeta}^{2}+2\,\zeta\,{\bar\zeta}^{3}-{\zeta}^{3}-11
	\,{\zeta}^{2}\bar\zeta-11\,\zeta\,{\bar\zeta}^{2}-{\bar\zeta}^{3
	}+2\,{\zeta}^{2}+8\,\zeta\,\bar\zeta+2\,{\bar\zeta}^{2} \right)
	f_{1}(\zeta,\bar\zeta)\cr
	-4\,i \left( 2\,{\zeta}^{3}\bar\zeta+4\,{\zeta}^{2}{\bar\zeta}^{2}+2\,\zeta\,{\bar\zeta}^{3}-{\zeta}^{3}-7\,{\zeta}^{2}\bar\zeta-7\,\zeta\,{\bar\zeta}^{2}-{\bar\zeta}^{3}+{\zeta}^{2}+6
	\,\zeta\,\bar\zeta+{\bar\zeta}^{2} \right) D(\zeta,\bar\zeta)\cr
	- \left( 2
	\,\zeta\,\bar\zeta-\zeta-\bar\zeta \right)  \left( \zeta-\bar\zeta
	\right)  \left( \zeta+\bar\zeta \right) \ln  \left(
	\zeta\,\bar\zeta \right) \cr
	+ \left( 2\,\zeta\,\bar\zeta-\zeta-\bar\zeta
	\right)  \left( \zeta-\bar\zeta \right)  \left( \zeta+\bar\zeta-2
	\right) \ln  \left(  \left( 1-\zeta \right)  \left( 1-\bar\zeta
	\right)  \right) +2\, \left( \zeta-\bar\zeta \right) ^{3}
\end{multline}}
  
\section{Recurring expressions}\label{sec:recurring_exp}
In this appendix we collect the recurring expressions entering the
evaluation of the Witten diagrams. These expressions are single-valued
multiple polylogarithms.
The evaluation of the parametric form of the Witten diagram is done
using \texttt{HyperInt}~\cite{Panzer:2014caa}. We will the conventions
of this work for the multiple polylogarithms
\begin{equation}
    \mathrm{Li}_{s_1, \dots,  s_k}(x_1, \dots ,x_k):=\sum\limits_{0<p_1 <\cdots <p_k}^\infty\frac{x_1^{p_1}}{p_1^{s_1}} \cdots \frac{x_k^{p_k}}{p_k^{s_k}}\quad\text{for }\abs{x_1\cdots x_i}<1,\quad\forall i\in\{1,..,k\}\,.
\end{equation}
The sum $s_1+s_2+ \dots +s_k$ is referred to as the weight of the multiple polylogarithm.

Some useful definitions and identities are
\begin{align}
    \Mpl{1}{x}&=-\log(1-x)\,,\\
    \Mpl{1,1}{y,x}&=\Mpl{2}{\frac{x(y-1)}{1-x}}-\Mpl{2}{\frac{x}{x-1}}-\Mpl{2}{xy}\,,
\end{align}
and the (single-valued) Bloch-Wigner dilogarithm given by:
	\begin{equation}
		\label{eq:BlochWigner}
		D(\zeta,\zetab)=\frac{1}{2i}\left(\Mpl{2}{\zeta}-\Mpl{2}{\zetab}-\frac12\log(\zeta\zetab)\left(\Mpl{1}{\zeta}-\Mpl{1}{\zetab}\right)\right).
	\end{equation}
Some other recurring expressions are weight 3 single-valued multiple-polylogarithms
\begin{multline}\label{e:f1}
    f_1(\zeta,\zetab)=\Mpl{3}{\zeta}-\Mpl{3}{\bar\zeta}+
		\Mpl{2,1}{1,\zeta}-\Mpl{2,1}{1,\bar\zeta}\cr
		+2\,\Mpl{2,1}{\zeta,{\frac {\bar\zeta}{\zeta}}}-2\,
		\Mpl{2,1}{\bar\zeta,{\frac {\zeta}{\bar\zeta}}}+
		\Mpl{1,2}{\zeta,{\frac {\bar\zeta}{\zeta}}}
		-\Mpl{1,2}{\bar\zeta,{\frac {\zeta}{\bar\zeta}}}\cr
		-2\,\Mpl{1}{{\frac {\bar\zeta}{\zeta}}}\Mpl{2}{\zeta}-
		\Mpl{2}{{\frac {\bar\zeta}{\zeta}}}\Mpl{1}{\zeta}+2\,
		\Mpl{1}{{\frac {\zeta}{\bar\zeta}}}\Mpl{2}{\bar\zeta}
		+\Mpl{1}{\bar\zeta}\Mpl{2}{{\frac
                    {\zeta}{\bar\zeta}}}\cr
+                \log(\zeta\zetab)\left(\Mpl{1,1}{\bar\zeta,{\zeta\over\bar\zeta}}-\Mpl{1,1}{\zeta,{\bar\zeta\over\zeta}}
		+\Mpl{1}{\zeta}\Mpl{1}{{\bar\zeta\over\zeta}}-\Mpl{1}{\bar\zeta}\Mpl{1}{{\zeta\over\bar\zeta}}\right),
  \end{multline}
  \begin{multline}
    \label{eq:Omega2}
		f_2(\zeta,\zetab)=\frac{1}{2}\left(\Mpl{2}{\zeta} \Mpl{1}{\zetab}-\Mpl{2}{\zetab} \Mpl{1}{\zeta}\right)+\Mpl{1,2}{1,\zeta}-\Mpl{1,2}{1,\zetab}\cr
		+\frac{1}{2}\left(\Mpl{2,1}{1,\zeta}-\Mpl{2,1}{1,\zetab}\right)+\frac{1}{2}\log\left(\zeta\zetab\right)\left(-\Mpl{1,1}{1,\zeta}+\Mpl{1,1}{1,\zetab}\right)
\end{multline}
  \begin{multline}                
		f_3(\zeta,\zetab)=-(\Mpl{1,1}{1,\zeta}+\Mpl{1,1}{1,\zetab})+\frac12\left(\log(\zeta\zetab)+4\right)\log((1-\zeta)(1-\zetab))\cr
		-\log(1-\zeta)\log(1-\zetab)
              \end{multline}
                   \begin{equation} 	f_4(\zeta,\zetab)=
               -4i{\zeta+\bar\zeta\over\zeta-\bar\zeta}D(\zeta,\bar\zeta)-\log((1-\zeta)(1-\bar\zeta))\log\left((1-\zeta)(1-\bar\zeta)\over\zeta\bar\zeta\right)
             \end{equation}
             \begin{equation}
		f_5(\zeta,\zetab)={4i
			(\zeta+\bar  \zeta
			-2\zeta\bar\zeta)\over  \zeta -\bar
			\zeta }
		D(\zeta,\bar\zeta)-\log(\zeta\bar\zeta)\log((1-\zeta)(1-\bar\zeta)),
	\end{equation}
For a detailed discussion of these functions and their properties we refer the interested reader to~\cite{Zagier:2007knq,goncharov2001multiple,Ablinger:2011te,Remiddi:1999ew}.

\section{OPE coefficients and conformal blocks}
\label{sec:OPE_CB}
The squared OPE coefficients for a canonically normalized double trace operator $[\Op_i\Op_j]_{n,l}$ in an OPE between $\Op_i$ and $\Op_j$ for a generalized field has been calculated in~\cite{Fitzpatrick:2011dm} and is given by
{\footnotesize\begin{align}
    \label{OPE_coeffs_GFF}
    A^{i,j}_{[\Op_i\Op_j]_{n,l}}&=\frac{(-1)^l\left(\Delta_i-\frac{d}{2}+1\right)_n\left(\Delta_j-\frac{d}{2}+1\right)_n(\Delta_i)_{l+n}(\Delta_j)_{l+n}}
    {l!n!\left(l+\frac{d}{2}\right)_n(\Delta_i+\Delta_j+n-d+1)_n(\Delta_i+\Delta_j+2n+l-1)_l\left(\Delta_i+\Delta_j+n+l-\frac{d}{2}\right)_n}\,,
\end{align}}
where $(x)_n:=\frac{\Gamma(x+n)}{\Gamma(x)}$ is the Pochhammer symbol.
The conformal block for a multiplet of dimension $\Delta$ and spin $l$ in a four-point function with external dimensions $\Delta_1,\Delta_2,\Delta_3$ and $\Delta_4$ in $d=3$ space-time dimensions has been calculated in~\cite{Li:2019cwm} and is given by
\begin{equation}
	    G^{a,b}_{\Delta,l}(\tilde v,\tilde Y)=\sum\limits_{k=0}^\infty \tilde v^{\frac{\Delta-l}{2}+k}\sum\limits_{m=0}^{2k}A_{k,m}f_{k,m}(\tilde Y),
	\end{equation}
	with
	\begin{equation}
	    f_{k,m}(\tilde Y)=\tilde Y^{l-m}
	    {}_2F_1\left({\Delta+l\over2}+k-m-a,{\Delta+l\over2}+k-m+b, \Delta+l+2k-2m;\tilde Y\right),
	    \end{equation}
	    and
	    \begin{multline}
	    A_{k,m}(\Delta)=\sum\limits_{m_1,m_2=0}^{\lfloor\frac{m}{2}\rfloor}(-1)^{m+m_1+1}4^{m_1+m_2}
	    \frac{(-l)_m(-\lfloor m/2\rfloor))_{m_1+m_2}(k-\lfloor m/2\rfloor)+1/2)_{m_1}}{m! m_1! m_2!(k-m+m_1)!}\cr
	    \times\frac{(\Delta-1)_{2k-m}(3/2-\Delta)_{m-k-m_1-m_2}(l-\Delta+2)_{2(\lfloor m/2\rfloor-m_2)-n}}{(\Delta+l-m-1)_{2k-m}(\Delta+l)_{2(k+m1-\lfloor m/2\rfloor)-m}}\cr
	    \times\prod\limits_{\alpha\in\{\pm a,\pm b\}}\left(\left(\frac12(\Delta+l)+\alpha\right)_{k-m+m_1}\left(\frac12(\Delta-l-1)+\alpha\right)_{m_2}\right)(1+(4ab-1)(n\text{ mod } 2)).
	\end{multline}
	where $a=\frac{\Delta_1-\Delta_2}{2}$ and $b=\frac{\Delta_3-\Delta_4}{2}$ and the conformal cross ratios are defined in a slightly different way as
	\begin{equation}
	    \tilde{v}=\frac{v}{1-Y};\qquad 1-\tilde Y=\frac{1}{1- Y}\,.
	\end{equation}
	Note that we use a slightly different normalization compared to~\cite{Li:2019cwm}.

        \bibliographystyle{JHEP}

\providecommand{\href}[2]{#2}\begingroup\raggedright\endgroup

\end{document}